%% file: main.tex
\documentclass[a4paper, 11pt]{article}

\renewcommand{\baselinestretch}{1.2}
\input{Preamble.tex}

\begin{document}

\thispagestyle{empty}
\fontsize{12pt}{20pt}
\hfill 
\vspace{13mm}
\begin{center}
{\huge Lieb-Schultz-Mattis constraints from 
\\\vspace{6pt}
stratified anomalies of modulated symmetries}
\\[13mm]
{\large Salvatore D. Pace$^{1}$ and Daniel Bulmash$^{2}$ 
}

\bigskip
{\it 
$^1$ Department of Physics, Massachusetts Institute of Technology, Cambridge, MA 02139, USA \\
$^2$ Department of Physics, United States Naval Academy, Annapolis, MD
21402, USA \\[.6em]	}

\bigskip
\today
\end{center}

\bigskip

\begin{abstract}
\noindent

We introduce stratified symmetry operators and stratified anomalies in quantum lattice systems as generalizations of onsite symmetry operators and onsite projective representations. 
A stratified symmetry operator is a symmetry operator that factorizes into mutually independent subsystem symmetry operators; its stratified anomaly is defined as the collection of anomalies associated with these subsystem operators. 
We develop a cellular chain complex formalism for stratified anomalies of internal symmetries and show that, in the presence of crystalline symmetries, they give rise to Lieb-Schultz-Mattis (LSM) constraints. 
This includes LSM anomalies and SPT-LSM theorems. 
We apply this framework to modulated $G$ symmetries, which are symmetries whose total symmetry group is ${G_\mathrm{tot} = G \rtimes G_\mathrm{s}}$, with $G_\mathrm{s}$ the crystalline symmetry group. 
Notably, a nonzero stratified anomaly within a fundamental domain of $G_\mathrm{s}$ (e.g., a unit cell) does not always imply an LSM anomaly for modulated symmetries. 
Instead, the existence of an LSM anomaly also depends on how $G_\mathrm{s}$ acts on $G$. When $G_\mathrm{s}$ is the lattice translation group, we find an explicit criterion for when a stratified anomaly causes an LSM anomaly, and classify LSM anomalies using homology groups of $G_\mathrm{s}$-invariant cellular chains.
We illustrate this through examples of exponential and dipole symmetries with stratified anomalies, both in ${(1+1)}$D and ${(2+1)}$D, and construct a stabilizer code model of a modulated SPT subject to an SPT-LSM theorem.

\end{abstract}

\vfill

\newpage

\pagenumbering{arabic}
\setcounter{page}{1}
\setcounter{footnote}{0}

{\renewcommand{\baselinestretch}{.88} \parskip=0pt
\setcounter{tocdepth}{3}
\tableofcontents}

\vspace{20pt}
\hrule width\textwidth height .8pt
\vspace{13pt}

\section{Introduction}\label{IntroductionSection}

Anomalies of global symmetries are a fundamental structural aspect of a quantum theory. They are part of the identifying information---the ``DNA''---of a quantum theory, and are independent of a specific Hamiltonian/Lagrangian presentation of the system. Anomalies can arise in both quantum field theories and lattice systems, and their consequences can be equally powerful in both settings. 

There are many inequivalent, but related, definitions of a symmetry anomaly. Under most definitions, an anomaly precludes a symmetry-preserving, short-range-entangled gapped phase---an SPT.\footnote{We define a symmetry protected topological phase (SPT) as a gapped phase with a unique, symmetric ground state on all spaces without boundary. Therefore, a gapped phase with a symmetric product state ground state is an SPT---it may (non-canonically) be called the trivial SPT, but it is nonetheless still an SPT.} This characteristic is so distinctive that it can be taken as the very definition of an anomaly, both in lattice systems and continuum field theories~\cite{CLS180204445, W181202517, TW191202817, KLW200514178, CCH211101139, CRS230509713, CPS240912220, SZJ250721267}. In this paper, we will adopt this ``obstruction to an SPT'' definition of an anomaly and consider anomalies of bosonic quantum lattice systems whose total symmetry is described by a group $G_\mathrm{tot}$. Suppose the lattice system has Hilbert space $\scrH$ and its $G_\mathrm{tot}$ symmetry operators are unitary operators ${U_{g_\mathrm{tot}}}$ with ${g_\mathrm{tot}\in G_\mathrm{tot}}$. We then define the anomaly of the $G_\mathrm{tot}$ symmetry as an equivalence class of its symmetry operators under:
\begin{enumerate}
    \item Tensoring $\scrH$ and $U_{g_\mathrm{tot}}$, respectively, by an ancilla Hilbert space ${\scrV}$ and $G_\mathrm{tot}$ symmetry operator $V_{g_\mathrm{tot}}$ that acts on $\scrV$ and admits an SPT.\footnote{The $G_\mathrm{tot}$ symmetry operator of the ancillas need not be onsite. This is especially important since $G_\mathrm{tot}$ may include crystalline symmetries which, as a matter of definition, cannot have onsite symmetry operators.} This causes
    \begin{equation}
    \begin{aligned}
        \scrH &\mapsto \scrH\otimes \scrV,\\
        U_{g_\mathrm{tot}} &\mapsto U_{g_\mathrm{tot}}\otimes V_{g_\mathrm{tot}}.
    \end{aligned}
    \end{equation}
    The ancilla Hilbert space ${\scrV}$ and $G_\mathrm{tot}$ symmetry operator $V_{g_\mathrm{tot}}$ must correspond to the Hilbert space and symmetry operators of an allowed, standalone quantum lattice system. Furthermore, the group elements of $G_\mathrm{tot}$ corresponding to crystalline symmetries of the original symmetry must also act as crystalline symmetries for the ancillas.
    \item Conjugating each $G_\mathrm{tot}$ symmetry operator ${U_{g_\mathrm{tot}}\otimes V_{g_\mathrm{tot}}}$ by the same locality-preserving unitary operator acting on ${\scrH\otimes \scrV}$.
\end{enumerate}
We interpret step 1 as stacking the original $G_\mathrm{tot}$ symmetric theory with another $G_\mathrm{tot}$ symmetric theory that admits an SPT. We interpret step 2 as a locality-preserving change of basis. We say a $G_\mathrm{tot}$ symmetry is anomaly-free iff it admits an SPT (after possibly adding ancillas). 

In this paper, we will focus on Lieb-Schultz-Mattis (LSM) anomalies, which are anomalies in the above sense that involve crystalline symmetries.\footnote{What we call LSM anomalies are sometimes referred to as generalized LSM theorems. Furthermore, we emphasize that our definition of LSM anomalies does not include filling constraints~\cite{OYA9610168, O9911137, Hastings_2005}, which are often referred to as a type of LSM theorem.} That is, they are anomalies of $G_\mathrm{tot}$ such that explicitly breaking all crystalline symmetries leads to an anomaly-free internal symmetry. We use the modifier ``LSM'' in LSM anomaly to emphasize that it involves crystalline symmetries, which are both quantitatively and qualitatively different from internal symmetries. The original LSM anomaly applied to half-integer quantum spin chains with spin rotation and lattice translation symmetries~\cite{Lieb61, Affleck:1986pq}. Since then, LSM anomalies have been found to arise in generic quantum lattice systems with various internal and crystalline symmetries~\cite{H0305505, NS0608046, CGW10083745, HHG160408591, PWJ170306882, HSH170509243, C180410122, OT180808740, SXG181000801, ET190708204, OTT200406458, YO201009244, ATM210208389, YF211008819, YGH211112097, YLO230709843, KS240102533, SSS240112281, LYZ240514929, M240514949, PLA240918113, LY241003607, LYZ251006555}. Furthermore, LSM anomalies share many similarities with 't Hooft anomalies of internal symmetries~\cite{CZB151102263, JBX170500012, AMF230800743, S230805151, PAL250702036, EHC251018689, OSE260121625}, and can match 't Hooft anomalies in the IR quantum field theory of a lattice system~\cite{CRH170503892, MT170707686, LF220705092, CS221112543, SS230702534, PKC250504684, SSZ250817115,
SZ260101191}.

The most commonly studied LSM anomalies are those in which the total symmetry group $G_{\mathrm{tot}}$ factorizes as a direct product ${G \times G_{\mathrm{s}}}$, where $G$ is the (possibly trivial) internal symmetry group and $G_{\mathrm{s}}$ is the crystalline symmetry group. However, as we review in Section~\ref{CrystallineSymmetrySection}, the total symmetry group more generally fits into the group extension
\begin{equation}\label{grp ext intro}
1\to G \to G_{\mathrm{tot}} \to G_\mathrm{s} \to 1.
\end{equation}
That is, the total symmetry group generally features an interplay between the internal and crystalline symmetries. The total symmetry group factorizes as ${G \times G_\mathrm{s}}$ when~\eqref{grp ext intro} is a trivial extension. Just as a ${G \times G_\mathrm{s}}$ symmetry may have LSM anomalies, the more general $G_{\mathrm{tot}}$ symmetry described by~\eqref{grp ext intro} may have LSM anomalies, too.

It is not well understood how the existence and characterization of LSM anomalies depend on the extension data of~\eqref{grp ext intro}. In this paper, we will investigate this dependency in one of the simplest scenarios where interplays between $G$ and $G_\mathrm{s}$ can affect LSM anomalies and lead to interesting consequences. In particular, we will explore the LSM anomalies of symmetries with total symmetry group
\begin{equation}\label{semi direc prod G tot intro}
G_\mathrm{tot} = G \rtimes G_\mathrm{s}.
\end{equation}
This corresponds to~\eqref{grp ext intro} being a split extension. The semi-direct product in~\eqref{semi direc prod G tot intro} implies that the crystalline symmetry group $G_\mathrm{s}$ acts on the internal symmetry group $G$. This action is described by a group homomorphism
\begin{equation}
\begin{aligned}
\rho \colon G_\mathrm{s}&\to \Aut(G)\\
s&\mapsto \rho_s,
\end{aligned}
\end{equation}
where $\Aut(G)$ is the automorphism group of $G$. The elements of ${G_\mathrm{tot} = G \rtimes G_\mathrm{s}}$ can be decomposed as ${(g,s)}$, with ${g\in G}$ and ${s\in G_\mathrm{s}}$, and the $G_\mathrm{tot}$ group law is ${(g_1,s_1) \cdot (g_2, s_2) = (g_1 \rho_{s_1}(g_2), \,s_1 s_2)}$. When $\rho$ is trivial,~\eqref{semi direc prod G tot intro} becomes the direct product group ${G \times G_\mathrm{s}}$.

A symmetry with total symmetry group~\eqref{semi direc prod G tot intro} and nontrivial $\rho$ is called a modulated symmetry.\footnote{We will always assume that the crystalline symmetry group $G_\mathrm{s}$ is nontrivial when discussing modulated symmetries.} The ${G \rtimes G_\mathrm{s}}$ group law implies that each internal symmetry operator $U_g$ and crystalline symmetry operator $U_{s}$ satisfies
\begin{equation}\label{s action on g op form}
U_{s}\, U_g\, U_{s}^\dag = U_{\rho_{s}(g)}.
\end{equation}
Modulated symmetries get their name from the fact that if $U_{s}$ acts by permuting degrees of freedom throughout the lattice without any additional unitary transformations, then $U_g$ must act in a spatially modulated way to be compatible with~\eqref{s action on g op form}. We will refer to symmetries whose symmetry group~\eqref{semi direc prod G tot intro} has trivial $\rho$ as uniform symmetries. Modulated symmetries have recently attracted substantial attention and have been studied in diverse contexts. This includes spontaneous symmetry breaking~\cite{GGH13085967, BHH211103668, SLN211108041, LHS220104132, YCY220108597, KS220809056, LLH221002470, LS230208499, SQG230309573, AB230401181, ACH230412911, AB240319601, PAL250702036}, SPTs~\cite{SNB180608780, MH210300008, HLL230910036, L231104962, LHY240313880, EHN240604919, YO240708786, KLH250315834, PAL250702036, KYH250702324, B250806604, SCH250909244}, filling constraints~\cite{HWP191210520,DMH200104477, BMP230816241}, Kramers–Wannier type dualities~\cite{YL240316017, CLY240605454, PDL240612962, EH240916744, SCS241104182}, slow thermalization and Hilbert space fragmentation~\cite{SRV190404266, KHN190404815, MM210810324, SLR211008302, HLR230403276, OFL230413028, SYH230708761}, gauge theory~\cite{P160405329, BB180210099, MHC180210108, SPP180700827, SS200310466, SS200400015, GLS200704904, OKM211002658, GLS220110589, OKH220401279, PW220407111, E230203747, H230502492, DCY230617121, DY231009490, EHN240110677, H240308158}, and generalized global symmetries~\cite{HYA220700854, OPH230104706, EHN231006701, DLR231013794, EHN240604919, PCS241218606}. Across all of these settings, the spatial modulation---the nontrivial $\rho$---gives rise to structures and phenomena that do not occur for ${G\times G_\mathrm{s}}$ symmetries.

\setlength{\tabcolsep}{8pt} \renewcommand{\arraystretch}{1.5} 
\begin{table}
\begin{center}
\begin{tabular}{|c|c|c|} 
\hline  
  \quad Spatial modulation ($\rho$) \quad   &  \quad  Stratified anomaly \quad  & \quad  LSM anomaly \quad  \\ 
\hhline{|=|=|=|}
Trivial  & Zero &  No \\
\hline 
Trivial  & Nonzero &  Yes \\
\hline 
Nontrivial  & Zero &  No \\
\hline 
Nontrivial  & Nonzero &  Maybe \\
\hline 
\end{tabular}
\caption{\label{LSMtable} 
Suppose the spatial lattice is a Bravais lattice and the crystalline symmetry includes lattice translations. In this case, for uniform symmetries, a nonzero stratified anomaly of an internal symmetry always leads to an LSM anomaly. For modulated symmetries, however, whether a given nonzero stratified anomaly leads to an LSM anomaly also depends on $\rho$.
}
\end{center}
\end{table} 
\renewcommand{\arraystretch}{1}

Anomalies of modulated symmetries fit into the equivariant generalized homology classification of general $G_\mathrm{tot}$ symmetry anomalies proposed in~\cite{SXG181000801}. To the best of our knowledge, however, anomalies of modulated symmetries---particularly their LSM manifestations---have received little explicit attention in the literature. The purpose of this paper is to fill this gap. In particular, we will explore a characterization of LSM anomalies based on what we call stratified anomalies and apply it to understand LSM anomalies of modulated symmetries.

Stratified anomalies are generalizations of onsite projective representations, which are ubiquitous in characterizing many LSM anomalies of uniform symmetries. Recall that onsite projective representations are based on onsite symmetry operators $\prod_{\bm{r}}U_g^{(\bm{r})}$, where $\bm{r}$ denotes a lattice site. The onsite projective representation at $\bm{r}$ is the projective representation of $U_g^{(\bm{r})}$, and it can be understood as the anomaly of the ${(0+1)}$D symmetry operator $U_g^{(\bm{r})}$. Stratified anomalies are based on what we call stratified symmetry operators, which are symmetry operators that factorize as $\prod_{i}U_g^{(\Si_i)}$ where $\{U_g^{(\Si_i)}\}$ are mutually independent subsystem symmetry operators for the subsystems $\{\Si_i\}$. The stratified anomaly is the collection of anomalies associated with these subsystem operators. We will show how and when stratified anomalies of modulated symmetries give rise to LSM anomalies. Interestingly, as summarized in Table~\ref{LSMtable}, a nonzero stratified anomaly of a modulated symmetry does not guarantee an LSM anomaly. Furthermore, as we discuss in Appendix~\ref{AHSS app}, the equivariant generalized homology classification is naturally formulated as the classification of stratified anomalies using the Atiyah-Hirzebruch spectral sequence discussed in~\cite{SXG181000801}.

\subsection{Outline}

The rest of this paper is structured as follows. There are four sections of the main text:
\begin{itemize}
\item In Section~\ref{StratifiedSymmetrySection}, we introduce in greater detail the definitions of stratified symmetry operators and stratified anomalies. This section considers internal $G$ symmetries only, and lays out much of the formal framework whose physical consequences will be explored in later sections. We consider simple examples of stratified symmetry operators and anomalies in Section~\ref{stratified syms and anomaly examples} to help contextualize our definitions and terminology. We also present a cellular chain complex description of stratified anomalies in Section~\ref{CellComplexStratAnomSec}, which is used throughout the rest of the paper. Using this, we define an equivalence relation on stratified anomalies in Section~\ref{stratAnomEquiv} induced by tensoring SPT-compatible ancillas. The corresponding equivalence classes describe anomalies of $G$ that arise from stratified anomalies.
\item In Section~\ref{CrystallineSymmetrySection}, we incorporate crystalline symmetries into the story. We first review some basic aspects of crystalline symmetries and their interplay with internal symmetries through the group extension~\eqref{grp ext intro}. We then discuss how crystalline symmetries can affect stratified anomalies. First, we show how crystalline symmetries can cause stratified anomalies of internal symmetries to give rise to LSM anomalies. Then we explain how crystalline symmetries, too, can have stratified anomalies.
\item In Section~\ref{ModulatedSymmetrySection}, we discuss stratified anomalies of modulated symmetries and their consequences. We specialize to modulated symmetries whose crystalline symmetry is lattice translations. In Section~\ref{LSMconstSection}, we discuss the LSM constraints that arise from stratified anomalies of these modulated symmetries. We derive an explicit criterion based on the cellular chain complex description from Section~\ref{CellComplexStratAnomSec} for when a stratified anomaly leads to an LSM anomaly, and discuss the related classification of stratified anomalies. When a stratified anomaly does not lead to an LSM anomaly, we show that it gives rise to an SPT-LSM theorem, which is an obstruction to a trivial SPT~\cite{L170504691, YJV170505421, LRO170509298, ET190708204, JCQ190708596, PLA240918113}. 

We demonstrate these general results in Section~\ref{ExampleSection} using simple examples of ${(1+1)}$D and ${(2+1)}$D quantum lattice systems. In ${(1+1)}$D, we show that stratified anomalies of $\Z_N$ dipole symmetries always lead to LSM anomalies. We then show that, in contrast, stratified anomalies for exponential symmetries do not always lead to LSM anomalies. In this case, we classify the LSM anomalies. We then present a stabilizer code SPT model for an exponential symmetry whose stratified anomaly leads to an SPT-LSM theorem rather than an LSM anomaly. In ${(2+1)}$D, we consider stratified anomalies of $\Z_N$ dipole symmetry and show that, unlike its ${(1+1)}$D counterpart, its stratified anomalies can, but do not always lead to an LSM anomaly.
\item In Section~\ref{OutlookSection}, we provide an outlook on our results, highlighting some interesting follow-up directions that arise from our work.
\end{itemize}
There are also four Appendix sections of the paper:
\begin{itemize}
\item In Appendix~\ref{AHSS app}, we discuss how stratified anomalies fit into the Atiyah-Hirzebruch spectral sequence for the equivariant generalized homology classification of anomalies~\cite{SXG181000801}. This mathematical machinery is especially useful for stratified anomalies of more general modulated symmetries than those discussed in Section~\ref{ModulatedSymmetrySection}, as well as stratified anomalies of more general $G_\mathrm{tot}$ symmetries. As a demonstration, we use it in Appendix~\ref{pmSpaceGrpEx} to classify stratified anomalies of modulated symmetries in ${(1+1)}$D whose crystalline symmetry group is the $pm$ space group---the group of one-dimensional lattice translations and reflections.
\item In Appendix~\ref{cohomologyApp}, we review basic aspects of group cohomology. We particularly focus on reviewing restriction and corestriction maps of group cohomologies, which are used in Appendix~\ref{AHSS app}.
\item In Appendix~\ref{def net mod spt app}, we review the real-space construction of bosonic modulated SPTs~\cite{B250806604}. This is used in Section~\ref{ModulatedSymmetrySection} to formulate the LSM anomaly criterion for stratified anomalies.
\item In Appendix~\ref{ExpSPTModelDerivation}, we derive the Hamiltonian for the exponential SPT subject to an SPT-LSM theorem discussed in Section~\ref{ModulatedSymmetrySection}. This is done by gauging an exponential symmetry and constructing a related, generalized Kennedy-Tasaki transformation.
\end{itemize}

\section{Stratified symmetries and anomalies}\label{StratifiedSymmetrySection}

Consider a ${(d+1)}$D bosonic quantum lattice system whose spatial lattice $\La$ forms a cellulation of a path-connected $d$-dimensional topological space $X_d$. We assume the system's Hilbert space $\scrH$ satisfies the tensor product factorization 
\begin{equation}\label{tensor product Hilb main}
    \scrH = \bigotimes_{c \in \La} \scrH_{c},
\end{equation}
where $\scrH_{c}$ is a finite-dimensional local Hilbert space associated to the cell $c$ of $\La$.\footnote{Quantum lattice systems on infinite lattices are most precisely formulated using the algebra of local operators constructed from the onsite Hilbert spaces $\scrH_c$ of~\eqref{tensor product Hilb main}~\cite{N13112717}. In this formulation, symmetries are described by automorphisms of this algebra. In contrast, we will be imprecise when working with infinite-size systems and refer to the Hilbert space~\eqref{tensor product Hilb main} and describe symmetries using unitary operators acting on~\eqref{tensor product Hilb main}. We do not expect this mathematical imprecision to affect any of our results.} A unitary $G$ symmetry is described by unitary operators ${\{U_g\}_{g\in G}}$ that act on $\scrH$, transform every local operator to another local operator (at a possible different position in space), and satisfy the $G$ group law up to a possible U(1) phase:
\begin{equation}\label{U G grp law}
    U_{g_1} U_{g_2} U_{g_1 g_2}^\dag \in \Uone.
\end{equation}
A Hamiltonian $H$ acting on $\scrH$ is $G$ symmetric iff it commutes with each $U_g$. Importantly, symmetries can be discussed and explored directly using their symmetry operators, without any reference to a specific symmetric Hamiltonian.

In this section, we assume that the $G$ symmetry is an internal symmetry and, unless specified otherwise, that $G$ is a finite group for simplicity. In what follows, we introduce the concept of stratified symmetry operators and stratified anomalies. As we will see, a stratified symmetry operator is a symmetry operator that factorizes into mutually independent subsystem symmetry operators, and a stratified anomaly is a collection of anomalies of these subsystem symmetry operators. 
We summarize the terminology introduced in this Section in Table~\ref{TerminologyTable}.
As we discuss in Sections~\ref{CrystallineSymmetrySection} and~\ref{ModulatedSymmetrySection}, stratified anomalies of internal symmetries can give rise to LSM anomalies.

\setlength{\tabcolsep}{8pt} \renewcommand{\arraystretch}{1.5} 
\begin{table}
\begin{center}
\begin{tabular}{|c||c|c|c|c|} 
\hline  
  Object  
  & $\prod_{i\in I}U_g^{(\Si_i)}$ 
  & $U_g^{(\Si_i)}$  
  & $\{[\nu_{\Si_i}]\}_{i\in I}$
  & $[\nu_{\Si_i}]$
  \\ 
\hline
Name  
& Stratified operator 
&  $d_i$-stratum operator
& Stratified anomaly 
&  $d_i$-stratum anomaly
\\
\hline 
\end{tabular}
\caption{\label{TerminologyTable} 
Terminology introduced in Section~\ref{StratifiedSymmetrySection} with $d_i$ the spatial dimension of the subsystem $\Si_i$. We will sometimes call $d_i$-stratum operators and $d_i$-stratum anomalies just stratum operators and stratum anomalies.
}
\end{center}
\end{table} 
\renewcommand{\arraystretch}{1}

\subsection{Stratified symmetry operators}

To motivate stratified symmetry operators, consider the onsite $G$ symmetry operator
\begin{equation}\label{Ug onsite}
    U_g  = \prod_{c \in \La} U_g^{(c)},
\end{equation}
where each $U_g^{(c)}$ is a unitary operator acting nontrivially only on $\scrH_{c}$ and satisfying ${U_{g_1}^{(c)}U_{g_2}^{(c)} = U_{g_1g_2}^{(c)}}$. There is a fruitful way to interpret the onsite symmetry operator~\eqref{Ug onsite}. Each local $G$ symmetry operator ${\{U_g^{(c)}\}_{g\in G}}$ in~\eqref{Ug onsite} corresponds to a $G$ subsystem symmetry operator acting on  $\scrH_{c}$.\footnote{See \Rf{M220403045} for a review of subsystem symmetries.} All together, the subsystem symmetry operators ${\{U_g^{(c)}\}_{g\in G, c\in\La}}$ faithfully represent the group $\prod_{c\in \La} G_c$: the element ${(\,g_c\mid c\in\La)\in\prod_{c\in \La} G_c}$ is represented by $\prod_{c \in \La} U_{g_c}^{(c)}$. The onsite symmetry operator~\eqref{Ug onsite} then represents the diagonal subgroup of ${\prod_{c\in \La} G_c}$ formed by the group elements ${(\,g\mid c\in\La)}$. This perspective gives rise to a natural generalization of onsite symmetry operators to stratified symmetry operators by using more general subsystems than single cells. We will discuss stratified symmetry operators generally here. In Section~\ref{stratified syms and anomaly examples}, we will discuss examples.

Consider a locally finite family of distinct subsystems 
\begin{equation}
    \Si = \{\Si_i\mid i\in I\},
\end{equation}
where each $\Si_i$ is a subcomplex of $\La$ and $I$ is some label set. We allow ${\Si_i\cap\Si_j\neq\emptyset}$ for ${i\neq j}$. Given these subsystems, consider the ${G_{\Si_i}\cong G}$ subsystem symmetry operators ${\{U_g^{(\Si_i)}\}_{g\in G, i\in I}}$, where each $U_g^{(\Si_i)}$ acts only on ${\bigotimes_{c\in \Si_i}\scrH_c\subset \scrH}$. In the spirit of onsite symmetry operators, we require each ${\{U_g^{(\Si_i)}\}_{g\in G}}$ to act on independent degrees of freedom. That is, if ${I(c) := \{i\in I \mid c\in \Si_i\}}$\footnote{The set $I(c)$ is always finite because we assume the family $\Si$ is locally finite.} is non-empty, then the local Hilbert space ${\scrH_c}$ must admit the factorization
\begin{equation}
    \scrH_c = \bigotimes_{i\in I(c)} \scrH_c^{(i)},    
\end{equation}
where each ${\{U_g^{(\Si_i)}\}_{g\in G}}$ acts nontrivially on at most one subspace $\scrH_c^{(i)}$. Two consequences of this are:
\begin{enumerate}
    \item ${[U_{g}^{(\Si_i)},U_{h}^{(\Si_j)}]=0}$ for all ${g,h\in G}$ when ${i\neq j}$,\\
    \item the unitary operators ${\prod_{i\in I}U_{g_i}^{(\Si_i)}}$ form a faithful representation of ${\prod_{i\in I}G_{\Si_i}}$. In other words, they are ${\prod_{i\in I}G_{\Si_i}}$ subsystem symmetry operators.
\end{enumerate}

The diagonal subgroup of ${\prod_{i\in I}G_{\Si_i}}$ is represented by 
\begin{equation}\label{stratified G op}
    U_g = \prod_{i\in I}U_{g}^{(\Si_i)}.
\end{equation}
Because each subsystem symmetry operator $U_{g}^{(\Si_i)}$ is an independent and ${G}$ symmetry operator,~\eqref{stratified G op} is a $G$ symmetry operator. We call~\eqref{stratified G op} a stratified $G$ symmetry operator with respect to $\Si$ because it decomposes a $G$ symmetry operator into a collection of independent $G$ subsystem symmetry operators.\footnote{We use the terminology "stratified symmetry operator" by analogy to a stratified space in topology, which is a topological space with a particular decomposition into subspaces called strata. We note that a given symmetry operator may be a stratified operator with respect to multiple families of subsystems.} We call the subsystem operators $U_g^{(\Si_i)}$ its stratum operators.
The onsite symmetry operator~\eqref{Ug onsite} is a stratified symmetry operator with ${\Si = \{c\mid c\in\La\}}$. The unitary~\eqref{stratified G op} is a $G$ 0-form symmetry if the union ${\bigcup_{i\in I}\Si_i}$ of all subsystems densely fills $\La$---if there exists an ${R\sim\cO(\text{lattice spacing})}$ such that every $d$‑ball of radius $R$ in $X_d$ intersects some $\Si_i$. Since we focus on 0-form symmetries in this paper, we will always assume that ${\bigcup_{i\in I}\Si_i}$ densely fills $\La$.

There are many different types of allowed subsystems ${\Si_i}$. For example, $\Si_i$ could be fractal-like, in which case ${\{U_g^{(\Si_i)}\}_{g\in G}}$ would represent a fractal subsystem symmetry~\cite{W160305182, DYB180504097, DW180604663, SNB180608780, D181202721, MLJ211002237, CDG231012894, CDY240619275}. In this paper, we will only consider $\Si_i$ that are orientable, boundaryless subcomplexes. Such $\Si_i$ can be expressed as the ${d_i\equiv \mathrm{dim}(\Si_i)}$ cycle
\begin{equation}\label{di cycle expression}
    \sum_{c_{d_i}} 
    \si_{c_{d_i}}(\Si_i)\,
    c_{d_i}\in Z_{d_i}(\La),
\end{equation}
where ${\si_{c_{d_i}}(\Si_i) = 0,\pm1}$.
Under this simplification, each stratum operator $U_{g}^{(\Si_i)}$ will correspond to a ${(d_i + 1)}$D $G$ symmetry operator. When we need to emphasize the spatial dimension of a stratum operator $U_{g}^{(\Si_i)}$, we will refer to it as a $d_i$-stratum operator. We will always assume that the dimension of each subsystem $\Si_i$ is less than the spatial dimension: ${d_i<d}$.

\subsection{Stratified anomalies}\label{strat anomalies sec}

The parent $\prod_{i\in I}G_{\Si_i}$ subsystem symmetry of a stratified $G$ symmetry operator can be anomalous. We call the anomaly of the ${G}$ subsystem symmetry represented by ${\{U_{g}^{(\Si_i)}\}_{g\in G}}$ a stratum anomaly. When we need to emphasize the spatial dimension of $\Si_i$, we will call the stratum anomaly of ${\{U_{g}^{(\Si_i)}\}_{g\in G}}$ a $d_i$-stratum anomaly. We call the collection of all stratum anomalies the stratified anomaly of the stratified operator~\eqref{stratified G op}. In this paper, we will consider only anomalies captured by group cohomology for simplicity,\footnote{We review basic aspects of group cohomology in Appendix~\ref{cohomologyApp}.} and ignore beyond-cohomology anomalies that arise in high spatial dimension~\cite{K14031467}. Then, since each stratum operator $U_{g}^{(\Si_i)}$ is a ${(d_i + 1)}$D symmetry operator, its corresponding stratum anomaly is specified by a cohomology class\footnote{Strictly speaking, this classification of anomalies of finite $G$ symmetries is for 't Hooft anomalies in the continuum. It has been proven to hold for the lattice anomalies we discuss here in ${(1+1)}$D when the $G$ symmetry operator is a QCA~\cite{KS240102533}. We are not aware of any proofs in higher dimensions, and will assume that the classification of $G$-anomalies in ${(d+1)}$D always includes a $\cH^{d+2}(G,\Uone)$ subgroup.}
\begin{equation}\label{stratum anom class sec 2}
    [\nu_{\Si_i}]\in \cH^{d_i+2}(G,\Uone).
\end{equation}
The U(1) coefficient module in~\eqref{stratum anom class sec 2} is a trivial $G$-module since we assume that the $G$ symmetry is a unitary symmetry.

Stratum anomalies are closely related to defect anomalies in quantum field theory (see~\cite{KPR250814963} for an overview). In particular, an $n$-stratum anomaly at the $n$-dimensional subsystem $\Si_i$ can be interpreted as the defect anomaly for an $n$-dimensional defect localized about $\Si_i$. The sub-Hilbert space on which the stratum operator $U_g^{(\Si_i)}$ acts encodes the defect’s internal degrees of freedom, and $U_g^{(\Si_i)}$ plays the role of the defect’s $n$-dimensional junction operator with the $G$ symmetry operator.

The stratified anomaly of a ${(d+1)}$D symmetry operator has a ${(d+2)}$D anomaly-inflow theory. It is a defect network of SPTs formed by the respective inflow theories of each $d_i$-stratum anomaly. In particular, each $G_{\Si_i}$ subsystem symmetry of~\eqref{stratified G op} contributes a ${(d_i+2)}$D SPT ${[\nu_{\Si_i}]}$ on ${\Si_i\times 
\ell\subset \La\times \ell}$, where $\ell$ points along the inflow direction. Turning on a background $G$ gauge field activates the same background $G_{\Si_i}$ gauge field within each ${\Si_i\times 
\ell}$.

A stratified anomaly need not coincide with an anomaly of the $G$ 0-form symmetry. It is only sometimes the case that a stratified anomaly causes the total symmetry to be anomalous. To determine whether a stratified anomaly leads to an actual anomaly, one must use the equivalence relation discussed in Section~\ref{IntroductionSection}. We will do this in Section~\ref{stratAnomEquiv}. An alternative approach, which we will not prove is equivalent but believe it to be, is to use the stratified anomaly-inflow theory. In particular, if the stratified anomaly-inflow theory can be made trivial under smooth deformations of the SPT defect network,\footnote{For example, under a smooth deformation of the SPT defect network, two ${(d_i+2)}$D SPTs may be canceled under parallel fusion.} then the stratified anomaly does not coincide with an anomaly of the $G$ 0-form symmetry. This implies that if the underlying topological space $X_d$ of the lattice is non-compact, e.g., ${X_d = \R^d}$, then a stratified anomaly never leads to an actual anomaly since its inflow theory can be made trivial by deforming each ${(d_i+2)}$D SPT to infinity. In other words, a stratified anomaly can cause the internal $G$ symmetry to be anomalous only on a finite lattice. In Section~\ref{stratAnomEquiv}, we will show that this result also follows from the equivalence relation definition of anomalies. We will show in Sections~\ref{CrystallineSymmetrySection} and~\ref{ModulatedSymmetrySection} that once crystalline symmetries are present, stratified anomalies of internal symmetries are much more powerful and can lead to LSM anomalies, which apply to both finite and infinite lattices.

\subsection{Examples}\label{stratified syms and anomaly examples}

Having introduced stratified symmetry operators and stratified anomalies abstractly, we now present some simple examples to highlight both concepts.

\subsubsection{\texorpdfstring{$0$}{0}-stratum anomaly}\label{0 stratum anomalies ex sec}

In this first example, we discuss a familiar class of symmetries whose operators factorize as~\eqref{Ug onsite} but furnish onsite projective representations. This will be an example of 0-stratum anomalies. 

To start, consider a lattice system whose $d$-dimensional lattice $\La$ has a quantum spin-$s_{\bm{r}}$ at each lattice site (0-cell) $\bm{r}$. We allow $s_{\bm{r}}$ to be site-dependent. The total Hilbert space is ${\scrH = \bigotimes_{\bm{r}\in\La} \C^{2s_{\bm{r}}+1}}$. Consider the ${\prod_{\bm{r}\in\La}\mathrm{SO}(3)}$ subsystem symmetry operators $\ee^{\ii \theta \,\bm{\hat{n}}\cdot \bm{S}_{\bm{r}}}$, where $\bm{S}_{\bm{r}} = (S^x_{\bm{r}},S^y_{\bm{r}},S^z_{\bm{r}})$ is the spin-$s_{\bm{r}}$ operator acting on $\C^{2s_{\bm{r}}+1}$ at site $\bm{r}$ and $\bm{\hat{n}}$ a unit 3-vector. The diagonal subgroup of ${\prod_{\bm{r}\in\La}\mathrm{SO}(3)}$ is described by the stratified SO(3) symmetry operator
\begin{equation}\label{SO(3)spinRotOp}
    U_{\theta, \bm{\hat{n}}} = \prod_{\bm{r}\in\La} \ee^{\ii \theta \,\bm{\hat{n}}\cdot \bm{S}_{\bm{r}}}.
\end{equation}
This SO(3) spin rotation symmetry operator is ubiquitous in quantum spin systems. The stratum operator $\ee^{\ii \theta \,\bm{\hat{n}}\cdot \bm{S}_{\bm{r}}}$ furnishes the spin-$s_{\bm{r}}$ representation of SO(3) and satisfies
\begin{equation}
    \ee^{2\pi\ii  \,\bm{\hat{n}}\cdot \bm{S}_{\bm{r}}} = (-1)^{2s_{\bm{r}}}.
\end{equation}
This is a linear representation of SO(3) for integer spins and a projective representation of SO(3) for half-integer spins. Therefore, if $s_{\bm{r}}$ is half-integer, then the stratum operator $\ee^{\ii \theta \,\bm{\hat{n}}\cdot \bm{S}_{\bm{r}}}$ is anomalous and the stratified SO(3) symmetry operator~\eqref{SO(3)spinRotOp} has a 0-stratum anomaly at the site $\bm{r}$.\footnote{Recall that the anomalies of a $G$ symmetry in ${(0+1)}$D are classified by the projective representations of $G$. See \cite[Section~1.1]{CS221112543} for a review.}

It is straightforward to generalize this SO(3) symmetry example to a general unitary $G$ symmetry. For a quantum lattice system with Hilbert space ${\scrH = \bigotimes_{\bm{r}\in\La}\scrH_{\bm{r}}}$, consider the $\prod_{\bm{r}\in\La}G$ subsystem symmetry operators $U_g^{(\bm{r})}$ whose diagonal symmetry is the stratified operator
\begin{equation}\label{GonSiteOp}
    U_g = \prod_{\bm{r}\in\La} U_g^{(\bm{r})}.
\end{equation}
Its 0-stratum operators satisfy
\begin{equation}\label{local G proj alg}
    U_{g_1}^{(\bm{r})}U_{g_2}^{(\bm{r})} = \nu_{\bm{r}}(g_1,g_2)\,U_{g_1g_2}^{(\bm{r})},
\end{equation}
where the 2-cocycle ${\nu_{\bm{r}}\colon G\times G\to \Uone}$ specifies a $G$ projective representation by its cohomology class ${[\nu_{\bm{r}}]\in \cH^2(G,\Uone)}$. The stratum operator at $\bm{r}$ is anomalous iff ${[\nu_{\bm{r}}]\neq [1]}$, in which case~\eqref{GonSiteOp} has a 0-stratum anomaly at $\bm{r}$. The stratified anomaly-inflow theory is a network of parallel ${(1+1)}$D SPTs labeled by $[\nu_{\bm{r}}]$ for all ${\bm{r}\in\La}$. For example, when ${d=1}$, the defect network with a boundary looks like (in space, not spacetime)
\begin{equation}\label{0-stratum anomaly ex inflow thy}
\begin{tikzpicture}[scale = 0.5, baseline = {([yshift=-.5ex]current bounding box.center)}]
\node (s1) at (0,0) {};
\node (s2) at (4,0) {};
\node (s3) at (8,0) {};
\node (s1Top) at (0,4) {};
\node (s2Top) at (4,4) {};
\node (s3Top) at (8,4) {};
\node (s1Mid) at (0,2) {};
\node (s2Mid) at (4,2) {};
\node (s3Mid) at (8,2) {};
\draw[anom, line width=0.03in] (s1) -- (s1Top);
\draw[anom, line width=0.03in] (s2) -- (s2Top);
\draw[anom, line width=0.03in] (s3) -- (s3Top);
\filldraw[black] (s1) circle (7pt);
\filldraw[black] (s2) circle (7pt);
\filldraw[black] (s3) circle (7pt);
\node[anom, left] at (s1Mid) {\normalsize $[\nu_{j-1}]$};
\node[anom,left] at (s2Mid) {\normalsize $[\nu_{j}]$};
\node[anom,left] at (s3Mid) {\normalsize $[\nu_{j+1}]$};
\node[black, below, yshift=-4pt] at (s1) {\normalsize $j-1$};
\node[black, below, yshift=-4pt] at (s2) {\normalsize $j$};
\node[black, below, yshift=-4pt] at (s3) {\normalsize $j+1$};
\node[black, right, xshift=10pt] at (s3Mid) {\normalsize $\cdots$};
\node[black, left, xshift=-30pt] at (s1Mid) {\normalsize $\cdots$};
\end{tikzpicture}.
\end{equation}
The red lines denote the ${(1+1)}$D SPTs in two-dimensional space, and the black disks denote the boundary of each defect.

Suppose the lattice $\La$ is finite. The 0-stratum anomalies can then cause the total $G$ symmetry to be anomalous. The local algebra~\eqref{local G proj alg} implies that
\begin{equation}
    U_{g_1}U_{g_2} = \left(\prod_{\,\bm{r}\in\La}\nu_{\bm{r}}(g_1,g_2)\right)\,U_{g_1g_2}.
\end{equation}
Therefore, the $G$ symmetry is anomalous if the cohomology class ${[\prod_{\,\bm{r}\in\La}\nu_{\bm{r}}]\neq [1]}$. For example, the SO(3) symmetry~\eqref{SO(3)spinRotOp} is anomalous if there is an odd number of half-integer spins. When this is the case, the total inflow theory~\eqref{0-stratum anomaly ex inflow thy} cannot be trivialized by smooth deformations of the defect network.

\subsubsection{\texorpdfstring{$1$}{1}-stratum anomaly}

We now present a simple example of a stratified symmetry operator in ${(2+1)}$D that has 1-stratum anomalies. Examples of 1-stratum anomalies in fermionic systems were also discussed in~\cite{FVM180408628, C180410122} using the ${(3+1)}$D anomaly-inflow theory.

Let $\La$ be an ${L_x\times L_y}$ square lattice on a torus with a single qubit at each site (0-cell) ${\bm{r} = (r_x,r_y)}$. The total Hilbert space is ${\scrH = \bigotimes_{\bm{r}\in\La} \scrH_{\bm{r}}}$ with ${ \scrH_{\bm{r}} = \C^2}$.  The Pauli operators ${X_{\bm{r}},Z_{\bm{r}}}$ acting on the qubit at site $\bm{r}$ satisfy periodic boundary conditions
\begin{equation}
    X_{\bm{r}} = X_{\bm{r} + L_x \hat{x}} = X_{\bm{r} + L_y \hat{y}},
    \qquad
    Z_{\bm{r}} = Z_{\bm{r} + L_x \hat{x}} = Z_{\bm{r} + L_y \hat{y}}.
\end{equation}
Consider the $\prod_{x=1}^{L_x} \Z_2$ subsystem symmetry generated by
\begin{equation}\label{subsys CZX ops}
    U^{(x)} = \prod_{y=1}^{L_y} X_{x,y}\, \prod_{y=1}^{L_y}\mathsf{CZ}_{x;y,y+1},
\end{equation}
where ${\mathsf{CZ}_{x;y_1,y_2} = \frac12(1+Z_{x,y_1}) + \frac12(1 - Z_{x,y_1}) Z_{x,y_2}}$ is a controlled-Z gate. Each operator $U^{(x)}$ is unitary, satisfies ${(U^{(x)})^2 = 1}$, and acts on the sub-Hilbert space ${\scrH_x=\bigotimes_{\bm{r}\in\Si_x}\scrH_{\bm{r}}}$ where $\Si_x$ is the one-dimensional subcomplex formed by all 0 and 1-cells of $\La$ with fixed ${r_x = x}$. 

The diagonal subgroup of this $\prod_{x=1}^{L_x} \Z_2$ subsystem symmetry is generated by the stratified $\Z_2$ symmetry operator
\begin{equation}\label{1anomZ2SymOp}
    U = \prod_{x=1}^{L_x} U^{(x)}.
\end{equation}
It satisfies ${U^2 = 1}$ and acts on the Pauli operators by
\begin{equation}
    U X_{\bm{r}} U^\dag = Z_{\bm{r}-\hat{y}}\,X_{\bm{r}}\,Z_{\bm{r}+\hat{y}},
    \qquad
    U Z_{\bm{r}} U^\dag = -Z_{\bm{r}}.
\end{equation}
The 1-stratum operator $U^{(x)}$ generates an anomalous $\Z_2$ symmetry in ${(1+1)}$D~\cite{CLW11064752, LG12023120}. Therefore, the stratified symmetry operator~\eqref{1anomZ2SymOp} has 1-stratum anomalies at each subsystem $\Si_x$.

Because ${\cH^3(\Z_2,\Uone) = \Z_2}$, the anomaly of $U^{(x)}$ is order 2. Therefore, we do not expect the 1-stratum anomalies to cause the $\Z_2$ 0-form symmetry to be anomalous when there is an even number of subsystems $\Si_x$---when $L_x$ is even. Indeed, when $L_x$ is even, the inflow theory of each 1-stratum anomaly can be pairwise canceled, leading to a trivial stratified anomaly-inflow theory. On the other hand, when $L_x$ is odd, the 1-stratum anomaly inflow theories cannot be globally canceled, and we expect the $\Z_2$ 0-form symmetry to be anomalous.

We can explicitly demonstrate that $U$ is anomaly-free when $L_x$ is even. Suppose $L_x$ is even, and consider the Hamiltonian
\begin{equation}\label{1anomZ2SPT}
   -\sum_{n=1}^{L_x/2}\sum_{y=1}^{L_y} \left(A_{(n,y)} + B_{(n,y)}(1+A_{(n,y-1)}
    A_{(n,y+1)})\right),
\end{equation}
where the mutually commuting operators
\begin{equation}
    A_{(n,y)} = Z_{(2n,y)}Z_{(2n+1,y)},
    \qquad
    B_{(n,y)} = X_{(2n,y)}X_{(2n+1,y)}.
\end{equation}
The Hamiltonian~\eqref{1anomZ2SPT} commutes with the $\Z_2$ symmetry operator~\eqref{1anomZ2SymOp} since
\begin{equation}
    U A_{(n,y)} U^\dag = A_{(n,y)},\qquad
    U B_{(n,y)} U^\dag = A_{(n,y-1)}\,B_{(n,y)}\,A_{(n,y+1)}.
\end{equation}
Furthermore, this Hamiltonian has a unique gapped ground state satisfying ${A_{(n,y)}=B_{(n,y)}=1}$. It is unique because the ${L_xL_y}$ commuting stabilizers $A_{(n,y)}$ and $B_{(n,y)}$ are independent (no nontrivial product yields the identity). Therefore,~\eqref{1anomZ2SPT} is in a $\Z_2$-SPT phase and~\eqref{1anomZ2SymOp} is anomaly-free when $L_x$ is even. 

It is straightforward to generalize this $\Z_2$ example to similar stratified $G$ symmetry operators in ${(2+1)}$D with 1-stratum anomalies. Keeping $\La$ as a finite square lattice on a torus, let $U^{(x)}_g$ be a ${(1+1)}$D ${G}$ symmetry operator acting on $\scrH_x$ with anomaly ${[\nu_{x}] \in \cH^3(G,\Uone)}$. The ${(2+1)}$D stratified $G$ symmetry operator ${U_g = \prod_{x=1}^{L_x} U_g^{(x)}}$ has a 1-stratum anomaly at $\Si_{x}$ if ${[\nu_{x}]\neq [1]}$. We expect this to cause the $G$ symmetry to be anomalous when ${[\prod_{x=1}^{L_x}\nu_{x}]\neq [1]}$. The ${(3+1)}$D inflow theory for this stratified anomaly is a network of parallel $({2+1})$D $G$ SPTs labeled by $[\nu_x]$. This defect network looks like
\begin{equation}
\begin{tikzpicture}[
    x={(1cm,0cm)},      
    y={(0.4cm,0.7cm)},  
    z={(0cm,1cm)},      
    line width=0.03in, 
    baseline = {([yshift=-.5ex]current bounding box.center)},
    scale = 1.5
]

\foreach \i in {1,...,2} {
    \foreach \j in {0,...,1} {
        \draw (\j,\i,0) -- (\j+1,\i,0);
    }
}

\foreach \i in {0,...,2} {
    \fill[fill=anom, draw=anom, fill opacity=0.15]
        (\i,0,0) -- (\i,2,0) -- (\i,2,1) -- (\i,0,1)  -- cycle;
}

\foreach \j in {0,...,1} {
    \draw (\j-.025,0,0) -- (\j+1.025,0,0);
}
\foreach \j in {0,...,1} {
    \draw[white, line width=0.04in] (\j-.025,-.085,0) -- (\j+1.045,-.085,0);
}

\foreach \i in {0,...,2} {
    \foreach \j in {0,...,1} {
        \draw (\i,\j,0) -- (\i,\j+1,0);
    }
}

\node[black, below] at (0,0) {\normalsize $x'-1$};
\node[black, below] at (1,0) {\normalsize $x'$};
\node[black, below] at (2,0) {\normalsize $x'+1$};

\node[anom, below] at (-1.25,2.75) {\normalsize $[\nu_{x'-1}]$};
\node[anom, below] at (0,2.75) {\normalsize $[\nu_{x'}]$};\node[anom, below] at (2.1,2.75) {\normalsize $[\nu_{x'+1}]$};

\end{tikzpicture}.
\end{equation}

\subsubsection{Coexisting 0 and 1-stratum anomalies}

The examples presented thus far have discussed stratified anomalies with a single degree of stratum anomaly (e.g., only 0-stratum anomalies or only 1-stratum anomalies). Different degree stratum anomalies can appear simultaneously. Such a stratified anomaly can be constructed by stacking theories with different degrees of stratum anomalies. For example, stacking a theory whose $G$ symmetry operator has $n$-stratum anomalies with a theory whose $G$ symmetry operator has ${m\neq n}$ stratum anomalies yields a symmetry operator whose stratified anomaly includes $n$-stratum and $m$-stratum anomalies.

Here, we consider a simple example of a stratified symmetry operator on a finite ${L_x\times L_y}$ square lattice $\La$ with both 0 and 1-stratum anomalies. We place three qubits at each lattice site ${\bm{r}}$, which are respectively acted on by the Pauli operators ${X_{\bm{r}}^{(i)},Z_{\bm{r}}^{(i)}}$ (${i=1,2,3}$), which satisfy periodic boundary conditions. This makes the local Hilbert space ${\scrH_{\bm{r}} = \bigotimes_{i=1}^3 \scrH_{\bm{r}}^{(i)}}$ with each ${\scrH_{\bm{r}}^{(i)} = \C^2}$.

Let $\Si_{x}$ and $\Si_{y}$ be one-dimensional subsystems respectively formed by all lattice sites with fixed ${r_x = x}$ and ${r_y = y}$, and $\Si_{(x,y)}$ is the lattice site at ${\bm{r} = (x,y)}$. Consider the family of subsystems
\begin{equation}
    \Si = \{\Si_{x}, \Si_{y}, \Si_{(x,y)}\mid \forall \,x,y\in\Z_{L_x}\times\Z_{L_y}\}.
\end{equation}
Using these, we construct the ${\prod_{x=1}^{L_x}(\Z_2\times\Z_2)\times \prod_{y=1}^{L_y}(\Z_2\times\Z_2)\times \prod_{\bm{r}\in\La}(\Z_2\times\Z_2)}$ subsystem symmetry operators
\begin{align}
    &U_{n,m}^{(x)} = \prod_{y=1}^{L_y} (X^{(1)}_{x,y})^n\,\prod_{y=1}^{L_y}(\mathsf{CZ}^{(1)}_{x;y,y+1})^m,\\
    &U_{n,m}^{(y)} = \prod_{x=1}^{L_x} (X^{(2)}_{x,y})^n \,\prod_{x=1}^{L_x}(\mathsf{CZ}^{(2)}_{x,x+1;y})^m,\\
    &U^{(\bm{r})}_{n,m} = (X^{(3)}_{\bm{r}})^n \,(Z^{(3)}_{\bm{r}})^m,
\end{align}
where ${n,m\in\{0,1\}\cong\Z_2}$ and the controlled-Z gates
\begin{align}
    \mathsf{CZ}^{(1)}_{x;y_1,y_2} &= \frac12(1+Z^{(1)}_{x,y_1}) + \frac12(1 - Z^{(1)}_{x,y_1}) Z^{(1)}_{x,y_2},\\
    \mathsf{CZ}^{(2)}_{x_1,x_2;y} &= \frac12(1+Z^{(2)}_{x_1,y}) + \frac12(1 - Z^{(2)}_{x_1,y}) Z^{(2)}_{x_2,y}.
\end{align}

The diagonal of this subsystem symmetry is the stratified ${\Z_2\times\Z_2}$ 0-form symmetry operator
\begin{equation}\label{Umn 0 and 1 stratum}
    U_{n,m} = \prod_{x=1}^{L_x} U^{(x)}_{n,m}\, \prod_{y=1}^{L_y} U^{(y)}_{n,m}\, \prod_{\bm{r}\in\La} U^{(\bm{r})}_{n,m}.
\end{equation}
Each stratum operator in $U_{n,m}$ has a stratum anomaly. Indeed, the 1-stratum operators $U^{(x)}_{n,m}$ and $U^{(y)}_{n,m}$ are well-known anomalous ${(1+1)}$D ${\Z_2\times\Z_2}$ symmetry operators, and the 0-stratum operators $U^{(\bm{r})}_{n,m}$ furnish a projective representation of ${\Z_2\times\Z_2}$. Therefore, the stratified ${\Z_2\times\Z_2}$ symmetry operator~\eqref{Umn 0 and 1 stratum} has 0 and 1-stratum anomalies. The stratified anomaly-inflow theory is formed by the defect network
\begin{equation}
\begin{tikzpicture}[
    x={(1cm,0cm)},      
    y={(0.4cm,0.7cm)},  
    z={(0cm,1cm)},      
    line width=0.03in, 
    baseline = {([yshift=-.5ex]current bounding box.center)},
    scale = 1.5
]

\foreach \i in {0,...,2} {
    \fill[fill=anom, draw=anom, fill opacity=0.15]
        (0,\i,0) -- (2,\i,0) -- (2,\i,1) -- (0,\i,1)  -- cycle;
}

\foreach \i in {1,...,2} {
    \foreach \j in {0,...,1} {
        \draw (\j,\i,0) -- (\j+1,\i,0);
    }
}

\foreach \i in {0,...,2} {
    \fill[fill=anom, draw=anom, fill opacity=0.15]
        (\i,0,0) -- (\i,2,0) -- (\i,2,1) -- (\i,0,1)  -- cycle;
}

\foreach \j in {0,...,1} {
    \draw (\j-.025,0,0) -- (\j+1.025,0,0);
}
\foreach \j in {0,...,1} {
    \draw[white, line width=0.04in] (\j-.025,-.085,0) -- (\j+1.045,-.085,0);
}

\foreach \i in {0,...,2} {
    \foreach \j in {0,...,1} {
        \draw (\i,\j,0) -- (\i,\j+1,0);
    }
}

\foreach \i in {0,1,2} {
    \foreach \j in {0,1,2} {
        \draw[spt] (\j,\i,0) -- (\j,\i,1);
    }
}

\foreach \i in {0,1,2} {
    \foreach \j in {0,1,2} {
        \fill (\i,\j) circle (2pt);
    }
}

\foreach \i in {0,...,2} {
    \foreach \j in {0,...,1} {
        \draw[anom] (\i,\j,1) -- (\i,\j+1,1);
    }
}

\foreach \i in {0,...,1} {
    \foreach \j in {0,...,2} {
        \draw[anom] (\i,\j,1) -- (\i+1,\j,1);
    }
}

\node[black, below] at (0,0) {\normalsize $x'-1$};
\node[black, below] at (1,0) {\normalsize $x'$};
\node[black, below] at (2,0) {\normalsize $x'+1$};

\node[black, right] at (2,0) {\normalsize $y'-1$};
\node[black, right] at (2,1) {\normalsize $y'$};
\node[black, right] at (2,2) {\normalsize $y'+1$};
\end{tikzpicture}.
\end{equation}
Each red plane is a ${(2+1)}$D ${\Z_2\times\Z_2}$ SPT ${(-1)^{\int a\,\smile\, \mathrm{Bock}(b)}}$ and each blue line is a ${(1+1)}$D ${\Z_2\times\Z_2}$ SPT ${(-1)^{\int a\,\smile\, b}}$, where $a$ and $b$ are $\Z_2$ gauge fields, $\smile$ is the cup product, and Bock is a Bockstein homomorphism. When both $L_x$ and $L_y$ are even, these SPTs can be trivialized via fusion, and we expect the stratified anomaly not to cause the total ${\Z_2\times\Z_2}$ 0-form symmetry to be anomalous. Otherwise, they cannot be canceled, and we expect the stratified anomaly to cause the ${\Z_2\times\Z_2}$ symmetry to be anomalous.

\subsection{Cellular chain complex formulation}\label{CellComplexStratAnomSec}

Consider an internal $G$ symmetry and corresponding stratified symmetry operator~\eqref{stratified G op}. As discussed in Section~\ref{strat anomalies sec}, for each $d_i$-dimensional subsystem $\Si_i$, there is a possible $d_i$-stratum anomaly specified by ${[\nu_{\Si_i}]\in \cH^{d_i+2}(G,\Uone)}$. 
We now discuss a description of stratified anomalies where cohomological data is assigned to the cells of the lattice $\La$ rather than the subsystems $\Si$ of the stratified operator. This is a type of cellular chain complex formulation of stratified anomalies. It will be used throughout the rest of the paper, and is how we relate stratified anomalies to the Atiyah-Hirzebruch spectral sequence of~\cite{SXG181000801} in Appendix~\ref{AHSS app}.

To repackage a stratified anomaly from data based on subsystems $\Si$ to data based on cells of the lattice $\La$, we will describe $n$-stratum anomaly data using $n$-cells of the lattice. Without a loss of generality, we assume a cellulation $\La$ of space such that each 0-stratum anomaly belongs to a 0-cell of $\La$. Consider an $n$-cell $c_n$ of $\La$ and the cohomology class\footnote{The superscript in $[\nu^{(n+2)}_{c_{n}}]$ is included to emphasize that $[\nu^{(n+2)}_{c_{n}}]$ is a degree ${n+2}$ cohomology class. It will be useful in later sections where there may be multiple cohomological classes of differing degrees assigned to each cell.} 
\begin{equation}\label{n-cell anom data}
    [\nu^{(n+2)}_{c_{n}}] := \prod_{i\in I\mid d_i = n} [\nu_{\Si_i}]^{\si_{c_n}(\Si_i)}\in \cH^{n+2}(G,\Uone),
\end{equation}
where ${\si_{c_n}(\Si_i) = 0,\pm 1}$ is the coefficient appearing in~\eqref{di cycle expression}. The cohomology class $[\nu^{(n+2)}_{c_{n}}]$ describes the total $n$-stratum anomaly that passes through $c_n$. Accordingly, we will refer to $[\nu^{(n+2)}_{c_{n}}]$ as describing the $n$-stratum anomaly at the $n$-cell $c_n$. If no $n$-dimensional subsystems contain an $n$-cell $c_n$, then ${[\nu^{(n+2)}_{c_{n}}] = [1]}$ for that $n$-cell.

Eq.~\eqref{n-cell anom data} is the key starting point for describing $n$-stratum anomalies using $n$-cells rather than $n$-dimensional subsystems. It specifies the maps (${n=0,1,\cdots, d-1}$)
\begin{equation}
\begin{aligned}
    \prod_{i\in I\mid d_i = n}\cH^{n+2}(G,\Uone) &\to \prod_{c_n \in \La} \cH^{n+2}(G,\Uone)\\
    \bigg\{[\nu_{\Si_i}]\mid i\in I, d_i = n\bigg\} &\mapsto \bigg\{\prod_{i\in I\mid d_i = n} [\nu_{\Si_i}]^{\si_{c_n}(\Si_i)}\mid c_n \in \La\bigg\}.
\end{aligned}
\end{equation}
For ${n=0}$, this is the constant map since each $\Si_i$ with ${d_i = 0}$ is assumed to be a $0$-cell.
For ${n>0}$, each map is neither injective nor surjective. The failure of injectivity means that two distinct stratified anomalies of a $G$ symmetry can lead to the same cellular description. This happens when two stratified anomalies differ for superficial reasons and lead to the same consequences. The failure of surjectivity means that not every collection of cohomology classes $\{[\nu_{c_n}^{(n+2)}]\mid c_n\in\La\}$ corresponds to an allowed stratified anomaly when ${n>0}$. For each ${(n-1)}$-cell $c_{n-1}$ with ${0\leq n-1 <d-1}$, the $n$-stratum anomaly cellular data $\{[\nu_{c_n}^{(n+2)}]\}$ must satisfy
\begin{equation}\label{higheranomCondGenTransl}
    \prod_{c_{n}}\,[\nu^{(n+2)}_{c_{n}}]^{\si_{c_{n-1}}(\pp c_n)} = [1],
\end{equation}
where ${\si_{c_{n-1}}(\pp c_n) = 0,\pm 1}$ are the coefficients appearing in ${\pp_n c_{n} = \sum_{c_{n-1}} \si_{c_{n-1}}(\pp c_n)\, c_{n-1}}$. This is automatically satisfied by~\eqref{n-cell anom data} because each $\Si_i$ is a boundaryless subcomplex of $\La$. For example, this constraint for ${n=1}$ and $\La$ a ${d=2}$ square lattice is
\begin{equation}
    \begin{tikzpicture}[scale = 0.5, baseline = {([yshift=-.5ex]current bounding box.center)}]
        \coordinate (L) at (0,0);
        \coordinate (R) at (6,0);
        \coordinate (B) at (3,-3);
        \coordinate (T) at (3,3);
        \coordinate (mid) at (3,0);
        \coordinate (midL) at (1.5,0);
        \coordinate (midR) at (4.5,0);
        \coordinate (midB) at (3,-1.5);
        \coordinate (midT) at (3,1.5);
        \draw[line width=0.015in, black] (L) -- (R);
        \draw[line width=0.015in, black] (B) -- (T);
        \fill[white] (1.125+3,0) circle (24pt);
        \node[anom, yshift = 3pt] at (midR) {\normalsize $[\nu^{(3)}_{3}]^{-1}$};
        \fill[white] (midL) circle (24pt);
        \node[anom, yshift = 3pt] at (midL) {\normalsize $[\nu^{(3)}_{1}]$};
        \fill[white] (midT) circle (18pt);
        \node[anom, xshift = +5.35pt] at (midT) {\normalsize $[\nu^{(3)}_{4}]^{-1}$};
        \fill[white] (midB) circle (18pt);
        \node[anom] at (midB) {\normalsize $[\nu^{(3)}_{2}]$};
    \end{tikzpicture}
    \equiv\, [\nu^{(3)}_{1}][\nu^{(3)}_{2}][\nu^{(3)}_{3}]^{-1}[\nu^{(3)}_{4}]^{-1} = [1],
\end{equation}
for each lattice site.

The $n$-stratum anomaly cellular data admits a useful chain complex description. The collection $\{[\nu^{(n+2)}_{c_n}]\}$ of $n$-stratum anomalies defines an ${\cH^{n+2}(G,\Uone)}$-valued cellular $n$-chain by
\begin{equation}\label{cellular n chain def}
    \nu^{(n+2)}_{n\text{-chain}} := \sum_{c_n} [\nu^{(n+2)}_{c_n}]\, c_n \in C_n(\La; \cH^{n+2}(G,\Uone)).
\end{equation}
While we have thus far used multiplicative notation for ${\cH^{n+2}(G,\Uone)}$, we will use additive notation for ${\cH^{n+2}(G,\Uone)}$ as the coefficient group of $C_n(\La; \cH^{n+2}(G,\Uone))$.\footnote{In the additive notation, the condition~\eqref{higheranomCondGenTransl} is 
\begin{equation}\label{higheranomCondGenTranslADDITIVE}
    \sum_{c_{n}}\si_{c_{n-1}}(\pp c_n)\, [\nu^{(n+2)}_{c_{n}}] = 0, 
\end{equation}
where $0$ is the trivial $n$-chain in the additive notation.} The formal sum~\eqref{cellular n chain def} can be infinite but is always locally finite. The image of $\nu^{(n+2)}_{n\text{-chain}}$ under the boundary map ${\pp_n}$ is
\begin{equation}
    \pp_n \nu^{(n+2)}_{n\text{-chain}} = \sum_{c_n} \,[\nu^{(n+2)}_{c_n}]\, \pp_n c_n
    =
    \sum_{c_{n-1}} \bigg(\sum_{c_{n}}\si_{c_{n-1}}(\pp c_n)\, [\nu^{(n+2)}_{c_{n}}]\bigg) \,c_{n-1}.
\end{equation}
Comparing this to~\eqref{higheranomCondGenTransl}, which is given by~\eqref{higheranomCondGenTranslADDITIVE} in the additive notation, we see that the condition~\eqref{higheranomCondGenTransl} is
\begin{equation}\label{ncycle strat anom cond}
    \pp_n \nu^{(n+2)}_{n\text{-chain}} = 0.
\end{equation}
Therefore, the allowed cellular $n$-stratum anomalies correspond to the $n$-cycles
\begin{equation}\label{n-stratum anomalie as n-cycles}
    \ker(\pp_n) := Z_n(\La; \cH^{n+2}(G,\Uone)).
\end{equation}
From here on, we will assume that ${\nu^{(n+2)}_{n\text{-chain}}\in \ker(\pp_n)}$ unless stated otherwise.

\subsection{Stratified anomaly equivalence relation}\label{stratAnomEquiv}

The stratification structure of a symmetry operator is not an invariant property of the symmetry. For instance, the subsystems $\Si$ can be changed by permuting degrees of freedom on the lattice $\La$. Recall from Section~\ref{IntroductionSection} that the anomaly of a $G$ symmetry operator is an equivalence class of symmetry operators up to tensoring ancillas with SPT-compatible $G$ symmetry operators and performing a locality-preserving unitary transformation. This procedure can change the stratification structure of a symmetry operator. In particular, it can change its stratified anomaly. Therefore, this procedure defines an equivalence relation on stratified anomalies. Equivalent stratified anomalies lead to the same anomaly class of the symmetry operators. Here, we will identify what this equivalence relation is in the cellular chain complex formalism introduced in Section~\ref{CellComplexStratAnomSec}, and show that the classification of $n$-stratum anomalies up to this equivalence relation is given by the Borel-Moore homology group ${H_n(\La; \cH^{n+2}(G,\Uone)}$.

Consider a $G$ stratified symmetry operator $U_g$ that has the family of subsystems $\Si$ and that acts on ${\scrH = \bigotimes_{c\in\La}\scrH_c}$. The stratum anomaly at $\Si_i$ is described by $[\nu_{\Si_i}]$. The first step in the anomaly equivalence relation is to add ancillas. Without a loss of generality, we place the ancillas on the 0-cells of $\La$. The ancilla Hilbert space is then ${\scrV = \bigotimes_{c_0\in\La}\scrV_{c_0}}$. We do not require the ancillas to be placed all throughout the lattice. Suppose the $G$ symmetry operator acting on the ancilla Hilbert space is the stratified operator
\begin{equation}\label{ancilla strat operatornew}
    V_g = \prod_{i\in I_a} V_{g}^{(\Xi_i)},
\end{equation}
specified by the family of subsystems ${\Xi = \{\Xi_i\mid i\in I_a\}}$. Each $\Xi_i$ is an orientable, boundaryless subcomplex of $\La$, and $\Xi$ is not necessarily related to $\Si$ of the original stratified operator. After tensoring in these ancillas, the total symmetry operator is a $G$ stratified operator with respect to the family of subsystems ${\Si \cup \Xi}$.

The ancilla $G$ symmetry operator~\eqref{ancilla strat operatornew} must admit an SPT.
However, this does not imply that each stratum anomaly ${[\mu_{\Xi_i}]\in \cH^{\mathrm{dim}(\Xi_i)+2}(G,\Uone)}$ of $V_g$ must be trivial.
The internal $G$ symmetry operator~\eqref{ancilla strat operatornew} is anomaly-free iff its stratum anomalies can cancel among themselves.
This was suggested using the stratified anomaly-inflow theory in Section~\ref{strat anomalies sec} and observed in the examples in Section~\ref{stratified syms and anomaly examples}. It is true in general since every such stratified anomaly can be created by starting with a manifestly anomaly-free stratified symmetry with stratum operators ${U^{(\Si_i)}_g = U^{(\Si_i;1)}_g\otimes U^{(\Si_i;2)}_g}$ and stratum anomalies ${[\nu_{\Si_i}] = [\nu_{\Si_i;1}][\nu_{\Si_i;2}] = [1]}$ and then deforming the ${[\nu_{\Si_i;1}]}$ stratum anomalies elsewhere in the lattice using ancillas and locality-preserving unitaries.
The stratum anomalies can globally cancel in this way iff for each ${(d-n)}$-cycle $C^\vee_{d-n}$ on the dual lattice of $\La$, the $n$-stratum anomaly data satisfies
\begin{equation}\label{anom free gen ancillasnew}
    \prod_{i\in I_a \,\mid\, \mathrm{dim}(\Xi_i) = n} [\mu_{\Xi_i}]^{\#(\Xi_i, C^\vee_{d-n})} = [1],
\end{equation}
where ${\#(\Xi_i, C^\vee_{d-n})\in\Z}$ is the intersection number of $\Xi_i$ and $C^\vee_{d-n}$. For example, when ${n=0}$,~\eqref{anom free gen ancillasnew} becomes ${\prod_{c_0} [\mu_{c_0}] = [1]}$, which is the same condition found in example~\ref{0 stratum anomalies ex sec}. When $\Xi_i$ is contractible, a nontrivial $[\mu_{\Xi_i}]$ does not cause $V_{g}^{(\Xi_i)}$ to be anomalous as a ${(\mathrm{dim}\,3\Xi_i+1)}$D symmetry operator embedded in $d$-dimensional space, nor does it contribute to the product in~\eqref{anom free gen ancillasnew} since ${\#(\Xi_i, C^\vee_{d-n}) = 0}$ for contractible $\Xi_i$. This is because $V_{g}^{(\Xi_i)}$ for contractible $\Xi_i$ is equivalent to a ${(0+1)}$D operator, and the only way for a ${(0+1)}$D symmetry operator to be anomalous is to furnish a projective representation. This can also be argued for using the stratified anomaly-inflow theory. A contractible $\Xi_i$ contributes a homologically-trivial SPT defect embedded in the ${(d+2)}$D inflow theory. Since this SPT defect is an invertible topological defect, it can be deformed to a trivial defect. Thus, a nontrivial $[\mu_{\Xi_i}]$ for contractible $\Xi_i$ does not affect the anomaly-inflow theory and, therefore, does not give rise to an anomaly in ${(d+1)}$D.

Let us pass to the cellular chain complex formulation from Section~\ref{CellComplexStratAnomSec}. The stratified anomaly of $V_g$ is now described by the classes $\{[\mu^{(n+2)}_{c_{n}}]\}_{c_n\in\La}$. The condition~\eqref{anom free gen ancillasnew} becomes
\begin{equation}\label{anom free gen ancillas2}
    \prod_{c_{n}\in C^\vee_{d-n}} [\mu^{(n+2)}_{c_{n}}]^{\si_{c_n}(C^\vee_{d-n})} = [1],
\end{equation}
where ${\si_{c_n}(C^\vee_{d-n})\in\Z}$ is the coefficient of the ${(d-n)}$-cell dual to $c_n$ in $C^\vee_{d-n}$, and the product ${\prod_{c_{n}\in C^\vee_{d-n}}}$ is over all $n$-cells ${c_n\in\La}$ whose dual ${(d-n)}$-cells lie in $C^\vee_{d-n}$. 
When the $n$-stratum anomalies satisfy~\eqref{anom free gen ancillas2}, the $\cH^{n+2}(G,\Uone)$-valued cellular ${n}$-chain ${\mu^{(n+2)}_{n\text{-chain}}:= \sum_{c_n} [\mu^{(n+2)}_{c_n}]\, c_n}$ can be decomposed as
\begin{equation}
    \mu^{(n+2)}_{n\text{-chain}} = \pp_{n+1}\ga^{(n+2)}_{n+1\text{-chain}},
\end{equation}
where ${\ga^{(n+2)}_{n+1\text{-chain}}}$ is an $\cH^{n+2}(G,\Uone)$-valued cellular ${(n+1)}$-chain. Therefore, tensoring these ancillas causes the stratified anomaly data to transform as
\begin{equation}\label{strat anom post ancilla gen}
    \nu^{(n+2)}_{n\text{-chain}}  \mapsto \nu^{(n+2)}_{n\text{-chain}} + \pp_{n+1} \ga^{(n+2)}_{n+1\text{-chain}}.
\end{equation}

The first step of the equivalence condition introduces new stratum anomalies via ancillas to change the original $n$-stratum anomalies by~\eqref{strat anom post ancilla gen}. The second step, on the other hand, does not add new stratum anomaly data. At most, a locality-preserving unitary that maps a stratified operator to another stratified operator will move already existing stratum anomalies throughout space. However, this effect on the stratified anomaly is already achieved using ancillas. For example, to move the stratum anomaly $[\nu_{\Si_i}]$ at $\Si_i$ to $\Si_f$ with ${\mathrm{dim}(\Si_i) = \mathrm{dim}(\Si_f)}$ and $\Si_i$ homologous to $\Si_f$, we use ancillas with stratum anomalies ${[\mu_{\Si_i}]= [\nu_{\Si_i}]^{-1}}$ and ${[\mu_{\Si_f}]=[\nu_{\Si_i}]}$. Therefore, the effect of step 2 in the equivalence relation on $n$-stratum anomalies leads to changes that also take the form of~\eqref{strat anom post ancilla gen}, and the total equivalence relation on $n$-stratum anomalies is
\begin{equation}
    \nu^{(n+2)}_{n\text{-chain}} \sim \nu^{(n+2)}_{n\text{-chain}} + \pp_{n+1} \ga^{(n+2)}_{n+1\text{-chain}}.
\end{equation}

Since a stratified anomaly has every ${\nu_{n\text{-chain}}^{(n+2)}\in \ker(\pp_n)}$, equivalent $n$-stratum anomalies are classified by the degree-$n$ homology classes
\begin{equation}\label{gen strat anom class}
    \ker(\pp_n)/\operatorname{im}(\pp_{n+1}) := H_n(\La; \cH^{n+2}(G,\Uone)).
\end{equation}
A stratified anomaly obstructs a $G$-SPT iff there is an $n$ such that the homology class ${[\nu^{(n+2)}_{n\text{-chain}}] \neq [0]}$.
It is very important to note that because the $n$-chain~\eqref{cellular n chain def} can be an infinite sum,~\eqref{gen strat anom class} is the degree-$n$ Borel-Moore homology group of the underlying topological space $X_d$ of the lattice. When $X_d$ is compact, Borel-Moore homology reduces to singular homology. They are not the same, however, for non-compact spaces.\footnote{\label{Bm vs sing homolo footnote}For example, the Borel-Moore homology of $\R^d$ is
\begin{equation}
    H_n(\R^d;A) \equiv H^{\text{BM}}_n(\R^d;A) = \begin{cases}
        A \qquad &n = d,\\
        0 \qquad &n \neq d,
    \end{cases}
\end{equation}
whereas the singular homology of $\R^d$ is
\begin{equation}
    H^{\text{singular}}_n(\R^d;A) = \begin{cases}
        A \qquad &n = 0,\\
        0 \qquad &n \neq 0.
    \end{cases}
\end{equation}
}

The classification~\eqref{gen strat anom class} of stratified anomalies depends on the topology of space. Consider the following examples with ${d>0}$:
\begin{enumerate}
    \item \underline{${X_d = \R^d}$:} When $\La$ is a cellulation of $\R^d$, the classification of $n$-stratum anomalies~\eqref{gen strat anom class} is trivial for all ${n<d}$ (see footnote~\ref{Bm vs sing homolo footnote}). The fact that there are no nontrivial stratified anomalies on $\R^d$ agrees with the argument based on stratified anomaly-inflow theories from Section~\ref{strat anomalies sec}.
    \item \underline{${X_d = S^d}$:} When $\La$ is a cellulation of a $d$-sphere $S^d$, the classification of $n$-stratum anomalies is $\cH^{2}(G,\Uone)$ for ${n=0}$ and trivial for ${0<n<d}$. In this case, a simple representative 0-cycle of ${[\nu_{0\text{-chain}}^{(2)}]\in \cH^{2}(G,\Uone)}$ is ${\nu^{(2)}_{0\text{-chain}} = [\nu_{0\text{-chain}}^{(2)}]\, c_0}$ for an arbitrary 0-cell $c_0$.
    \item \underline{${X_d = T^d}$:} When $\La$ is a cellulation of a $d$-torus $T^d$, the classification of $n$-stratum anomalies is ${\bigoplus_{i=1}^{\binom{d}{n}} \cH^{n+2}(G,\Uone)}$, where the binomial coefficient ${\binom{d}{n} := \frac{d!}{n!(d-n)!}}$.
\end{enumerate}

\section{Crystalline symmetries}\label{CrystallineSymmetrySection}

In Section~\ref{StratifiedSymmetrySection}, we introduced stratified symmetry operators and stratified anomalies for internal symmetries. In addition to formalizing the general notion of stratified symmetries, a main purpose of this paper is to demonstrate when and how LSM anomalies arise from stratified anomalies in the presence of crystalline symmetries. In this section, we will review the basics of crystalline symmetries and discuss their interplay with stratified symmetries at a high level. In Section~\ref{ModulatedSymmetrySection}, we will discuss these physical consequences in greater detail for modulated symmetries.

\subsection{Crystalline symmetries and their group extension}

Crystalline symmetries are symmetries whose transformations implement a nontrivial automorphism of the underlying lattice $\La$. Let us denote by $G$ the internal symmetry group, $G_\mathrm{s}$ the crystalline symmetry group, and $G_{\mathrm{tot}}$ the total symmetry group formed by $G_\mathrm{s}$ and $G$. In this section, we will assume that the symmetry operators furnish a linear representation of $G_{\mathrm{tot}}$.

When a quantum lattice system has a crystalline symmetry, its Hilbert space $\scrH$ is isomorphic to ${\bigotimes_{c\in\La}\scrH_c}$ with ${\scrH_c = \scrH_{s\triangleright c}}$ for all ${s\in G_\mathrm{s}}$. Here, ${s\triangleright c}$ denotes the image of a cell $c$ of $\La$ under the lattice-automorphism induced by ${s\in G_\mathrm{s}}$. In this presentation of the Hilbert space, every crystalline symmetry operator $U_s$ admits the canonical decomposition
\begin{equation}\label{Us decomposition}
    U_s = U_s^{\mathrm{int}}\,U_s^{\La},
\end{equation}
where the unitary operator $U_s^{\La}$ satisfies
\begin{equation}\label{UsLa transformation}
    U_s^\La \bigotimes_{c\in\La}\ket{\psi_c}
= \bigotimes_{c\in\La}\ket{\psi_{s^{-1}\triangleright c}}\qquad\forall~\ket{\psi_c}\in\scrH_c.
\end{equation}
We follow the convention where lattice automorphisms implement passive transformations on the Hilbert space, which is why $s^{-1}$ appears on the right-hand side of~\eqref{UsLa transformation}. The unitary operator $U_s^{\mathrm{int}}:= U_s\,(U_s^{\La})^\dag$ in~\eqref{Us decomposition} does not transform the lattice. 

Using~\eqref{Us decomposition}, the crystalline symmetry operators satisfy ${U_{s_1}U_{s_2} U_{s_1 s_2}^\dag = 
U_{s_1}^{\mathrm{int}}\,U_{s_1}^{\La}
U_{s_2}^{\mathrm{int}}\,(U_{s_1}^{\La})^\dag (U_{s_1 s_2}^{\mathrm{int}})^\dag}$. It follows from~\eqref{UsLa transformation} that the operator on the right-hand side of this equation does not transform the lattice and, therefore, $U_{s_1}U_{s_2} U_{s_1 s_2}^\dag$ must be an internal $G$ symmetry operator:
\begin{equation}\label{extension class}
    U_{s_1}U_{s_2} U_{s_1 s_2}^\dag = U_{f(s_1,s_2)},\qquad f\colon G_\mathrm{s}\times G_\mathrm{s} \to G.
\end{equation}
A similar argument shows that ${U_{s}\,U_{g}\, U_{s}^\dag}$ must be an internal symmetry operator. In particular,
\begin{equation}\label{s action on g general}
    U_{s}\,U_{g}\, U_{s}^\dag = U_{\rho_s(g)},\qquad \rho\colon G_\mathrm{s} \to\Aut(G).
\end{equation}
The maps $f$ and $\rho$ are not generally independent. Firstly, associativity of the $U_{s}$ symmetry operators requires
\begin{equation}\label{nonabelian cocycle}
    \rho_{s_1}(f(s_2,s_3))f(s_1,s_2s_3) = f(s_1,s_2)f(s_1s_2,s_3).
\end{equation}
for all ${s_1,s_2,s_3\in G_\mathrm{s}}$. Secondly, conjugating~\eqref{s action on g general} by a crystalline symmetry operator, we find
\begin{equation}\label{geneal rho composition}
    \rho_{s_1}\circ \rho_{s_2} =  \mathrm{conj}_{f(s_1,s_2)}\circ \rho_{s_1 s_2},
\end{equation}
where $\mathrm{conj}_g$ is the $G$-automorphism satisfying ${\mathrm{conj}_{g_1}(g_2) := g_1 g_2 g_1^{-1}}$.

The total symmetry group $G_{\mathrm{tot}}$ fits into the group extension\footnote{See \cite[Appendix B]{PAL250702036} for further discussion of group extensions in the case of abelian $G$.}
\begin{equation}\label{GtotGrpExt}
    1\to G \to G_{\mathrm{tot}} \to G_\mathrm{s} \to 1.
\end{equation}
Indeed, $G$ is always a normal subgroup of $G_{\mathrm{tot}}$ because of~\eqref{s action on g general}, which makes $G_\mathrm{s}$ isomorphic to the quotient group $G_{\mathrm{tot}}/G$. The different interplays between $G$ and $G_\mathrm{s}$ that characterize the total symmetry group $G_\mathrm{tot}$ are classified by the different group extensions~\eqref{GtotGrpExt}. In particular, they are classified by the maps $f$ and $\rho$ satisfying~\eqref{nonabelian cocycle} and~\eqref{geneal rho composition} and subject to the equivalence relations
\begin{equation}
\begin{aligned}
    f(s_1,s_2)&\sim b(s_1)\,\rho_{s_1}(b(s_2))\,f(s_1,s_2) \,b(s_1s_2)^{-1},\\
    \rho_s &\sim \mathrm{conj}_{b(s)}\circ \rho_s,
\end{aligned}
\end{equation}
where ${b\colon G_\mathrm{s}\to G}$. These equivalence relations arise from choosing a different section of the extension, which corresponds to changing ${U_s \mapsto U_{b(s)}U_s}$.

\subsection{Crystalline symmetries and stratified anomalies}

Not all stratified symmetry operators are compatible with a crystalline symmetry. Suppose that the internal $G$ symmetry operator is a stratified symmetry operator $U_g$ with respect to the family of subsystems $\Si$. Since a $G_\mathrm{s}$ crystalline symmetry transforms $\La$ nontrivially, it will generically induce a nontrivial action on $\Si$. We will denote the image of $s$ acting on $\Si_i$ by ${s\triangleright\Si_i \equiv \Si_{s\triangleright i}}$. In order for the stratified symmetry operator to be compatible with the crystalline symmetry, the action of $G_\mathrm{s}$ on $\Si$ must be closed. That is, for each ${s\in G_\mathrm{s}}$ and ${i\in I}$, $s$ sends the subsystem $\Si_i$ to a subsystem $s\triangleright \Si_i$ that is also in $\Si$.

Crystalline symmetries also constrain the stratified anomalies of a stratified symmetry operator. In particular, the stratum anomaly of $U_{g}^{(\Si_i)}$ must be the same as (the inverse of) the stratum anomaly of ${U_s\,U_{g}^{(\Si_i)}\,U_s^\dag}$ if $s$ preserves (reverses) the orientation of $\Si_i$.\footnote{It is easy to prove how the anomaly class transforms under spatial transformations in the continuum~\cite{K14031467}. To our knowledge, this is not proven on the lattice. We will not try to prove it here, but we believe it to be true in any unitary, local, quantum lattice system.} Suppose that
\begin{equation}\label{general s action on subsys op}
    U_s\, U^{(\Si_i)}_{g} \, U_s^\dag = U^{(s^{-1}\triangleright \Si_i)}_{\rho_{s}(g)}.
\end{equation}
This transformation of the stratum operators is compatible with the stratified symmetry operator satisfying~\eqref{s action on g general}, and implies that the anomaly of $U_{g}^{(\Si_i)}$ must be the same as/inverse of the anomaly of ${U^{(s^{-1}\triangleright \Si_i)}_{\rho_{s}(g)}}$. Therefore, letting ${o_{i}(s) = +1}$ ($-1$) if $s$ preserves (reverses) the orientation of $\Si_i$, the stratum anomalies must satisfy the $G_\mathrm{s}$-covariance condition\footnote{Recall that the pullback of a group homomorphism ${f\colon G\to K}$ yields a group homomorphism ${f^*\colon \cH^n(K,A)\to \cH^n(G,f^*A)}$, where $f^* A$ is the $G$-module with $G$-action ${g\triangleright_{f^*A} a :=f(g)\triangleright_{A} a}$. In particular, ${f^*\colon [\om]\mapsto f^*[\om] = [f^*\om]}$ for ${[\om]\in \cH^n(K,\Uone)}$, where $f^*$ acts on a representative $n$-cocycle ${\om\colon K^n \to \Uone}$ of $[\om]$ by pre‑composition:
\begin{equation}
(f^{*}\om)(g_1,g_2, \cdots, g_{n}) = \om(f(g_1), f(g_2),\cdots , f(g_{n})).
\end{equation}}  
\begin{equation}\label{gen anomal covariance cond}
    [\nu_{s\triangleright \Si_i}]^{o_{i}(s)} = \rho^*_s[\nu_{\Si_i}].
\end{equation}
While this covariance condition was found assuming~\eqref{general s action on subsys op}, we believe it holds generally. In terms of the cellular stratum anomaly data~\eqref{n-cell anom data}, it implies that\footnote{Each $\si_{c_n}(\Si_i)$ satisfies ${\si_{s\triangleright c_n}(s\triangleright \Si_i) = o_{c_n}(s) \,o_i(s)\, \si_{c_n}(\Si_i)}$.}
\begin{equation}\label{gen cell anomal covariance cond}
    [\nu^{(n+2)}_{s\triangleright c_n}]^{o_{c_n}(s)} = \rho^*_s[\nu^{(n+2)}_{c_n}],
\end{equation}
where ${o_{c_n}(s) = +1}$ ($-1$) if $s$ preserves (reverses) the orientation of $c_n$. Note that ${o_{c_0}(s) = +1}$ for all 0-cells $c_0$. A consequence of this covariance condition is that the stratified-anomaly inflow SPT defect network must be $G_\mathrm{s}$-covariant, which makes the anomaly-inflow theory a crystalline SPT. These stratified anomaly-inflow theories are generalizations of the inflow theories for LSM anomalies discussed in~\cite{CZB151102263, FVM180408628, C180410122, ET190708204, PAL250702036}.

In addition to limiting the possible stratified anomalies, crystalline symmetries also modify the equivalence relation and classification of stratified anomalies from Section~\ref{stratAnomEquiv}. The ancillas must now have a $G_\mathrm{tot}$ symmetry that is compatible with a $G_\mathrm{tot}$ SPT. This reduces the number of allowed ancillas since it requires the ancilla Hilbert space to be $G_\mathrm{s}$ symmetric (i.e., the ancillas must be added in $G_\mathrm{s}$-symmetric patterns). Furthermore, the ancillas' stratum anomalies must satisfy the covariance condition~\eqref{gen anomal covariance cond}. These restrictions decrease the number of equivalence relations on stratified anomalies and, thus, often increase the number of their equivalence classes as fewer stratified anomalies can be trivialized. These equivalence classes can describe LSM anomalies of the $G_\mathrm{tot}$ symmetry. In particular, a stratified anomaly leads to an LSM if it was equivalent to the zero stratified anomaly before enforcing the $G_\mathrm{s}$ crystalline symmetry, but inequivalent after.

Lastly, there is a sense in which crystalline symmetry operators can also be stratified symmetry operators. In particular, we say that the crystalline symmetry operator $U_s$ is a stratified symmetry operator if $U_s^{\mathrm{int}}$ in~\eqref{Us decomposition} is a stratified symmetry operator with respect to $s$-invariant subsystems. If a stratum operator of $U_s^{\mathrm{int}}$ acting on an $s$-invariant subsystem is anomalous, then $U_s$ has a stratified anomaly. A simple example of a crystalline symmetry operator with a stratified anomaly is a ${(1+1)}$D $\Z_2$ lattice reflection symmetry operator ${U_R = U_R^{\mathrm{int}}\,U_R^{\La}}$ where $U_R^{\mathrm{int}}$ is an onsite operator with a 0-stratum anomaly at the reflection center(s) of $U_R^{\La}$~\cite{PWJ170306882, OTT200406458, AMF230800743, PAL250702036, SSZ250817115}. Stratified anomalies of crystalline symmetries can also give rise to LSM anomalies. We discuss stratified anomalies of crystalline symmetries more in Appendix~\ref{AHSS app}.

\section{Modulated symmetries with stratified anomalies}\label{ModulatedSymmetrySection}

Thus far, we have introduced the notion of stratified symmetry operators and stratified anomalies in Section~\ref{StratifiedSymmetrySection} and have discussed their interplay with crystalline symmetries in Section~\ref{CrystallineSymmetrySection}. In this section, we will discuss the physical consequences of stratified anomalies for modulated symmetries. By definition, the group extension~\eqref{GtotGrpExt} splits for modulated symmetries, causing the total symmetry group to become the semi-direct product group
\begin{equation}\label{general mod sym group}
    G_\mathrm{tot} = G\rtimes G_\mathrm{s}.
\end{equation}
In this case, the function ${f}$ in~\eqref{extension class} satisfies ${f(s_1,s_2) = 1}$ for all ${s_1,s_2\in G_\mathrm{s}}$, and~\eqref{geneal rho composition} simplifies to 
\begin{equation}
    \rho_{s_1}\circ \rho_{s_2} =  \rho_{s_1 s_2}.
\end{equation}
Thus, the map ${\rho\colon G_\mathrm{s}\to\Aut(G)}$ becomes a group homomorphism for modulated symmetries.

In this section, for simplicity, we will consider the case where the crystalline symmetry is $d$-dimensional lattice translation symmetry. We discuss the more general case in Appendix~\ref{AHSS app}. For example, in Appendix~\ref{d1 formula app}, we classify 0-stratum anomalies in ${(1+1)}$D when $G_\mathrm{s}$ includes both lattice translations and spatial reflections. Here, we denote the group of $d$-dimensional discrete lattice translations of $\La$ by $\mathbb{T}_d$. A translation by the lattice vector ${\bm{v}\in \mathbb{T}_d}$ then acts on ${g\in G}$ by ${g\mapsto \rho_{\bm{v}}(g)}$. For uniform symmetries, $\rho_{\bm{v}}$ is the trivial $G$-automorphism for all $\bm{v}$. We will denote the ${\mathbb{T}_d}$ translation symmetry operators by $T_{\bm{v}}$.

\subsection{Modulated stratified symmetries}\label{modulated strat sym subsection}

As was made clear in Section~\ref{CrystallineSymmetrySection}, modulated symmetry operators can be stratified operators and have a stratified anomaly. Consider a modulated $G$ symmetry with stratum anomalies ${\{[\nu_{\Si_i}]\}_{i\in I}}$. Since lattice translations are orientation-preserving, the covariance condition~\eqref{gen anomal covariance cond} simplifies to
\begin{equation}\label{T action on strat anom}
    \rho_{\bm{v}}^*[\nu_{\Si_{i}}] =[\nu_{\bm{v}\triangleright \Si_{i}}].
\end{equation}
This implies that the stratum anomaly at $\Si_i$ must satisfy ${\rho_{\bm{v}}^*[\nu_{\Si_{i}}] = [\nu_{\Si_{i}}]}$ for every ${\bm{v}\in\mathbb{T}_d}$ satisfying ${\bm{v}\triangleright \Si_i = \Si_i}$. Furthermore, for translations, the covariance condition~\eqref{gen cell anomal covariance cond} for cellular stratum anomalies simplifies to
\begin{equation}\label{cellular translation covariance}
    [\nu^{(n+2)}_{\bm{v}\triangleright c_n}] = \rho_{\bm{v}}^*[\nu^{(n+2)}_{c_n}].
\end{equation}
Because of this, the entire cellular stratum anomaly data is fully specified by the cellular stratum anomaly data within a single unit cell of $\La$.

The covariance condition~\eqref{cellular translation covariance} is naturally formulated in terms of the cellular $n$-cycle~\eqref{cellular n chain def}. 
Consider the action of $\mathbb{T}_d$ on $C_n(\La; \cH^{n+2}(G,\Uone))$ defined by
\begin{equation}\label{Td action onf cellular chains}
    \bm{v}\triangleright\nu^{(n+2)}_{n\text{-chain}} 
    := \sum_{c_n} \rho^*_{\bm{v}}[\nu^{(n+2)}_{c_n}]\,(\bm{v}\triangleright c_n)
    = \sum_{c_n} \rho^*_{\bm{v}}[\nu^{(n+2)}_{-\bm{v}\triangleright c_n}]\,c_n
    .
\end{equation}
The boundary map $\pp_n$ is $\mathbb{T}_d$-equivariant and satisfies ${\pp_n(\bm{v}\triangleright\nu^{(n+2)}_{n\text{-chain}}) = \bm{v}\triangleright(\pp_n\nu^{(n+2)}_{n\text{-chain}})}$. Therefore, ${\bm{v}\triangleright\nu^{(n+2)}_{n\text{-chain}}}$ is always an $n$-cycle since ${\nu^{(n+2)}_{n\text{-chain}}}$ is an $n$-cycle, and ${\{{\bm{v}\triangleright\nu^{(n+2)}_{n\text{-chain}}}\}_{n=0}^{d-1}}$ corresponds to an allowed stratified anomaly. Using the $\mathbb{T}_d$-action~\eqref{Td action onf cellular chains}, the covariance condition~\eqref{cellular translation covariance} becomes the invariance condition
\begin{equation}\label{Td invariant cond}
\bm{v}\triangleright\nu^{(n+2)}_{n\text{-chain}} = \nu^{(n+2)}_{n\text{-chain}}.
\end{equation}
We denote the group of $\mathbb{T}_d$-invariant, $\cH^{n+2}(G,\Uone)$-valued $n$-chains by
\begin{equation}\label{Td invariant nchains}
    C^{\mathbb{T}_d}_n(\La; \cH^{n+2}(G,\Uone)) := \left\{
    \nu^{(n+2)}_{n\text{-chain}}\in C_n(\La; \cH^{n+2}(G,\Uone))
    \mid
    \bm{v}\triangleright\nu^{(n+2)}_{n\text{-chain}} = \nu^{(n+2)}_{n\text{-chain}}\right\}
    .
\end{equation}
The allowed $n$-stratum anomalies for modulated symmetries with lattice translation symmetry satisfy
\begin{equation}\label{Tinv ncycle strat anom cond}
    \pp^{\mathbb{T}_d}_{n}\nu^{(n+2)}_{n\text{-chain}} = 0,
\end{equation}
where $\pp^{\mathbb{T}_d}_{n}$ is the restriction of $\pp_{n}$ to $\mathbb{T}_d$-invariant ${n}$-chains. $\pp^{\mathbb{T}_d}_{n}$ is well defined because $\pp_n$ is $\mathbb{T}_d$-equivariant. We denote the group of $\mathbb{T}_d$-invariant $n$-cycles by
\begin{equation}\label{Td invariant ncycles}
    \ker(\pp^{\mathbb{T}_d}_{n}) := Z^{\mathbb{T}_d}_n(\La; \cH^{n+2}(G,\Uone)).
\end{equation}

\subsection{LSM constraints}\label{LSMconstSection}

The stratified anomaly of a modulated symmetry can cause the internal $G$ symmetry to be self-anomalous. However, even when the internal symmetry is anomaly-free, its stratified anomaly can still have consequences. In particular, it can lead to a Lieb-Schultz-Mattis (LSM) constraint.
This includes LSM anomalies, which are anomalies of the total ${G_\mathrm{tot} = G\rtimes \mathbb{T}_d}$ symmetry that involve translation symmetry. However, as we will show, a nonzero stratified anomaly does not guarantee an LSM anomaly (see Table~\ref{LSMtable}). When this occurs, the stratified anomaly leads to a different LSM constraint: an SPT-LSM theorem, which is an obstruction to a trivial SPT. In what follows, we will derive an explicit criterion for when a modulated symmetry with a stratified anomaly is LSM anomaly-free.

\subsubsection{LSM anomalies}\label{LSM section}

Recall from Section~\ref{IntroductionSection} that an LSM anomaly is an anomaly involving crystalline symmetries. In particular, the anomaly is an equivalence class of $G_\mathrm{tot}$ symmetry operators up to adding SPT-compatible ancillas and performing locality-preserving unitary transformations. Therefore, to deduce whether a stratified anomaly leads to an LSM anomaly, we must see whether tensoring SPT-compatible ancillas can trivialize the stratified anomaly. As in Section~\ref{stratAnomEquiv}, the ancillas will have their own $G_\mathrm{tot}$ symmetry operator and stratified anomaly that must be SPT-compatible. This then begs the question, which stratified anomalies of a ${G\rtimes\Z^d}$ symmetry are SPT-compatible?

Whether a stratified anomaly is SPT-compatible can be determined by attempting to construct an SPT in its presence and observing whether an obstruction arises. A common construction for SPTs is the real-space construction, in which a ${(d+1)}$D $G_\mathrm{tot}$ SPT is constructed from a network of ${(n+1)}$D SPTs (${n=0,1,\cdots, d}$) whose protecting symmetry groups are subgroups of $G_\mathrm{tot}$~\cite{SSF160408151, HSH170509243, SXG181000801, ET181010539, SFQ181011013, ET190708204, JCQ190708596, ZWY190905519, ZYQ201215657, ZNQ220413558, SAN240411650, SZQ250802661, B250806604}.\footnote{Real-space constructions have been applied to general gapped phases, and they are sometimes instead called defect network or block state constructions.} As we review in Appendix~\ref{AHSS app}, the real-space construction is a physical realization of the Atiyah–Hirzebruch spectral sequence for the equivariant generalized homology theory that classifies $G_\mathrm{tot}$ SPTs~\cite{SXG181000801}. Here, we assume that every modulated SPT has a real-space construction,\footnote{It is conjectured that every gapped phase can be constructed using a real-space construction~\cite{ABP200205166}.} and derive a criterion for SPT-compatible stratified anomalies based on this assumption. In particular, we will consider a stratified anomaly to be SPT-incompatible iff it obstructs the construction of a modulated SPT using the real-space construction. A similar perspective was used for $0$-stratum anomalies of uniform symmetries in~\cite{ET190708204}.

In Appendix~\ref{def net mod spt app}, we review the real-space construction for modulated SPTs~\cite{B250806604}. 
For a modulated SPT whose crystalline symmetry is translation symmetry, it is specified by a collection of ${\cH^{n+1}(G,\Uone)}$-valued cellular $n$-chains (${n=0,1,\cdots, d}$)
\begin{equation}\label{SPT chain data}
    \om^{(n+1)}_{n\text{-chain}} := \sum_{c_n} \,[\om^{(n+1)}_{c_{n}}]\, c_n,
\end{equation}
each of which satisfies
\begin{equation}
     \bm{v}\triangleright \om^{(n+1)}_{n\text{-chain}} :=
     \sum_{c_n} \,\rho_{\bm{v}}^*[\om^{(n+1)}_{c_{n}}]\, (\bm{v}\triangleright c_n)
     =\om^{(n+1)}_{n\text{-chain}}.
\end{equation}
The coefficient ${[\om^{(n+1)}_{c_{n}}]}$ in~\eqref{SPT chain data} is interpreted as an ${(n+1)}$D $G$-SPT decorating the $n$-cell $c_n$. Each of these ${(n+1)}$D SPTs contributes an inflow anomaly at the ${(n-1)}$-cells on their boundary. The net total of these inflow anomalies is described by the collection of ${\cH^{n+1}(G,\Uone)}$-valued cellular ${(n-1)}$-chains ${\{\pp^{\mathbb{T}_d}_{n}\om^{(n+1)}_{n\text{-chain}}\}_{n=1}^{d}}$.

For the real-space construction with input data ${\{\om^{(n+1)}_{n\text{-chain}}\}_{n=0}^{d}}$ to yield an SPT, all anomalous realizations of the symmetry must cancel. When there is no stratified anomaly, this implies that the inflow anomalies must all be trivial: ${\pp^{\mathbb{T}_d}_{n}\om^{(n+1)}_{n\text{-chain}} = 0}$. In the presence of a stratified anomaly, however, there are now two contributions of anomalies in the real-space construction: there are still the inflow anomalies, but there is now also the intrinsic stratified anomaly $\{\nu^{(n+2)}_{n\text{-chain}}\}_{n=0}^{d-1}$. For the real-space construction to be free of anomalies and produce an SPT, these two contributions must cancel. That is, for each ${n=0,1,\cdots,d-1}$, 
\begin{equation}\label{inflow anom cancel cond transl}
   \pp^{\mathbb{T}_d}_{n+1}\om^{(n+2)}_{n+1\text{-chain}} +  \nu^{(n+2)}_{n\text{-chain}} = 0.
\end{equation}
If there is an $\om^{(n+2)}_{n+1\text{-chain}}$ satisfying~\eqref{inflow anom cancel cond transl} for every $n$, then the stratified anomaly $\{\nu^{(n+2)}_{n\text{-chain}}\}_{n=0}^{d-1}$ is SPT-compatible. In other words, SPT-compatible stratified anomalies are those for which every
\begin{equation}\label{LSM anom cond mod sym transl}
    \nu^{(n+2)}_{n\text{-chain}}\in 
    \operatorname{im}(\pp^{\mathbb{T}_d}_{n+1}) := B^{\mathbb{T}_d}_n(\La; \cH^{n+2}(G,\Uone)).
\end{equation}
If there is an $n$ for which ${\nu^{(n+2)}_{n\text{-chain}}\not\in \operatorname{im}(\pp^{\mathbb{T}_d}_{n+1})}$, then the stratified anomaly is not SPT-compatible. Note that, from Section~\ref{stratAnomEquiv}, if a stratified anomaly satisfies~\eqref{LSM anom cond mod sym transl}, then it also does not give rise to an anomaly for the internal $G$ symmetry.

The condition~\eqref{LSM anom cond mod sym transl} shows that SPT-compatible ancillas can have nonzero $n$-stratum anomalies of the form $\pp^{\mathbb{T}_d}_{n+1}\ga^{(n+2)}_{n+1\text{-chain}}$. Changes from locality-preserving unitaries also take this form. Therefore, given a modulated stratified symmetry operator, its 
$n$-stratum anomalies $\nu^{(n+2)}_{n\text{-chain}}$ are subject to the equivalence relation
\begin{equation}\label{tinv Lat homotopy}
    \nu^{(n+2)}_{n\text{-chain}} \sim \nu^{(n+2)}_{n\text{-chain}} + \pp^{\mathbb{T}_d}_{n+1}\ga^{(n+2)}_{n+1\text{-chain}}.
\end{equation}
The classification of $n$-stratum anomalies under this equivalence relation is given by the $\mathbb{T}_d$-invariant Borel-Moore homology groups
\begin{equation}\label{Transl inv homology classes}
    H^{\mathbb{T}_d}_n(\La; \cH^{n+2}(G,\Uone)) := \ker(\pp^{\mathbb{T}_d}_{n})/\operatorname{im}(\pp^{\mathbb{T}_d}_{n+1}).
\end{equation}
While the Borel-Moore homology groups~\eqref{gen strat anom class} were always trivial for a lattice in $\R^d$, the $\mathbb{T}_d$-invariant Borel-Moore homology groups~\eqref{Transl inv homology classes} can be nontrivial for a lattice in $\R^d$. If every $\nu^{(n+2)}_{n\text{-chain}}$ is homologically trivial---if every ${\nu^{(n+2)}_{n\text{-chain}}\in \operatorname{im}(\pp^{\mathbb{T}_d}_{n+1})}$---then the stratified anomaly can be trivialized using SPT-compatible ancillas and, thus, the modulated symmetry is LSM anomaly-free. The classification of these LSM anomalies is given by ${\prod_{n = 0}^{d-1} H^{\mathbb{T}_d}_n(\La; \cH^{n+2}(G,\Uone))}$ as a set. For ${d=1}$, its classification as a group is ${H^{\mathbb{T}_1}_0(\La; \cH^{2}(G,\Uone))}$. For ${d>1}$, its group structure is more complicated and follows from a sequence of group extensions (see Appendix~\ref{AHSS app} for more details).

To better understand the LSM anomaly-free condition~\eqref{LSM anom cond mod sym transl} and the classification~\eqref{Transl inv homology classes}, let us specialize to a ${d=1}$ dimensional Bravais lattice with underlying topological space $\R$ and stratified anomaly $\nu^{(2)}_{0\text{-chain}}$. We label each 0-cell of $\La$ by ${j\in\Z}$, and the 1-cell connecting $j$ and ${j+1}$ by ${\<j,j+1\>}$. The translation symmetry group is $\Z$ and the element ${1\in \Z}$ acts on $\La$ as ${j\mapsto j+1}$. In this case,~\eqref{LSM anom cond mod sym transl} is nontrivial only for ${n=0}$. When it is satisfied, there exists a $\Z$-invariant, $\cH^2(G,\Uone)$-valued 1-chain $\ga^{(2)}_{1\text{-chain}}$ such that ${\nu^{(2)}_{0\text{-chain}} = \pp^{\mathbb{T}_d}_{1}\ga^{(2)}_{1\text{-chain}}}$. Using that 
\begin{align}
    \ga^{(2)}_{1\text{-chain}} := \sum_{j}[\ga^{(2)}_{\<j,j+1\>}]\<j,j+1\>,\qquad
    \nu^{(2)}_{0\text{-chain}} := \sum_{j} \,[\nu^{(2)}_{j}]\, j,
\end{align}
with ${[\ga^{(2)}_{\<j,j+1\>}] =  \rho_{j}^*[\ga^{(2)}]}$ and ${[\nu^{(2)}_{j}] =  \rho_{j}^*[\nu^{(2)}]}$ by translation-covariance, this is equivalent to
\begin{equation}\label{anomCancel1dTransl}
    [\nu^{(2)}] = [\ga^{(2)}]^{-1}\,\rho_{-1}^*[\ga^{(2)}].
\end{equation}
Thus, if there is no $\cH^2(G,\Uone)$ class ${[\ga^{(2)}]}$ satisfying~\eqref{anomCancel1dTransl} for fixed ${[\nu^{(2)}]}$, then the 0-stratum anomaly gives rise to an LSM anomaly for the modulated ${G\rtimes \Z}$ symmetry. Relatedly, the equivalence relation~\eqref{tinv Lat homotopy} in this case is
\begin{equation}\label{0-strat anom equiv cond}
    [\nu^{(2)}] \sim [\nu^{(2)}]\,
    [\ga^{(2)}]^{-1} \rho_{-1}^*[\ga^{(2)}],
\end{equation}
where ${[\ga^{(2)}]}$ is now any $\cH^2(G,\Uone)$ class. The $\cH^{2}(G,\Uone)$ classes $[\nu^{(2)}]$ subject to the equivalence relation~\eqref{0-strat anom equiv cond} forms the group of $\rho^*$-coinvariants in $\cH^{2}(G,\Uone)$. It is isomorphic to the cohomology group $\cH^1(\Z, \cH^{2}(G,\Uone)_{\rho^*})$ where $\cH^{2}(G,\Uone)_{\rho^*}$ is a $\Z$-module with underlying group $\cH^{2}(G,\Uone)$ and $\Z$ action $\rho^*$.\footnote{This isomorphism follows from the group cocycle condition ${\al(n+m) = \al(n)\rho_n^*(\al(m))}$ of a cocycle ${\al\in \cZ^1(\Z, \cH^{2}(G,\Uone))}$. Using it, the image of ${\al}$ for every ${n\in \Z}$ can be decomposed as ${\al(n) = \prod_{k=0}^{n-1} \rho_k^*\al(1)}$. This decomposition gives rise to the isomorphism ${\cZ^1(\Z, \cH^{2}(G,\Uone)) \cong \cH^{2}(G,\Uone)}$ where ${\al\mapsto \al(1)}$. The coboundary equivalence relation on $\al$ under this isomorphism becomes~\eqref{0-strat anom equiv cond}.} Therefore, the classification of 0-stratum anomalies in this example is isomorphic to
\begin{equation}\label{0-strat anom class transl}
    H^{\Z}_0(\La; \cH^{2}(G,\Uone))
    \cong \cH^1(\Z, \cH^{2}(G,\Uone)_{
\rho^*}).
\end{equation}
This is the same as the classification of LSM anomalies for the $G\rtimes \Z$ modulated symmetry. Notably, ${\cH^1(\Z, \cH^{2}(G,\Uone)_{\rho^*})}$ is a subgroup of ${\cH^3(G\rtimes \Z, \Uone)}$,\footnote{This follows from the Lyndon-Hochschild-Serre (LHS) spectral sequence of ${\cH^3(G\rtimes \Z, \Uone)}$. In particular, because ${\cH^n(\Z,M) = 0}$ for every ${n>1}$ and $\Z$-module $M$, this LHS spectral sequence stabilizes at the $E_2$-page and ${\cH^3(G\rtimes \Z, \Uone)}$ fits into the short exact sequence
\begin{equation}
        0 
        \to 
        \cH^1(\Z, \cH^{2}(G,\Uone)_{
    \rho^*}) 
        \to 
        \cH^3(G\rtimes \Z, \Uone)
        \to
        \cH^0(\Z, \cH^{3}(G,\Uone)_{
    \rho^*}) 
        \to
        0.
\end{equation}
We note that $\cH^0(\Z, \cH^{3}(G,\Uone)_{\rho^*})$ is the group of $\rho^*$-invariants in $\cH^{3}(G,\Uone)$.} which is the classification of ${G\rtimes \Z}$ anomalies in ${(1+1)}$D.\footnote{It was proven in~\cite{KS240102533} that the anomaly of a $G_\mathrm{tot}$ symmetry with a locality-preserving action in a ${(1+1)}$D bosonic theory on an infinite lattice is classified by $\cH^{3}(G_\mathrm{tot},\Uone)$. This includes the modulated symmetry ${G_\mathrm{tot} = G\rtimes \Z}$, and agrees with expectations based on the crystalline equivalence principle~\cite{TE161200846}.}

If $G$ were a uniform symmetry in this example, then ${\rho_{-1} = \mathrm{id}_G}$ and~\eqref{anomCancel1dTransl} implies that there is an LSM anomaly iff ${[\nu^{(2)}]\neq[1]}$. Furthermore, in the case of a uniform $G$ symmetry, the equivalence condition~\eqref{0-strat anom equiv cond} is trivial, and the classification~\eqref{0-strat anom class transl} simplifies to $\cH^{2}(G,\Uone)$. When $G$ is a modulated symmetry, however, the total symmetry can be LSM anomaly-free even when ${[\nu^{(2)}]\neq[1]}$. As summarized in Table~\ref{LSMtable}, whether or not there is an LSM anomaly depends on both the 0-stratum anomaly data $[\nu^{(2)}]$ and spatial modulation $\rho$. While we focused on ${d=1}$ and $0$-stratum anomalies here, Table~\ref{LSMtable} applies generally to any $d$-dimensional Bravais lattice. For instance, if lattice translation symmetry were added to the examples from Section~\ref{stratified syms and anomaly examples} with trivial $\rho$, their stratified anomalies would give rise to LSM anomalies with lattice translation whenever they were translation-covariant. We will discuss examples for modulated symmetries in Section~\ref{ExampleSection}.

\subsubsection{SPT-LSM theorems}\label{SPTclassSec}

An interesting feature of modulated symmetries with translations is that they can be LSM anomaly-free even when their stratified anomaly is nonzero on a Bravais lattice. That is, there are modulated SPTs compatible with nonzero stratified anomalies. It is then natural to ask: What are the consequences of stratified anomalies for these modulated SPTs?

One possibility is that the stratified anomaly affects the classification of modulated SPTs. However, this is not the case. Stratified anomalies only affect the existence of an SPT, not its classification. Indeed, for an SPT-compatible stratified anomaly with anomaly data ${\{\nu^{(n+2)}_{n\text{-chain}}\}_{n=0}^{d-1}}$, there is a modulated SPT real-space construction whose data ${\{\t{\om}^{(n+1)}_{n\text{-chain}}\}_{n=0}^{d}}$ satisfies ${\pp^{\mathbb{T}_d}_{n+1}\t{\om}^{(n+2)}_{n+1\text{-chain}} = -\nu^{(n+2)}_{n\text{-chain}}}$. Once a single solution ${\{\t{\om}^{(n+1)}_{n\text{-chain}}\}_{n=0}^{d}}$ is identified, every other modulated SPT can be constructed by starting with the SPT with ${\{\t{\om}^{(n+1)}_{n\text{-chain}}\}_{n=0}^{d}}$ and then stacking with other modulated SPTs without a stratified anomaly. Therefore, the real-space construction data for modulated SPTs with a SPT-compatible stratified anomaly all take the form
\begin{equation}
    \{\t{\om}^{(n+1)}_{n\text{-chain}} + \xi^{(n+1)}_{n\text{-chain}}\}_{n=0}^{d},
\end{equation}
with ${\pp^{\mathbb{T}_d}_n \xi^{(n+1)}_{n\text{-chain}} = 0}$. Mathematically, this makes the classification of modulated SPTs compatible with a stratified anomaly a torsor over the usual modulated SPT classification group. However, this classification is still based on the $n$-cycles ${\xi^{(n+1)}_{n\text{-chain}}}$ and, thus, SPT-compatible stratified anomalies do not change the number of allowed modulated SPT.

While stratified anomalies do not affect the classification of modulated SPTs, they do give rise to SPT-LSM theorems. An SPT-LSM theorem is an obstruction to a trivial SPT~\cite{L170504691, YJV170505421, LRO170509298, ET190708204, JCQ190708596, PLA240918113}, which is an SPT whose ground state responds trivially to all symmetry-based probes (e.g., gauge fields, symmetry defects and fluxes, etc.).\footnote{It is sometimes said that there is a unique trivial SPT, and it is defined as the SPT with a trivial symmetric boundary. This, however, is generally not true and is an inadequate perspective for fully characterizing and distinguishing SPTs. Firstly, it is possible for two different SPTs to both admit trivial symmetric boundary conditions (e.g., see~\cite[Appendix G]{SS240401369}). Secondly, introducing a spatial boundary often explicitly breaks crystalline symmetries.}\textsuperscript{,\,}\footnote{For SPTs of only internal symmetries, an SPT that includes a product state $\bigotimes_{c\in\La}\ket{\psi_c}$ as a ground state is a trivial SPT. With crystalline symmetries, however, a nontrivial SPT can have a product state as a ground state, and there can be multiple symmetric product states that belong to different SPT phases (e.g., weak SPTs in ${(1+1)}$D).} In the real-space construction, a trivial SPT corresponds to an SPT with ${\om^{(n+1)}_{n\text{-chain}} = 0}$ for all $n$. When there is an SPT-compatible stratified anomaly with ${\nu^{(n+2)}_{n\text{-chain}}\neq 0}$, the trivial SPT is not allowed since ${\om^{(n+1)}_{n\text{-chain}} = 0}$ would not satisfy the anomaly-cancellation condition~\eqref{inflow anom cancel cond transl}. Thus, every modulated SPT with ${\nu^{(n+2)}_{n\text{-chain}}\neq 0}$ obeys an SPT-LSM theorem. However, the notion of a trivial SPT is non-canonical,\footnote{For example, a trivial SPT can be made into a nontrivial SPT using an SPT entangler, which, as a finite-depth quantum circuit, can be viewed as simply changing the basis of the Hilbert space. More drastically, one can imagine starting with two different SPTs, one of which is called trivial, and then applying a finite-depth random quantum circuit. Afterwards, which of the two SPTs was initially trivial is unclear, demonstrating that there is no invariant notion of a trivial SPT.} and SPT-LSM theorems are basis and Hilbert space-dependent. For example, an SPT-LSM theorem is not robust under tensoring ancillas. Indeed, if a stratified anomaly is SPT-compatible, then each ${\nu^{(n+2)}_{n\text{-chain}}}$ is equivalent to $0$ by adding ancillas. Once the stratified anomaly is zero, however, the trivial SPT ${\om^{(n+1)}_{n\text{-chain}} = 0}$ is allowed.

\subsection{Examples}\label{ExampleSection}

We now discuss examples of stratified anomalies of modulated symmetries and whether they give rise to LSM anomalies with lattice translations. We will assume that the underlying topological space of the $d$-dimensional lattice is $\R^d$ and the lattice translation group is $\Z^d$.

\subsubsection{Dipole symmetry in \texorpdfstring{${(1+1)}$}{(1+1)}D}\label{1+1d dipole ex sec}

The first example we consider is a $\Z_N$ dipole symmetry in ${(1+1)}$D with $\Z$ lattice translations. This is a modulated ${\Z_N\times \Z_N}$ symmetry. We use additive notation for both groups $\Z$ and ${\Z_N\times \Z_N}$, and denote their elements by ${n\in \Z}$ and ${g\equiv (g_0,g_x) \in \Z_N\times\Z_N}$. The action of the translation ${n\in \Z}$ on ${(g_0,g_x) \in \Z_N\times\Z_N}$ is
\begin{equation}\label{1ddipoleSymAction}
    \rho_n(\,(g_0,g_x)\,) = (g_0+n\,g_x, g_x),
\end{equation}
where ${n\,g_x}$ is shorthand for ${\sum_{i=1}^n g_x}$ when ${n>0}$ and ${\sum_{i=1}^{-n} -g_x}$ when ${n<0}$. If we instead considered $\La$ to be a finite size ring of $L$ sites, in order for the dipole symmetry modulation $\rho_n$ to satisfy ${\rho_L = \mathrm{id}_{\Z_N\times \Z_N}}$, the number of sites $L$ must satisfy ${L = 0~\bmod N}$.

Consider a stratified $\Z_N$ dipole symmetry operator whose 0-stratum anomaly at site $j$ is described by 
\begin{equation}
    [\nu^{(2)}_j] = \rho_{j}^*[\nu^{(2)}]\in \cH^2(\Z_N\times\Z_N,\Uone)\cong\Z_N.
\end{equation}
 $\rho_{j}^*$ is the pullback of $\rho_{n=j}$ from~\eqref{1ddipoleSymAction}. It defines a ${\cH^2(\Z_N\times\Z_N,\Uone)}$-automorphism. However, $\rho_{j}^*$ is the trivial automorphism for all $j$. This is easily seen at the level of 2-cocycles. Every class ${[\nu^{(2)}]}$ has a representative 2-cocycle that satisfies ${\nu^{(2)}(g,h) = \ee^{\frac{2\pi k_{\nu} \ii}{N}  g_0h_x}}$ with ${k_{\nu}\in\Z_N}$. The cohomology class $\rho_{j}^*[\nu^{(2)}]$ has a representative 2-cocycle ${\rho_{j}^*\nu^{(2)}(g,h) = \nu^{(2)}(g,h)\ee^{\frac{2\pi k_{\nu} j\, \ii}{N}  g_xh_x}}$. However, $\ee^{\frac{2\pi k_{\nu} j\, \ii}{N}  g_xh_x}$ is a 2-coboundary:
\begin{equation}
    \ee^{\frac{2\pi k_{\nu} j\, \ii}{N}  g_xh_x} = \del b(g,h) \equiv \frac{b(g) b(h)}{b(gh)},
    \qquad
    b(g) = 
    \begin{cases}
    \ee^{-\frac{\pi k_{\nu} j\,\ii}{N} g_x^2} \qquad &N~\text{even},\\ 
    \ee^{-\frac{\pi k_{\nu} j\,\ii}{N} g_x(g_x-1)} \qquad &N~\text{odd}.
    \end{cases}
\end{equation}
Therefore, the 2-cocycle $\rho_{j}^*\nu^{(2)}$ is cohomologous to $\nu^{(2)}$, and ${\rho_{j}^*[\nu^{(2)}] = [\nu^{(2)}]}$. Thus, the 0-stratum anomalies of a $\Z_N$ dipole symmetry in ${(1+1)}$D satisfy
\begin{equation}\label{uniform 0 strat anom}
    [\nu^{(2)}_j] = [\nu^{(2)}].
\end{equation}
Because $\rho_{j}^*$ is trivial, the consequences and classification of stratified anomalies of a $\Z_N$ dipole symmetry are the same as those for a uniform ${\Z_N\times\Z_N}$ symmetry. In particular, from~\eqref{anomCancel1dTransl}, there is an LSM anomaly whenever ${[\nu^{(2)}]\neq [1]}$, and the classification~\eqref{0-strat anom class transl} of LSM anomalies simplifies to ${\cH^{2}(\Z_N\times \Z_N,\Uone)\cong \Z_N}$. 

The simplest realization of this stratified dipole symmetry with stratified anomaly is in a system with local Hilbert space ${\scrH_j = \C^{N}\otimes \C^{N}}$. We work in the basis ${\ket{n,\t n}\equiv \ket{n}\otimes \ket{\t n}}$ of $\scrH_j$ where ${n,\t n \in \Z_N}$, and use the generalized Pauli operators
\begin{equation}\label{Zn generalized paulis}
\begin{aligned}
    X\ket{n,\t n} &= \ket{n+1,\t n},\qquad Z\ket{n,\t n} = \ee^{\frac{2\pi\ii}{N} n} \ket{n,\t n},\\
    \t X\ket{n,\t n} &= \ket{n,\t n+1},\qquad \t Z\ket{n,\t n} = \ee^{\frac{2\pi\ii}{N}\t n} \ket{n,\t n}.
\end{aligned}
\end{equation}
Suppose the translation symmetry element ${n\in \Z}$ is represented by the operator $T_n$ that satisfies
\begin{equation}\label{vanilla transl znzn qudits}
    T_n X_j T_n^\dag = X_{j-n},\qquad T_n Z_j T_n^\dag = Z_{j-n},\qquad
    T_n \t{X}_j T_n^\dag = \t{X}_{j-n},\qquad T_n \t{Z}_j T_n^\dag = \t{Z}_{j-n}.
\end{equation}
Then, a $\Z_N$ dipole symmetry operator with 0-stratum anomaly~\eqref{uniform 0 strat anom} is ${U_g = U_{(1,0)}^{g_0}U_{(0,1)}^{g_x}}$ where\footnote{The operators $\t{X}_j$ are included in $U_{(0,1)}$ to ensure that each 0-stratum operator furnishes a faithful representation of ${\Z_N\times \Z_N}$. When $N$ is prime, $Z^{-k_\nu}_j$ is an order $N$ operator for all $k_\nu$, and then the operators $\t{X}_j$ and corresponding qudits are not needed.}
\begin{equation}
    U_{(1,0)} = \prod_j X_j,
    \qquad
    U_{(0,1)} = \prod_j Z^{-k_\nu}_j X^{j}_j\,\t{X}_j.
\end{equation}
Indeed, this symmetry operator for a general group element $g$ satisfies
\begin{equation}
    T_n U_g T_{n}^\dag = 
    U_{(1,0)}^{g_0}(U_{(1,0)}^{n}U_{(0,1)})^{g_x}
    =U_{\rho_n(g)}.
\end{equation}
Furthermore, the stratum operator at site $j$ is ${U_{g}^{(j)}= Z^{-k_\nu\, g_x}_j X^{g_0 + j\, g_x}_j \t{X}^{g_x}_j}$. It is acted on by $T$ as
\begin{equation}
    T_n U_{g}^{(j)} T^\dag_n = U_{\rho_n(g)}^{(j-n)}.
\end{equation}
$U_{g}^{(j)}$ is a faithful $G$-representation and satisfies
\begin{equation}
    U_{g}^{(j)}U_{h}^{(j)} (U_{g+h}^{(j)})^\dag
    =
    \ee^{\frac{2\pi\ii}{N}k_\nu h_x(g_0 + j g_x)},
\end{equation}
which is the aforementioned 2-cocycle $\rho_j^*\nu^{(2)}(g,h)$ whose cohomology class is $[\nu^{(2)}]$.

\subsubsection{Exponential symmetry in \texorpdfstring{${(1+1)}$}{(1+1)}D}

In the dipole symmetry example, we found that the spatial modulation did not affect LSM anomalies. We now consider an example where spatial modulation does have an effect. Consider the modulated ${\Z_N\times\Z_N}$ symmetry in ${1+1}$D whose spatial modulation is described by
\begin{equation}\label{exponentialSymAction}
    \rho_n(\,(g_1,g_2)\,) = (a^n\,g_1, b^n\,g_2).
\end{equation}
The nonzero integers $a$ and $b$ must be coprime to $N$ to ensure that their multiplicative inverses $a^{-1}$ and $b^{-1}$ exist modulo $N$. This modulated symmetry is called a ${\Z_N\times\Z_N}$ exponential symmetry. As before, we assume there is one site per unit cell and that the lattice is infinite. If the lattice was instead a finite ring of $L$ sites, the integers $a$ and $b$ must satisfy ${a^L = b^L = 1~\bmod N}$ in order for ${\rho_L = \mathrm{id}_{\Z_N\times \Z_N}}$.

Consider a stratified exponential symmetry operator whose 0-stratum anomaly at site $j$ is described by ${[\nu^{(2)}_j] = \rho_{j}^*[\nu^{(2)}]\in \cH^2(\Z_N\times\Z_N,\Uone)}$. The automorphism $\rho_{j}^*$ acts on a representative 2-cocycle ${\nu^{(2)}(g,h) = \ee^{\frac{2\pi k_{\nu} \ii}{N}  g_1h_2}}$ of $[\nu^{(2)}]$ by ${\rho_j^* \nu^{(2)}(g,h) = \ee^{\frac{2\pi k_\nu \ii  }{N} a^j b^j\, g_1 h_2}}$. Therefore, the action of $\rho_{j}^*$ on $[\nu^{(2)}]$, and equivalently the 0-stratum anomaly at site $j$, is
\begin{equation}\label{exp sym 0 stratum anom}
    [\nu^{(2)}_j] = \rho_{j}^*[\nu^{(2)}] = [\nu^{(2)}]^{a^j b^j}.
\end{equation}

To determine if this 0-stratum anomaly leads to an LSM anomaly with translations, we can use the condition~\eqref{anomCancel1dTransl} from Section~\ref{LSM section}. That is, the exponential symmetry is LSM anomaly-free iff there exists a $[\ga^{(2)}]\in \cH^2(\Z_N\times \Z_N,\Uone)$ such that
\begin{equation}\label{nu to om exp sym ex}
    [\nu^{(2)}] = [\ga^{(2)}]^{-1} \,\rho_{-1}^*[\ga^{(2)}] = [\ga^{(2)}]^{a^{-1}b^{-1}-1}.
\end{equation}
To proceed, we write $[\nu^{(2)}]$ and $[\ga^{(2)}]$ as ${[\nu^{(2)}] = [\al]^{k_\nu}}$ and ${[\ga^{(2)}]= [\al]^{k_\ga}}$, where ${[\al]}$ is the generator of ${\cH^2(\Z_N\times \Z_N,\Uone)}$ with representative 2-cocycle ${\al(g,h) = \ee^{\frac{2\pi\ii}{N}g_1 h_2}}$. Using this decomposition, the LSM anomaly-free condition is equivalent to finding a $k_\ga$ such that
\begin{equation}\label{linear cong exp sym ex}
    k_\nu = k_\ga (a^{-1}b^{-1}-1)~\bmod N.
\end{equation}
Using that $a$ and $b$ are coprime to $N$ and defining the shorthand
\begin{equation}
    c:=\gcd(ab-1,N),
\end{equation}
the solutions to the linear congruence~\eqref{linear cong exp sym ex} are
\begin{equation}\label{komega solution}
    k_\ga =
    \begin{cases}
        \frac{k_\nu}{c}ab\left(\frac{1-ab}{c}\right)_{N/c}^{-1} + \kappa \frac{N}{c} \qquad &\text{if }c\mid k_\nu,\\
     \text{does not exist} \qquad &\text{if }c\nmid k_\nu,
    \end{cases}
\end{equation}
where ${\kappa\in \{0,1,\cdots,c-1\}}$, ${(\,\cdots\,)_{N/c}^{-1}}$ denotes the multiplicative inverse modulo $N/c$, and ${c\mid k_\nu}$ means that $c$ divides $k_\nu$. Therefore, the stratified anomaly leads to an LSM anomaly iff ${c\nmid k_\nu}$.

Whether or not $c$ divides $k_\nu$, and thus whether or not there is an LSM anomaly, depends sensitively on the spatial modulation and 0-stratum anomaly, i.e., on $a$, $b$, and $k_\nu$. For example, when:
\begin{itemize}
    \item \underline{${ab = 1~\bmod N}$:} In this case, ${c = N}$, so ${c\mid k_\nu}$ iff ${k_\nu = 0 ~\bmod N}$. Therefore, there is an LSM anomaly for all ${k_\nu\neq 0 ~\bmod N}$ when ${ab = 1~\bmod N}$. This happens because $\rho^*_n$ is trivial when ${ab = 1~\bmod N}$, which is similar to the example of a $\Z_N$ dipole symmetry.
    \item \underline{${c = 1}$:} In this case, $c$ always divides $k_\nu$, so there is never an LSM anomaly. Instead, there is always an SPT with ${k_\om = k_\nu\,ab
    \,(1-ab)^{-1}}$ and, therefore, an SPT-LSM theorem when ${k_\nu\neq 0 ~\bmod N}$.
    \item \underline{${\gcd(k_\nu,N) = 1}$:} When $k_\nu$ is coprime to $N$, then ${c\mid k_\nu}$ iff ${c=1}$. Therefore, in this case, there is an LSM anomaly whenever ${c\neq 1}$ and an SPT-LSM theorem otherwise.
\end{itemize}
The equivalence~\eqref{0-strat anom equiv cond} introduces the 0-stratum anomaly equivalence condition
\begin{equation}
    [\nu^{(2)}] \sim [\nu^{(2)}]\,[\ga^{(2)}]^{a^{-1}b^{-1}-1}.
\end{equation}
This implies that ${ k_\nu \sim k_\nu + a^{-1}b^{-1}-1}$. Combining this with the fact that ${ k_\nu \sim k_\nu + N}$, it implies that ${ k_\nu \sim k_\nu + c}$, from which we find that the classification of 0-stratum anomalies is
\begin{equation}
    \cH^1(\Z, \cH^{2}(G,\Uone)_{\rho^*}) = \Z_{c}.
\end{equation}

To emphasize that a nontrivial 0‑stratum anomaly can nevertheless be compatible with a translation-symmetric SPT phase, we now discuss an explicit commuting‑projector model of an SPT whose symmetry operators have ${k_\nu=1}$. In this case, the LSM anomaly vanishes iff ${c:=\gcd(ab-1,N)=1}$. For simplicity, we assume $N$ is a prime integer, which we emphasize by setting ${N = p}$. Then, ${c=1}$ for all $a$ and $b$ such that ${ab \neq 1 \bmod p}$.

The model will have a tensor product Hilbert space with local Hilbert space ${\scrH_j = \C^{p}\otimes \C^{p}}$, generalized Pauli operators~\eqref{Zn generalized paulis} with ${N=p}$, and translation operator~\eqref{vanilla transl znzn qudits}. The ${\Z_p\times\Z_p}$ exponential symmetry with 0-stratum anomaly~\eqref{exp sym 0 stratum anom} is realized in this Hilbert space as ${U_g = U_{(1,0)}^{g_1}U_{(0,1)}^{g_2}}$ where
\begin{equation}\label{double exp sym ops2}
    U_{(1,0)} = \prod_{j} (X_j)^{a^{j}},\qquad U_{(0,1)} = \prod_{j} (Z_j^{\dag} \t{Z}_j)^{b^j}.
\end{equation}
The symmetry operator for a general group element $g$ satisfies
\begin{equation}
    T_n U_g T_{n}^\dag = (U_{(1,0)}^{a^n})^{g_1}(U_{(0,1)}^{b^n })^{g_2}
    =U_{\rho_n(g)}.
\end{equation}
Furthermore, the stratum operator at site $j$ is ${U_{g}^{(j)}= (X_j)^{a^{j}g_1}(Z_j^{\dag} \t{Z}_j)^{b^j g_2}}$, which is a faithful ${\Z_p\times\Z_p}$ representation since $p$ is prime. It satisfies
\begin{equation}
    T_n U_{g}^{(j)} T_n^\dag= U_{\rho_n (g)}^{(j-n)},
\end{equation}
and
\begin{equation}
    U_{g}^{(j)}U_{h}^{(j)} (U_{g+h}^{(j)})^\dag
    = \ee^{-\frac{2\pi a^{j} b^j \ii}{p}  g_2 h_1}
    ,
\end{equation}
which is a 2-cocycle whose cohomology class is $[\nu^{(2)}]^{a^j b^j}$ with ${k_\nu = 1}$. Thus, it has the 0-stratum anomaly~\eqref{exp sym 0 stratum anom}.

Consider the Hamiltonian\footnote{We refer the reader to Appendix~\ref{ExpSPTModelDerivation} for a derivation of this Hamiltonian~\eqref{exp sym spt stab code} using gauging.}
\begin{equation}\label{exp sym spt stab code}
    H = -\sum_{j} \left(A_j + A_j^\dag + B_j + B_j^\dag\right) ,
\end{equation}
where 
\begin{equation}
    A_j = \t{X}_j^{\dag}\, X_j^{ (ab-1) }\, \t{X}^{a}_{j+1},\qquad
    B_j = Z_{j-1}^{-a} \, \t{Z}_j^{(ab-1)}\, Z_j.
\end{equation}
This Hamiltonian commutes with $T_n$, $U_{(1,0)}$, and $U_{(0,1)}$ and, therefore, has the exponential symmetry. 
It is a stabilizer code model, made from the mutually commuting stabilizers ${A_j}$ and ${B_j}$. Therefore, $H$ is a gapped Hamiltonian and exactly solvable. The stabilizers are independent and $H$ has a unique ground state
\begin{equation}\label{expSPTstate}
    \ket{\psi} \propto \sum_{\{m_j\}} \bigotimes_{j} \ket{\,(ab-1)\, m_j, a\, m_{j-1} - m_j \,},
\end{equation}
where ${m_{j}\in\{0,1,\cdots, p-1\}}$. This is a ground state because it satisfies ${A_j\ket{\psi} = B_j\ket{\psi} = \ket{\psi}}$. Furthermore, because
\begin{equation}\label{U10 A op}
    U_{(1,0)} = \prod_{j} A_j^{(ab-1)^{-1}\,a^j},
    \qquad
    U_{(0,1)} = \prod_{j} B_j^{(ab-1)^{-1}\,b^j},
\end{equation}
with ${(ab-1)^{-1}}$ the $\Z_p$ inverse of ${(ab-1)}$, the ground state also satisfies ${U_g\ket{\psi} = \ket{\psi}}$ for every ${g\in\Z_p\times \Z_p}$.

The domain wall decoration pattern~\cite{CLV13034301} for this SPT can be found by truncating in space the $U_{(1,0)}$ and $U_{(0,1)}$ operators written as~\eqref{U10 A op}. Doing so yields the string operators for the SPT:
\begin{align}
    \label{AstringOp}\prod_{j\geq I} A_j^{(ab-1)^{-1}\,a^j} &= \t{X}_I^{-(ab-1)^{-1}\,a^I} \, \prod_{j\geq I}  (X_j)^{a^j},\\
    \label{Bstringop}\prod_{j\geq I} B_j^{(ab-1)^{-1}\,b^j} &= Z_{I-1}^{-(ab-1)^{-1}ab^{I}}\,\prod_{j\geq I}  (Z^\dag_{j}\t{Z}_j)^{b^j}
\end{align}
The string operator~\eqref{AstringOp} creates a $U_{(1,0)}$ symmetry defect at ${\<I-1,I\>}$ but also acts $\t{X}_I^{-(ab-1)^{-1}\,a^I}$ which carries ${-(ab-1)^{-1}\,a^Ib^I}$ units of $U_{(0,1)}$ symmetry charge. Similarly, the string operator~\eqref{Bstringop} creates a $U_{(0,1)}$ symmetry defect at ${\<I-1,I\>}$ but also acts $Z_{I-1}^{-(ab-1)^{-1}\,ab^{I}}$ which carries ${(ab-1)^{-1}\,a^Ib^I}$ units of $U_{(1,0)}$ symmetry charge. Since both~\eqref{AstringOp} and~\eqref{Bstringop} act as the identity on the ground state, the ground state proliferates the ${U_{(1,0)}}$ and ${U_{(0,1)}}$ domain walls dressed by position-dependent symmetry charges ${-(ab-1)^{-1}\,a^Ib^I}$ and ${(ab-1)^{-1}\,a^Ib^I}$ of ${U_{(0,1)}}$ and ${U_{(1,0)}}$, respectively. This is the same decoration pattern described by the real-space construction of this SPT:
\begin{equation}
    \relax[\om^{(2)}_{\<I-1,I\>}] = \rho_{I-1}^*[\om^{(2)}] = [\om^{(2)}]^{a^{I-1} b^{I-1}} = [\nu^{(2)}]^{(ab-1)^{-1}a^{I} b^{I}},
\end{equation}
which follows from~\eqref{inflow anom cancel cond transl}.

\subsubsection{Dipole symmetry in \texorpdfstring{${(2+1)}$}{(2+1)}D}

For the final example, we consider a modulated symmetry in ${(2+1)}$D with both 0-stratum and 1-stratum anomalies. The spatial lattice is taken to be an infinite square lattice with one site per unit cell. 

Before starting, let us first generalize the LSM anomaly-free condition~\eqref{anomCancel1dTransl} of a ${d=1}$ Bravais lattice to a ${d=2}$ infinite square lattice. We denote the 0-cells of the lattice by ${\bm{r}\in \Z^2}$, the horizontal and vertical 1-cells by ${\<\bm{r}, \bm{r}+\bm{\hat{x}}\> \equiv (\bm{r},x)}$ and ${\<\bm{r}, \bm{r}+\bm{\hat{y}}\> \equiv (\bm{r},y)}$, respectively, and the $2$-cell whose center is displaced from $\bm{r}$ by ${\frac12(\bm{\hat{x}} + \bm{\hat{y}})}$ as $\Box_{\bm{r}}$. The $\Z^2$ translation symmetry acts on the lattice by ${\bm{r}\mapsto \bm{r}+\bm{v}}$ for a lattice vector ${\bm{v}\in \Z^2}$. For this setup, the LSM anomaly-free condition~\eqref{LSM anom cond mod sym transl} is nontrivial only when ${n=0}$ and ${n=1}$. The coefficients of $\nu^{(n+2)}_{n\text{-chain}}$ and $\ga^{(n+2)}_{n+1\text{-chain}}$ satisfy the translation-covariance conditions
\begin{gather}
    \label{0startumanomdip2dtranlcov}[\nu^{(2)}_{\bm{r}}] = \rho^*_{\bm{r}}[\nu^{(2)}],
    \qquad[\nu^{(3)}_{\bm{r},x}] = \rho^*_{\bm{r}}[\nu^{(3)}_h],
    \qquad
    [\nu^{(3)}_{\bm{r},y}] = \rho^*_{\bm{r}}[\nu^{(3)}_v],\\
    [\ga^{(2)}_{\bm{r},x}] = \rho^*_{\bm{r}}[\ga^{(2)}_h],
    \qquad
    [\ga^{(2)}_{\bm{r},y}] = \rho^*_{\bm{r}}[\ga^{(2)}_v],\qquad
    [\ga^{(3)}_{\Box_{\bm{r}}}] = \rho^*_{\bm{r}}[\ga^{(3)}],
\end{gather}
where ${[\nu^{(2)}]}$, ${[\ga^{(2)}_h]}$, and ${[\ga^{(2)}_v]}$ are ${\cH^2(G,\Uone)}$ classes, and ${[\nu^{(3)}_h]}$, ${[\nu^{(3)}_v]}$, and ${[\ga^{(3)}]}$ are ${\cH^3(G,\Uone)}$ classes. Using these, the LSM anomaly-free condition for 0-stratum anomalies is equivalent to\footnote{We use the orientation convention for the square lattice where 
\begin{align}
    \pp_1 (\bm{r},x) &= (\bm{r}+\bm{\hat{x}}) - (\bm{r}),
    \qquad
    \pp_1 (\bm{r},y) = (\bm{r}+\bm{\hat{y}}) - (\bm{r}),\qquad
    \pp_2 \Box_{\bm{r}} = (\bm{r},x) + (\bm{r}+\bm{\hat{x}},y) - (\bm{r}+\bm{\hat{y}},x) - (\bm{r},y).
\end{align}
}
\begin{equation}\label{0anomCancel2dTransl}
    [\nu^{(2)}] = [\ga^{(2)}_h] \,\rho_{-\bm{\hat{x}}}^*[\ga^{(2)}_h]^{-1}
    [\ga^{(2)}_v] \,\rho_{-\bm{\hat{y}}}^*[\ga^{(2)}_v]^{-1},
\end{equation}
and the LSM anomaly-free condition for the 1-stratum anomalies is equivalent to
\begin{equation}\label{1anomCancel2dTransl}
        \relax[\nu^{(3)}_h] = [\ga^{(3)}]^{-1} \,\rho_{-\bm{\hat{y}}}^* [\ga^{(3)}],\qquad
        [\nu^{(3)}_v] = [\ga^{(3)}] \,\rho_{-\bm{\hat{x}}}^* [\ga^{(3)}]^{-1}.
\end{equation}
With these conditions formulated, we now proceed with the example. 

Consider a $\Z_N$ dipole symmetry in ${(2+1)}$D, which is a $\Z_N^3$ modulated symmetry. Denoting its group elements by ${g\equiv (g_0,g_x,g_y)\in \Z_N^3}$, the action of ${\bm{v}\equiv(v_x,v_y) \in \Z^2}$ on $\Z_N^3$ is given by 
\begin{equation}
    \rho_{\bm{v}}(\,(g_0,\,g_x,\,g_y)\,) = (g_0 + v_x\, g_x + v_y\, g_y,\,g_x,\,g_y).
\end{equation}
We will now show that, unlike the ${(1+1)}$D $\Z_N$ dipole symmetry discussed in Section~\ref{1+1d dipole ex sec}, a nonzero stratified anomaly of a ${(2+1)}$D $\Z_N$ dipole symmetry does not always lead to an LSM anomaly with translations.

From Eq.~\eqref{0startumanomdip2dtranlcov}, the 0-stratum anomaly is specified by ${[\nu^{(2)}]\in \cH^2(\Z_N^3,\Uone)\cong \Z_N^3}$. Every degree-2 cohomology class ${[\al^{(2)}] \in \cH^2(\Z_N^3,\Uone)}$ can be written as
\begin{equation}\label{nu2 dipole 2d decomp}
    [\al^{(2)}] = [\al^{(0x)}]^{k^{(0x)}_\al}\,[\al^{(xy)}]^{k^{(xy)}_\al}\,[\al^{(y0)}]^{k^{(y0)}_\al},
\end{equation}
where $[\al^{(0x)}]$, $[\al^{(xy)}]$, and $[\al^{(y0)}]$ are the generators of $\cH^2(\Z_N^3,\Uone)$ with respective representative 2-cocycles ${ \al^{(ij)}(g,h) = \ee^{\frac{2\pi \ii}{N}g_ih_j}}$. A straightforward generalization of the argument from Section~\ref{1+1d dipole ex sec} shows that
\begin{align}
    \rho_{\bm{v}}^* [\al^{(0x)}] &= [\al^{(0x)}]\,[\al^{(xy)}]^{-v_y},\\
    \rho_{\bm{v}}^* [\al^{(xy)}] &= [\al^{(xy)}],\\
    \rho_{\bm{v}}^* [\al^{(y0)}] &= [\al^{(y0)}]\,[\al^{(xy)}]^{-v_x}.
\end{align}
Using the decomposition~\eqref{nu2 dipole 2d decomp} for $[\nu^{(2)}]$, $[\ga^{(2)}_h]$, and $[\ga^{(2)}_v]$, the LSM anomaly-free condition~\eqref{0anomCancel2dTransl} is equivalent to\footnote{The subscript on $k$ denotes which class it belongs to. For example, ${[\ga_{h}^{(2)}] = [\al^{(0x)}]^{k^{(0x)}_{\ga,h}}\,[\al^{(xy)}]^{k^{(xy)}_{\ga,h}}\,[\al^{(y0)}]^{k^{(y0)}_{\ga,h}}}$.}
\begin{align}
    k_{\nu}^{(0x)} &= 0~\bmod N,\\
    k_{\nu}^{(xy)} &= - k_{\ga,v}^{(0x)} -k_{\ga,h}^{(y0)}~\bmod N,\\
    k_{\nu}^{(y0)} &= 0~\bmod N.
\end{align}
Therefore, the 0-stratum anomalies lead to an LSM anomaly with translations iff ${k_{\nu}^{(0x)} \neq 0~\bmod N}$ and/or ${k_{\nu}^{(y0)} \neq 0~\bmod N}$. On the other hand, $k_{\nu}^{(xy)}$ can be any value of $\Z_N$ and will not lead to an LSM anomaly. 

The 1-stratum anomalies are specified by the ${\cH^3(\Z_N^3,\Uone)\cong \Z_N^7}$ cohomology classes $[\nu^{(3)}_h]$ and $[\nu^{(3)}_v]$. Every degree-3 cohomology class ${[\al^{(3)}] \in \cH^3(\Z_N^3,\Uone)}$ can be written as
\begin{equation}\label{nu3 dipole 2d decomp}
\begin{aligned}
    \relax[\al^{(3)}] = &\,[\al^{(0)}_{\mathrm{I}}]^{q_{\al}^{(0)}}\,
    [\al^{(x)}_{\mathrm{I}}]^{q_{\al}^{(x)}}\,
    [\al^{(y)}_{\mathrm{I}}]^{q_{\al}^{(y)}}\,[\al^{(0x)}_{\mathrm{II}}]^{q_{\al}^{(0x)}}\,[\al^{(xy)}_{\mathrm{II}}]^{q_{\al}^{(xy)}}\,[\al^{(y0)}_{\mathrm{II}}]^{q_{\al}^{(y0)}}\,[\al^{(0xy)}_{\mathrm{III}}]^{q_{\al}^{(0xy)}},
\end{aligned}
\end{equation}
where $[\al_{\mathrm{I}}^{(i)}]$, $[\al_{\mathrm{II}}^{(ij)}]$, and $[\al_{\mathrm{III}}^{(0xy)}]$ are the generators of $\cH^3(\Z_N^3,\Uone)$ with respective representative 3-cocycles
\begin{align}
    \label{type I 3cocycles}\al_{\mathrm{I}}^{(i)}(g,h,k) &= \ee^{\frac{2\pi \ii}{N^2}g_i\left(h_i+k_i-[h_i+k_i]_N\right)},\\
    \label{type II 3cocycles}\al_{\mathrm{II}}^{(ij)}(g,h,k) &= \ee^{\frac{2\pi \ii}{N^2}g_i\left(h_j+k_j-[h_j+k_j]_N\right)},\\
    \al_{\mathrm{III}}^{(0xy)}(g,h,k) &= \ee^{\frac{2\pi \ii}{N^2}g_0h_xk_y}.
\end{align}
In the type I and type II 3-cocycles~\eqref{type I 3cocycles} and~\eqref{type II 3cocycles}, respectively, addition in the exponent is only taken modulo $N$ when surrounded by the brackets $[\cdots]_N$. The action of $\rho_{\bm{v}}^*$ on these generators is~\cite{B250806604}
\begin{align}
    \rho^*_{\bm{v}}[\al_{\mathrm{I}}^{(0)}] &= [\al_{\mathrm{I}}^{(0)}]
    [\al_{\mathrm{II}}^{(0x)}]^{2 v_x}
    [\al_{\mathrm{II}}^{(y0)}]^{2 v_y}[\al_{\mathrm{I}}^{(x)}]^{v_x^2}
    [\al_{\mathrm{I}}^{(y)}]^{v_y^2}
    [\al_{\mathrm{II}}^{(xy)}]^{2v_x v_y},\\
    \rho^*_{\bm{v}}[\al_{\mathrm{I}}^{(x)}] &= [\al_{\mathrm{I}}^{(x)}],\\
    \rho^*_{\bm{v}}[\al_{\mathrm{I}}^{(y)}] &= [\al_{\mathrm{I}}^{(y)}] ,\\
    \rho^*_{\bm{v}}[\al_{\mathrm{II}}^{(0x)}] &= [\al_{\mathrm{II}}^{(0x)}]
    [\al_{\mathrm{I}}^{(x)}]^{v_x}
    [\al_{\mathrm{II}}^{(xy)}]^{v_y},\\
    \rho^*_{\bm{v}}[\al_{\mathrm{II}}^{(xy)}] &= [\al_{\mathrm{II}}^{(xy)}],\\
    \rho^*_{\bm{v}}[\al_{\mathrm{II}}^{(y0)}] &= [\al_{\mathrm{II}}^{(y0)}]
    [\al_{\mathrm{II}}^{(xy)}]^{v_x}
    [\al_{\mathrm{I}}^{(y)}]^{v_y},\\
    \rho^*_{\bm{v}}[\al_{\mathrm{III}}^{(0xy)}] &= 
    \begin{cases}
        [\al_{\mathrm{III}}^{(0xy)}][\al_{\mathrm{II}}^{(xy)}]^{\frac{N}{2}(v_x + v_y)} \qquad &N\text{ even},\\
        [\al_{\mathrm{III}}^{(0xy)}] \qquad &N\text{ odd}.
    \end{cases}
\end{align}
Using the decomposition~\eqref{nu3 dipole 2d decomp} for $[\nu^{(3)}_h]$, $[\nu^{(3)}_v]$, and $[\ga^{(3)}]$, the LSM anomaly-free condition~\eqref{1anomCancel2dTransl} is equivalent to (all equalities are understood modulo $N$)
\begin{align}
    \left( q_{\nu,h}^{(0)},\, q_{\nu,v}^{(0)} \right) &= \left( 0,\, 0 \right),\\
    \left( q_{\nu,h}^{(x)},\, q_{\nu,v}^{(x)} \right) &= \left( 0,\, q^{(0x)}_\ga - q^{(0)}_\ga \right),\\
    \left( q_{\nu,h}^{(y)},\, q_{\nu,v}^{(y)} \right) &= \left( q^{(0)}_{\ga}- q^{(y0)}_\ga,\, 0 \right),\\
    \left( q_{\nu,h}^{(0x)},\, q_{\nu,v}^{(0x)} \right) &= \left( 0,\, 2 q^{(0)}_\ga \right),\\
    \left( q_{\nu,h}^{(xy)}, q_{\nu,v}^{(xy)} \right) &= 
    \begin{cases}
        \left( \frac{N }2 q^{(0xy)}_{\ga} -q^{(0x)}_\ga, \frac{N }2 q^{(0xy)}_{\ga} 
        +q^{(y0)}_\ga \right) \qquad &N\text{ even},\vspace{5pt}\\
        \left( -q^{(0x)}_\ga,\, q^{(y0)}_\ga \right) \qquad &N\text{ odd},
    \end{cases}
    \\
    \left( q_{\nu,h}^{(y0)},\, q_{\nu,v}^{(y0)} \right) &= \left( -2 q^{(0)}_\ga,\, 0 \right),\\
    \left( q_{\nu,h}^{(0xy)},\, q_{\nu,v}^{(0xy)} \right) &= \left( 0,\, 0 \right).
\end{align}
Therefore, there are 1-stratum anomalies that can lead to an LSM anomaly (e.g., any 1-stratum anomaly with ${q_{\nu,h}^{(0)}\neq 0~\bmod N}$). However, there are also 1-stratum anomalies that are nonzero but LSM anomaly-free (e.g., a 1-stratum anomaly with ${q_{\nu,v}^{(x)}= 1}$, ${q_{\nu,h}^{(xy)}= -1}$, and all others zero are compatible with SPTs with ${q_{\om}^{(0x)}= 1}$).

In summary, the stratified anomaly of a $\Z_N$ dipole symmetry on a ${d=2}$ square lattice may or may not lead to an LSM anomaly with translations. An LSM anomaly can arise from either the 0-stratum or the 1-stratum anomalies. Furthermore, the total symmetry can be LSM anomaly-free even when either/both the 0-stratum and 1-stratum anomalies are nonzero, in which case there is an SPT-LSM theorem.

\section{Outlook}\label{OutlookSection}

In this paper, we introduced stratified symmetry operators and stratified anomalies, developed a cellular chain-complex framework for them, and showed how crystalline symmetries can promote stratified anomalies into Lieb-Schultz-Mattis (LSM) constraints. We then applied these ideas to characterize and classify LSM anomalies of modulated symmetries. There are many interesting follow-up directions that arise from this work:
\begin{enumerate}
    \item \txti{Generalizations}: In this paper, we focused on modulated symmetries with lattice translation symmetry in bosonic quantum lattice systems. It would be interesting to explore stratified anomalies of more general symmetries. This includes, for example, general crystalline symmetry groups, anti-unitary symmetries, general group extensions~\eqref{grp ext intro}, stratified anomalies of fermionic quantum lattice systems, and stratified anomalies of higher-group symmetries~\cite{KT13094721, CDI180204790, BCH180309336, BCH221111764}. Formulating stratified higher-group symmetry operators would generalize the notion of onsite higher-form symmetry operators~\cite{FKC250912304, FCH251023701}. These general stratified anomalies all fit into the Atiyah–Hirzebruch spectral sequence discussed in Appendix~\ref{AHSS app}. Relatedly, it would be interesting to explore stratified anomalies of Onsager symmetries, which are infinite-dimensional, non-compact Lie group symmetries of recent interest in the context of chiral symmetries~\cite{CPS240912220, GT250307708, PKC250504684, KPS251204150}. In particular, the charge operators of Onsager symmetries in dimensions greater than ${(1+1)}$D are Hamiltonians of foliated ${(1+1)}$D invertible fermionic phases, which causes their symmetry operators to naturally be stratified.   
    \item \txti{Stratifiability}: In this paper, we always started by assuming that a symmetry operator was stratified. It would be interesting to understand when a symmetry operator can be made into a stratified symmetry operator with respect to a fixed family of subsystems $\Si$. This is a generalization of onsiteability~\cite{SS250309717, KX250504719, KS250707430, TLE250721209, SZJ250721267, FCH251023701, CGT251202105}, which we call $\Si$-stratifiability. For example, it is known that an anomaly-free, finite internal symmetry is always onsiteable in ${(1+1)}$D~\cite{SS250309717}, but is not always onsiteable in higher dimensions~\cite{TLE250721209, SZJ250721267}. It would be interesting to explore whether similar relations between anomaly-free symmetries and more general stratifiability exist. Furthermore, it is also known that an onsiteable symmetry is always blend equivalent~\cite{FH190210285} to the identity operator. It would also be interesting to study the blend equivalence classes of stratifiable symmetry operators and see how this result, and similar ones, generalize.
    \item \txti{'t Hooft anomalies and gauging}: In this paper, we adopted the definition that an anomaly is an obstruction to an SPT. It would be interesting to connect the LSM anomalies of modulated symmetries we discussed with 't Hooft anomalies, which are obstructions to gauging internal symmetries. LSM and 't Hooft anomalies are known to be closely related, and it would be interesting to generalize the formalism of~\cite{S230805151} to modulated symmetries and $n$-stratum anomalies in higher dimensions. Relatedly, it would also be interesting to study which 't Hooft anomalies of quantum field theories can be matched by the LSM anomalies discussed here.
\end{enumerate}

\let\oldaddcontentsline\addcontentsline
\renewcommand{\addcontentsline}[3]{}
\section*{Acknowledgments}
\let\addcontentsline\oldaddcontentsline

We would like to thank 
{\"O}mer Aksoy,
Agn{\` e}s Beaudry,
Meng Cheng,
Kiyonori Gomi,
Yuan-Ming Lu,
Nathan Seiberg,
Sahand Seifnashri,
Shu-Heng Shao,
Ken Shiozaki,
Nikita Sopenko,
Xiao-Gang Wen,
and Carolyn Zhang
for discussions.
We further thank Arkya Chatterjee and Wucheng Zhang for feedback on the manuscript.
S.D.P. acknowledges support from the Simons Collaboration on Ultra-Quantum Matter, which is a grant from the Simons Foundation (651446, XGW). 
Part of this work was completed during the Okinawa Institute of Science and Technology (OIST) Thematic Program on ``Generalized Symmetries in Quantum Matter.''
The views expressed in this article are those of the authors and do not reflect the official policy or position of the U.S. Naval Academy, Department of the Navy, the Department of Defense, or the U.S. Government.

\appendix

\section{General theory for stratified anomalies}\label{AHSS app}

In this Appendix section, we discuss aspects of stratified anomalies for a general symmetry $G_\mathrm{tot}$ that fits into the group extension~\eqref{GtotGrpExt}. This includes stratified anomalies of crystalline symmetries, as well as symmetry groups $G_\mathrm{tot}$ for which~\eqref{GtotGrpExt} does not split. Anomalies of $G_\mathrm{tot}$ in $d$-dimensional space $X_d$ are conjectured to be classified by an equivariant generalized homology group denoted by $h^{G_\mathrm{tot}}_{-1}(X_d)$~\cite{SXG181000801}. There is a family of related equivariant generalized homology theories $h^{G_\mathrm{tot}}_{k}(X_d)$ labeled by the integer $k$. For example, while $h^{G_\mathrm{tot}}_{-1}(X_d)$ classifies the anomalies of $G_\mathrm{tot}$ on $X_d$, $h^{G_\mathrm{tot}}_{0}(X_d)$ classifies the SPTs of $G_\mathrm{tot}$ on $X_d$. In what follows, we will review the Atiyah–Hirzebruch spectral sequence (AHSS) used to compute $h^{G_\mathrm{tot}}_{-1}(X_d)$ and show how stratum anomalies arise as part of the input data for the AHSS. From here on, we will refer to the AHSS for $h^{G_\mathrm{tot}}_{k}(X_d)$ as just the AHSS.

\subsection{The Atiyah–Hirzebruch spectral sequence}

Consider a crystalline symmetry group ${G_\mathrm{s}}$ and the cellulation $\La$ of $X_d$ formed by the Wyckoff positions of $G_\mathrm{s}$ (see Appendix~\ref{def net mod spt app}). The AHSS~\cite{SXG181000801} starts with the collection of abelian groups (${n = 0,1,\cdots, d}$ and ${q\in\Z_{\geq 0}}$)
\begin{equation}\label{E1 page}
    E^1_{n,q} = \prod_{c_n\in \La/G_\mathrm{s}} h^{q+1}(G_{c_n}),
\end{equation}
where ${\La/ G_\mathrm{s}}$ denotes a choice of orbit representatives of cells under $G_{\mathrm s}$, and $G_{c_n}$ is the little group of $G_\mathrm{tot}$ at $c_n$. The abelian groups $h^{q+1}(G_{c_n})$ are the input of the AHSS. For example, $h^{n+2}(G_{c_n})$ is the group of possible $n$-stratum anomalies of $G_\mathrm{tot}$ at $c_n$. For bosonic, group cohomology anomalies, each ${h^{n+2}(G_{c_n}) = \cH^{n+2}(G_{c_n}, \Uone)}$. The groups $E^1_{n,q}$ are defined using a fundamental domain of $\La$ since $h^{q+1}(G_{c_n})$ data for $n$-cells outside of $\La/G_\mathrm{s}$ follow from those inside of $\La/G_\mathrm{s}$ by $G_\mathrm{s}$-covariance (e.g.,~\eqref{gen cell anomal covariance cond} for $n$-stratum anomalies).

The collection of groups~\eqref{E1 page} is called the $E^1$-page of the AHSS. The group $E^1_{n,n+1}$ describes possible $n$-stratum anomalies of $G_\mathrm{tot}$, and is a generalization of the cellular $n$-chain group~\eqref{Td invariant nchains}. The $E^1$-page includes more information than stratified anomalies. For example, the group ${E^1_{n,n}}$ describes decoration of $n$-cells by ${(n+1)}$D SPTs, and is the starting point for classifying $G_\mathrm{tot}$ SPTs using the AHSS. The group ${E^1_{n,n-1}}$, on the other hand, describes pumps of SPTs at $n$-cells.

Each element of $E^1_{n,q}$ corresponds to mutually-independent decorations of $n$-cells by elements of $h^{q+1}(G_{c_n})$. The decoration of each cell, however, must be compatible with the decorations on its neighboring cells. This introduces a compatibility condition for the $E^1$-page that is formulated using the group homomorphisms
\begin{equation}
    d^1_{n,q}\colon E^1_{n,q} \to E^1_{n-1,q}.
\end{equation}
They are defined for all ${n\geq 1}$ and called the first differentials. Each one is required to satisfy
\begin{equation}\label{d1 d^2=0}
    d^1_{n-1,q}\circ d^1_{n,q} = 0.
\end{equation}
The first differentials relate an $h^{q+1}(G_{c_n})$-valued decoration on $n$-cells to an $h^{q+1}(G_{c_{n-1}})$-valued decoration on ${(n-1)}$-cells. The locality-compatibility condition on the $E^1$-page is, for all ${e_{n,q}^1\in E_{n,q}^1}$,
\begin{equation}\label{d1 glue cond}
    d^1_{n,q} e_{n,q}^1 = 0.
\end{equation}
Elements $e_{n,q}^1$ satisfying~\eqref{d1 glue cond} have $n$-cell decorations that can be consistently glued together at ${(n-1)}$-cells.

Using~\eqref{d1 d^2=0}, the homology of the first differentials defines the $E^2$-page of the AHSS, which is made from the collection of abelian groups
\begin{equation}\label{E2 page def}
    E^2_{n,q} := \ker(d^1_{n,q})/\mathrm{im}(d^1_{n+1,q}).  
\end{equation}
Each element of $E^2_{n,q}$ corresponds to an ${e_{n,q}^1\in E_{n,q}^1}$ that satisfies~\eqref{d1 glue cond} and the equivalence relation
\begin{equation}\label{d1 equiv rel}
    e_{n,q}^1\sim e_{n,q}^1 + d^1_{n+1,q} e_{n+1,q}^1.
\end{equation}
Each element in ${\mathrm{im}(d^1_{n+1,q}) \subset E^1_{n,q}}$ corresponds to a decoration of $n$-cells that can arise from a decoration of ${(n+1)}$-cells. Thus, each group~\eqref{E2 page def} corresponds to decorations of $n$ cells that can be glued along ${(n-1)}$-cells when ${n >0}$ and cannot be trivialized by decorations on ${(n+1)}$-cells when ${n< d}$.

Let us relate this to $n$-stratum anomalies where ${q=n+1}$. The image of $d^1_{n,n+1}$ describes sources/sinks of $n$-stratum anomalies at ${(n-1)}$-cell. Such sources/sinks are ${(n-1)}$-cells at which an anomalous ${(n+1)}$D symmetry operator could be truncated. Such a truncation would violate the $G_\mathrm{tot}$ symmetry.\footnote{This follows from field theory arguments made in~\cite{JSY171007299, NY171209361, TW201215861} and lattice arguments based on blends made in~\cite{TLE250721209}.} The condition~\eqref{d1 glue cond} with ${q = n+1}$ ensures these sources/sinks do not occur. It is a generalization of~\eqref{Tinv ncycle strat anom cond} from the main text. On the other hand, the differential $d^1_{n+1,n+1}$ is interpreted as deforming SPT-compatible ancillas on $(n+1)$-cells with $n$-stratum anomalies to neighboring $n$-cells. Therefore, the equivalence relation~\eqref{d1 equiv rel} for ${q=n+1}$ is a generalization of the equivalence relation~\eqref{tinv Lat homotopy} from the main text. Thus, the abelian groups $E^2_{n,n+1}$ are generalizations of the homology groups~\eqref{Transl inv homology classes} of the main text.

For a general symmetry group $G_\mathrm{tot}$, the AHSS includes more than the $E^2$-page. In general, the AHSS includes a collection of $E^r$-pages with ${r = 1,2,\cdots, d+1}$, each of which is a collection of abelian groups $\{E^r_{n,q}\}$. The $E^{r+1}$-page is constructed from the $E^{r}$-page using $r$-th differentials. These are homomorphisms
\begin{equation}
    d^r_{n,q}\colon E^r_{n,q} \to E^r_{n-r,q+1-r},
\end{equation}
defined for ${n\geq r}$, which are required to satisfy
\begin{equation}
    d^r_{n-r,q+1-r}\circ d^r_{n,q} = 0.
\end{equation}
The groups in the $E^{r+1}$-page of the AHSS are defined by the homology of the $r$-th differentials:
\begin{equation}
    E^{r+1}_{n,q} = \ker(d^r_{n,q})/\mathrm{im}(d^r_{n+r,q+r-1}).
\end{equation}
An element in $E^{r+1}_{n,q}$ corresponds to a decoration of $n$-cells that can be glued together along $m$-cells with ${\max(0,n-r) \leq m \leq n-1}$ when ${n>0}$, and cannot be trivialized by decorations on $m$-cells with ${n+1 \leq m \leq \min(n+r,d)}$ when ${n<d}$. Therefore, $E^{d+1}_{n,q}$ describes decorations that can be glued together on all lower-dimensional cells and cannot be trivialized by any higher-dimensional decorations. The set ${\prod_{n=0}^{d} E^{d+1}_{n,n+k}}$ is equal to $h^{G_\mathrm{tot}}_{-k}(X_d)$ as a set. The group structure on $h^{G_\mathrm{tot}}_{-k}(X_d)$ is found by iteratively solving for the groups $F_{p} h_{-k} $ in the group extensions (${1\leq p \leq d}$)
\begin{equation}
    0 
    \to 
    F_{p-1} h_{-k} 
    \to
    F_{p} h_{-k} 
    \to E_{p,p+k}^{d+1} 
    \to
    0,
\end{equation}
where ${F_{0} h_{-k} = E_{0,k}^{d+1} }$ and ${F_{d} h_{-k} = h^{G_\mathrm{tot}}_{-k}(X_d)}$.

For a general symmetry group $G_\mathrm{tot}$, the set of $n$-stratum anomaly equivalence classes is given by $E^{d+1}_{n,n+1}$. In general, it is possible for an $n$-stratum anomaly class in the $E^r$-page with ${2\leq r \leq d}$ to lie outside of $\ker(d^r_{n,n+1})$ and thus not make it to the $E^{d+1}$-page. Such stratified anomalies cannot be realized by unitary operators furnishing a faithful representation of $G_\mathrm{tot}$. Similarly, it is possible for an $n$-stratum anomaly class to be nontrivial in the $E^r$-page with ${2\leq r \leq d}$, but be trivial in the $E^{d+1}$-page. Such an $n$-stratum anomaly can be realized by $G_\mathrm{tot}$ symmetry operators, but would not lead to an anomaly of $G_\mathrm{tot}$.

For the stratified anomalies described in Section~\ref{ModulatedSymmetrySection} in the main text, we expect that ${E^{d+1}_{n,n+1} = E^{2}_{n,n+1}}$ for all $d$---that the AHSS stabilizes at the $E^2$-page. To support this, recall that the differentials $d^r_{n,q}$ with ${r>1}$ become nontrivial when an $n$-cell decoration cannot be collapsed across a codimension‑$r$ cell. This arises from the codimension-$r$ cell being a high symmetry point and/or being decorated by a nontrivial symmetry defect/flux due to the extension class of $G_\mathrm{tot}$~\cite{SXG181000801}. In Section~\ref{ModulatedSymmetrySection}, the crystalline symmetries are lattice translations, which causes there to be no high symmetry points---causes the little groups ${G_{c_n} = G}$. Furthermore, Section~\ref{ModulatedSymmetrySection} considered bosonic systems with a split group extension for $G_\mathrm{tot}$. This causes the lattice not to be decorated by nontrivial symmetry defects/fluxes. Because of this, we expect $d^r_{n,q}$ with ${r>1}$ to be trivial for bosonic modulated symmetries with lattice translations, and, thus, ${E^{d+1}_{n,n+1} = E^{2}_{n,n+1}}$ for all $d$. We emphasize that this is a physical argument and mathematically conjectural. It would be interesting to prove that ${E^{d+1}_{n,n+1} = E^{2}_{n,n+1}}$ for the case discussed in Section~\ref{ModulatedSymmetrySection}.

\subsection{The first differential}\label{d1 formula app}

In the main text, whether a stratified anomaly led to an LSM anomaly was determined using an explicit formula based on the stratum anomaly data. To derive LSM anomalies for more general stratified anomalies using this approach, explicit formulas for the differentials of the AHSS are required. For the rest of this Appendix, we will discuss explicit expressions for the first-differentials of the AHSS for groups $G_\mathrm{tot}$ given by~\eqref{general mod sym group}. Instead of presenting the most general formula, we will focus on examples with ${d=1}$ and ${d=2}$. Generalizations to other dimensions and space groups are comparably straightforward. Finding general formulae for the 1st and higher differentials is interesting, but lies outside the scope of this work.

\subsubsection{\texorpdfstring{$pm$}{pm} space group}\label{pmSpaceGrpEx}

We first consider a one-dimensional Bravais lattice with both lattice translations and reflections. We denote the lattice sites by the integers ${j\in \Z}$ and have lattice translations $T$ act by ${j\mapsto j+1}$. We denote by $M$ the reflection ${j\to -j}$ centered about the site ${j=0}$. The lattice's space group is ${pm \cong \Z\rtimes\Z_2}$,\footnote{The space group $pm$ is sometimes denoted as $D_\infty$.} where ${\Z = \<T\>}$, ${\Z_2 = \<M\>}$, and the generators satisfy
\begin{equation}\label{pm group law}
    M^2 = 1, \qquad M T M = T^{-1}.
\end{equation}
We will also introduce the shorthand ${M_j \equiv T^{j} M T^{-j}}$ for reflections about the site $j$. The total symmetry group is ${G_\mathrm{tot} = G\rtimes pm}$, where $G$ is the possibly modulated internal symmetry group. The actions of $T^n$ and $M$ on $G$ are described by the $G$ automorphisms $\rho_n$ and $\rho_M$, respectively. From the $pm$ group law, they satisfy
\begin{equation}
    \rho_{n_1} \circ \rho_{n_2} = \rho_{n_1 + n_2},\qquad    
    \rho_M \circ \rho_n \circ \rho_M = \rho_{-n}.
\end{equation}

In the AHSS, we must first find the cellulation for the space group $pm$. There are two Wyckoff positions of $pm$ with nontrivial site-symmetry groups. The first are formed by the lattice sites $j$ and have the little group
\begin{equation}
    G_j = G\rtimes \<M_j\>.
\end{equation}
The second are the points ${j+1/2}$ in space, which are located at the center between neighboring lattice sites and have the little groups
\begin{equation}
    G_{j+1/2} = G\rtimes \<T M_j\>.
\end{equation}
The group ${\<T M_j\>\cong \Z_2}$ described reflections about the point ${j+1/2}$. Therefore, the 0-cells of the cellulation are the sites $j$ and edge centers ${j+1/2}$ of the lattice. The 1-cells connecting neighboring 0-cells come in two types: ${\<j-1/2,j\>}$ and ${\<j,j+1/2\>}$. We choose the orientations of ${\<j-1/2,j\>}$ and ${\<j,j+1/2\>}$ to always point away from $j$.

Having specified the cellulation, we now start with the $E^1$ page of the AHSS. Because the cellulation has lattice translation symmetry, all of the initial data for the AHSS is fully determined by the data in a unit cell, which is graphically represented by
\begin{equation}
\begin{tikzpicture}[scale = 0.5, baseline = {([yshift=-.5ex]current bounding box.center)}]
\node (L) at (0,0) {};
\node (R) at (10,0) {};
\node (midL) at (2.25,0) {};
\node (mid) at (5,0) {};
\node (midR) at (7.5,0) {};
\draw[link1, line width=0.03in,
      preaction={draw=black, line width=0.05in}] (L) -- (R);
\filldraw[fill=site1, draw=black, line width=0.8pt] (mid) circle (7pt);
\filldraw[fill=site2, draw=black, line width=0.8pt] (R) circle (7pt);
\node[black, above] at (midR) {\normalsize $\mathrm{id}$};
\node[black, above] at (midL) {\normalsize $M$};
\end{tikzpicture}
\end{equation}
The blue disk represents the 0-cell at the site ${j=0}$, and the green disk represents the 0-cell at ${j+1/2=1/2}$. The 1-cell labeled $M$ is related by ${M\in pm}$ to the fundamental domain 1-cell labeled $\mathrm{id}$.

The 0-cell data is packaged into the $E^1$-page by
\begin{align}
    ([\om^{(1)}_s], [\om^{(1)}_e]) &\in E^1_{0,0} = \cH^1(G_0,\Uone)\times \cH^1(G_{1/2},\Uone),\\
    ([\nu^{(2)}_s], [\nu^{(2)}_e]) &\in E^1_{0,1} = \cH^2(G_0,\Uone)\times \cH^2(G_{1/2},\Uone).\label{0stratdata pm ex}
\end{align}
The classes $[\om^{(1)}_s]$ and $[\om^{(1)}_e]$ are SPT data, while $[\nu^{(2)}_s]$ and $[\nu^{(2)}_e]$ describe the 0-stratum anomaly of $G_\mathrm{tot}$. Similarly, the 1-cell data is
\begin{align}
    [\mu^{(1)}]&\in E^1_{1,0} = \cH^1(G,\Uone),\\
    [\om^{(2)}]&\in E^1_{1,1} = \cH^2(G,\Uone),\\
    [\nu^{(3)}]&\in E^1_{1,2} = \cH^3(G,\Uone),
\end{align}
which respectively describe charge pump data, SPT data, and anomaly data. The nontrivial first differentials involving these groups are
\begin{align}
    d^1_{1,0}&\colon  E^1_{1,0} \to E^1_{0,0},\\
    d^1_{1,1}&\colon  E^1_{1,1} \to E^1_{0,1}.
\end{align}

The first differential $d^1_{1,0}$ physically corresponds to taking a symmetric pair of $G$ symmetry charges on each 1-cell and deforming them apart to neighboring 0-cells. However, the cohomology groups at each $1$-cell is different than those at each 0-cell since ${G_j \not\cong G}$ and ${G_{j+1/2}\not\cong G}$. The first differential must, therefore, put elements in $\cH^1(G,\Uone)$ on the same footing as those in $\cH^1(G_0,\Uone)$ and $\cH^1(G_{1/2},\Uone)$. To do so, we use that ${G\subset G_{c_0}}$ for every 0-cell $c_0$ and consider the corestriction map
\begin{equation}
    \cor_G^{G_{c_0}}\colon \cH^1(G,\Uone)\to \cH^1(G_{c_0},\Uone).
\end{equation}
We review restriction and corestriction maps in group cohomology in Appendix~\ref{cohomologyApp}. Denoting the group elements of $G_0$ by ${(g,M^{m})\in G_0}$, the image of a 1-cochain ${\mu^{(1)}\in\cC^1(G,\Uone)}$ under ${\cor_G^{G_{0}}}$ is the 1-cochain ${\cor_G^{G_0}\mu^{(1)}\in \cC^1(G_0,\Uone)}$ satisfying 
\begin{equation}\label{1dreflectionCorcochain00}
    \cor_G^{G_0}\mu^{(1)}(\,(g,M^{m})\,) = (\mu^{(1)}\rho_{M}^* \mu^{(1)})(g).
\end{equation}
A similar expression holds for ${\cor_G^{G_{1/2}}\mu^{(1)}}$, but with $M$ replaced by $TM$. The 1-cochain $\cor_G^{G_0}\mu^{(1)}$ has a simple interpretation: it arises from taking a symmetric pair of $G$ symmetry charges on each 1-cell and deforming them apart to neighboring 0-cells. This matches the physical interpretation of $d^1_{1,0}$. Therefore, we find that 
\begin{equation}
    d^1_{1,0}\left([\mu^{(1)}]\right) = \left(
    \cor_G^{G_0}[\mu^{(1)}]^{-1}
    ,\,
    \cor_G^{G_{1/2}}[\mu^{(1)}]
    \,\right).
\end{equation}
The image of 1-cochains under the corestriction map is particularly simple since the element $M^m$ does not appear on the right-hand side of~\eqref{1dreflectionCorcochain00} (see~\eqref{general cor map semi direc prodcs} for the general $n$-cochain expression).

The first differential $d^1_{1,1}$ maps the ${(1+1)}$D SPTs on 1-cells to their inflow anomaly on 0-cells. Similar to before, it must take into account that the SPT class $[\om^{(2)}]$ is not an element of $\cH^2(G_0,\Uone)$ or $\cH^2(G_{1/2},\Uone)$. This can again be achieved using corestriction maps. This time, we consider
\begin{equation}
    \cor_G^{G_{c_0}}\colon \cH^2(G,\Uone)\to \cH^2(G_{c_0},\Uone).
\end{equation}
The image of a 2-cochain ${\om^{(2)}\in\cC^2(G,\Uone)}$ under ${\cor_G^{G_{0}}}$ is the 2-cochain ${\cor_G^{G_0}\om^{(2)}\in \cC^2(G_0,\Uone)}$ satisfying
\begin{equation}\label{1dreflectionCorcochain}
    \cor_G^{G_0}\om^{(2)}(\,(g_1,M^{m_1}),\,(g_2,M^{m_2})\,) = (\om^{(2)}\rho_{M}^* \om^{(2)})(g_1, \rho_{M^{m_1}}(g_2)).
\end{equation}
A similar expression holds for ${\cor_G^{G_{1/2}}\om^{(2)}}$, but with $M$ replaced by $TM$. The cochain expression~\eqref{1dreflectionCorcochain} makes the physical meaning of $\cor_G^{G_0}\om^{(2)}$ transparent: it implements the folding trick about the reflection center ${j=0}$ to extract the inflow anomaly at ${j=0}$.\footnote{See \Rf{B250806604} for more discussion on this folding trick for modulated SPTs.} 
Therefore,  
\begin{equation}
    d^1_{1,1}\left([\om^{(2)}]\right) = \left(
    \cor_G^{G_0}[\om^{(2)}]^{-1}
    ,\,
    \cor_G^{G_{1/2}}[\om^{(2)}]
    \,\right).
\end{equation}

Using these expressions for the first differentials, we can construct the $E_2$-page of the AHSS and classify ${G\rtimes pm}$ SPTs and ${G\rtimes pm}$ anomalies in ${(1+1)}$D. This includes the 0-stratum anomalies of ${G\rtimes pm}$ and, thus, the LSM anomalies of ${G\rtimes pm}$.

The 0-stratum anomalies are classified by
\begin{equation}
    E^2_{0,1} = \ker(d^1_{0,1})/\mathrm{im}(d^1_{1,1}) = E^1_{0,1}/\mathrm{im}(d^1_{1,1}).
\end{equation}
Thus, the 0-stratum anomalies are classified by~\eqref{0stratdata pm ex} modulo the equivalence relation
\begin{equation}
\begin{aligned}
    \relax[\nu^{(2)}_s] &\sim [\nu^{(2)}_s] \,\cor_G^{G_0}[\ga^{(2)}]^{-1},\\
    [\nu^{(2)}_e] &\sim [\nu^{(2)}_e] \,\cor_G^{G_{1/2}}[\ga^{(2)}],
\end{aligned}
\end{equation}
for ${[\ga^{(2)}]\in\cH^2(G,\Uone)}$. Therefore, 0-stratum anomalies of ${G\rtimes pm}$ do not lead to an LSM anomaly iff there exists a $[\ga^{(2)}]$ such that
\begin{equation}\label{0 strat anom cancel pm ex}
\begin{aligned}
    \relax[\nu^{(2)}_s] &= \cor_G^{G_0}[\ga^{(2)}],\\
    [\nu^{(2)}_e] &= \cor_G^{G_{1/2}}[\ga^{(2)}]^{-1}.
\end{aligned}
\end{equation}

To better understand~\eqref{0 strat anom cancel pm ex}, we can ``forget'' the lattice reflection symmetry using the restriction map
\begin{equation}
    \res_G^{G_{c_0}}\colon \cH^2(G_{c_0},\Uone)\to \cH^2(G,\Uone).
\end{equation}
Its explicit action on a 2-cochain ${\nu^{(2)}_s\in\cC^2(G_0,\Uone)}$, for example, is
\begin{equation}
    \res_G^{G_0}\nu_s^{(2)}(g_1,g_2) = \nu_s^{(2)}((g_1,1),(g_2,1)).
\end{equation}
Acting this restriction map on~\eqref{0 strat anom cancel pm ex} and using~\eqref{res cor normalizer}, we find that
\begin{equation}\label{res 0 strat anom cancel pm ex}
\begin{aligned}
    \res_G^{G_0}[\nu_s^{(2)}] &= [\om^{(2)}]\,\rho_M^*[ \om^{(2)}],\\
    \res_G^{G_{1/2}}[\nu_e^{(2)}] &= [\,\om^{(2)}]^{-1}\,\rho_{TM}^* [\om^{(2)}]^{-1}.
\end{aligned}
\end{equation}
This is precisely the LSM anomaly-free conditions for this cellulation if the 0-stratum anomaly at $c_0$ was described by ${\res_G^{G_{c_{0}}}[\nu_{c_0}^{(2)}]}$ and if $G_\mathrm{s}$ included only lattice translations. The conditions~\eqref{res 0 strat anom cancel pm ex} are necessary but not sufficient conditions for ${G\rtimes pm}$ to be LSM anomaly-free. The conditions~\eqref{0 strat anom cancel pm ex}, on the other hand, are necessary and sufficient conditions.

\subsubsection{\texorpdfstring{$p4m$}{p4m} space group}

We next consider the square lattice, whose space group is ${p4m \cong \Z^2\rtimes D_8}$, where ${D_8\cong \Z_4\rtimes\Z_2}$ is the dihedral group of order 8. Denote the lattice sites by ${(x,y)\in \Z^2}$. This space group can be generated by lattice translations ${T_x\colon (x,y)\mapsto (x+1,y)}$ and ${T_y \colon (x,y)\mapsto (x,y+1)}$, a $\pi/2$ rotation $R$ centered at ${(x,y) = (0,0)}$, and the reflection ${M_y\colon (x,y)\mapsto (x,-y)}$. These generators satisfy
\begin{equation}
\begin{gathered}
    M_y^2 = R^4 = 1,\qquad M_y R M_y = R^3,\qquad
    M_y T_x M_y = T_x,\qquad M_y T_y M_y = T_y^{-1},\\
    R T_x R^3 = T_y,\qquad  R T_y R^3 = T^{-1}_x.
\end{gathered}
\end{equation}
It is convenient to also introduce notation for the reflection ${M_x \equiv R^2 M_y\colon (x,y)\mapsto (-x,y)}$ about the $y$-axis and the $\pi/2$ rotation ${R_p \equiv T_x R}$ about the point ${(1/2,1/2)}$. The total symmetry group is ${G_\mathrm{tot} = G\rtimes p4m}$, and the action of ${s\in p4m}$ on $G$ is described by ${\rho_{s}\in \Aut(G)}$.

To build the cellulation for the space group $p4m$, we first consider its Wyckoff positions. There are three inequivalent 0d Wyckoff patches, three inequivalent 1d Wyckoff patches, and one inequivalent 2d Wyckoff patch. The cellulation they form within the unit cell is 
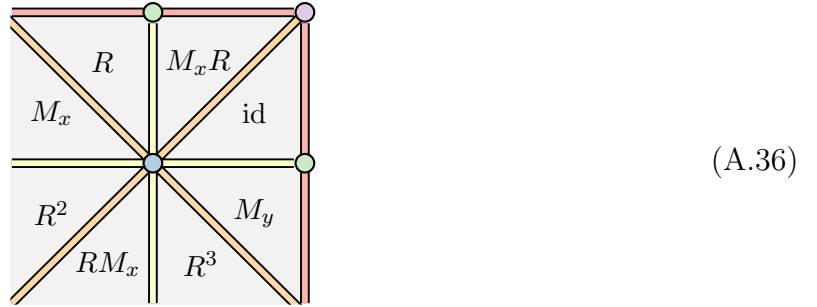
\begin{equation}\label{p4m unit cell}
\begin{tikzpicture}[scale = 0.5, baseline = {([yshift=-.5ex]current bounding box.center)}]
\node (bL) at (0,0) {};
\node (tL) at (0,8) {};
\node (tR) at (8,8) {};
\node (bR) at (8,0) {};
\node (midB) at (4,0) {};
\node (midT) at (4,8) {};
\node (midL) at (0,4) {};
\node (midR) at (8,4) {};
\node (center) at (4,4) {};
\fill[topCell] (center) +(-107pt,-107pt) rectangle +(114pt,114pt);
\draw[link2, line width=0.03in,
      preaction={draw=black, line width=0.05in}] (tL) -- (tR);
\draw[link1, line width=0.03in,
      preaction={draw=black, line width=0.05in}] (midB) -- (midT);
\draw[link1, line width=0.03in,
      preaction={draw=black, line width=0.05in}] (midL) -- (midR);
\draw[link2, line width=0.03in,
      preaction={draw=black, line width=0.05in}] (bR) -- (tR);
\draw[link3, line width=0.03in,
      preaction={draw=black, line width=0.05in}] (bL) -- ($(bL)!0.98!(tR)$);
\draw[link3, line width=0.03in,
      preaction={draw=black, line width=0.05in}] ($(bR)!0.98!(tL)$) -- ($(tL)!0.98!(bR)$);
\draw[white, line width=0.035in]   ($($(bR)!0.98!(bL)$)+(0,4.54pt)$) -- ($($(bL)!1.01!(bR)$)+(0,4.54pt)$);
\draw[white, line width=0.035in]   ($($(tL)!0.98!(bL)$)+(4.54pt,0)$) -- ($($(bL)!1.01!(tL)$)+(4.54pt,0)$);
\filldraw[fill=site3, draw=black, line width=0.8pt] (tR) circle (7pt);
\filldraw[fill=site2, draw=black, line width=0.8pt] (midT) circle (7pt);
\filldraw[fill=site2, draw=black, line width=0.8pt] (midR) circle (7pt);
\filldraw[fill=site1, draw=black, line width=0.8pt] (center) circle (7pt);
\node[xshift = 2pt] at (4-1.33, 4-2.67) {\normalsize $R M_x$};
\node[xshift = -1pt] at (4+1.33, 4-2.67) {\normalsize $R^{3}$};
\node at (4+2.67,4-1.33) {\normalsize $M_y$};
\node at (6.67,5.33) {\normalsize $\mathrm{id}$};
\node at (4-2.67,5.33) {\normalsize $M_x$};
\node at (4-2.67,4-1.33) {\normalsize $R^2$};
\node at (4-1.33,6.67) {\normalsize $R$};
\node[xshift = -2pt] at (5.33,6.67) {\normalsize $M_x R$};
\end{tikzpicture}
\end{equation}
Cells with the same color coding are equivalent Wyckoff patches (i.e., are related by a $p4m$ transformation). The blue 0-cell at the center is the ${(0,0)}$ site of the lattice. The $2$-cell labeled by $\mathrm{id}$ is a fundamental domain. The other 2-cells are labeled by the $p4m$ element that transforms the $\mathrm{id}$ 2-cell to that 2-cell.

For simplicity, here we will focus on the first differentials whose domains are the groups $E^1_{2,2}$ and $E^1_{1,1}$ in the $E^1$-page. That is, the first differentials
\begin{align}
    d^1_{2,2}&\colon E^1_{2,2} \to E^1_{1,2},\\
    d^1_{1,1}&\colon E^1_{1,1} \to E^1_{0,1}.
\end{align}
The elements of $E^1_{2,2}$ and $E^1_{1,1}$ are specified within the fundamental domain id of~\eqref{p4m unit cell}:
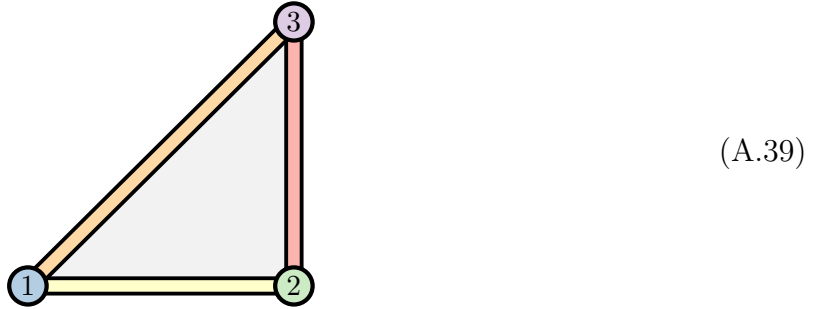
\begin{equation}\label{sqLatFundDom}
\begin{tikzpicture}[scale = 0.5, baseline = {([yshift=-.5ex]current bounding box.center)}]
\coordinate (bL) at (0,0);
\coordinate (bMid) at (3.5,0);
\coordinate (bR) at (7,0);
\coordinate (midR) at (7,3.5);
\coordinate (top) at (7,7);
\fill[topCell] (bL) -- (bR) -- (top) -- cycle;
\draw[link1, line width=0.06in,
      preaction={draw=black, line width=0.1in}] (bL) -- (bR);
\draw[link2, line width=0.06in,
      preaction={draw=black, line width=0.1in}] (bR) -- (top);
\draw[link3, line width=0.06in,
      preaction={draw=black, line width=0.1in}] (bL) -- (top);
\filldraw[fill=site1, draw=black, line width=1.6pt] (bL) circle (14pt);
\filldraw[fill=site2, draw=black, line width=1.6pt] (bR) circle (14pt);
\filldraw[fill=site3, draw=black, line width=1.6pt] (top) circle (14pt);
\node at (bL) {\normalsize $1$};
\node at (bR) {\normalsize $2$};
\node at (top) {\normalsize $3$};
\end{tikzpicture}
\end{equation}
where we have labeled the 0-cells by the integers $1$, $2$, and $3$. The orientation structure of this 2-cell specifies the orientation of the entire cellulation. We choose the orientation of the three 1-cells $[12]$, $[13]$, $[23]$ to always point from 0-cell $i$ to $j$ with ${i<j}$. The orientation of the 2-cell ${[123]}$ is determined by the right-hand rule and points out of the page. The little groups at these cells are
\begin{align}
    G_1 &= G\rtimes\<R,M_y\> \cong G\rtimes D_8,\\
    G_2 &= G\rtimes\<T_x M_x, M_y\> \cong G\rtimes (\Z_2\times \Z_2),\\
    G_3 &= G\rtimes\<R_p
    , T_x M_x\> \cong G\rtimes D_8,\\
    G_{12} &= G\rtimes\<M_y\> \cong G\rtimes \Z_2,\\
    G_{23} &= G\rtimes\<T_x M_x\> \cong G\rtimes \Z_2,\\
    G_{13} &= G\rtimes\<M_x R\> \cong G\rtimes \Z_2.
\end{align}
The group elements in $E^1_{2,2}$ and $E^1_{1,1}$ are
\begin{align}
    [\om^{(3)}] &\in E^1_{2,2} = \cH^3(G,\Uone),\\
    \left([\om^{(2)}_{12}], [\om^{(2)}_{23}], [\om^{(2)}_{13}]\right) &\in E^1_{1,1} = \prod_{i<j}\cH^2(G_{ij},\Uone),
\end{align}
which are SPT data on 2 and 1-cells, respectively. The first differentials $d^1_{2,2}$ and $d^1_{1,1}$ map this SPT data to their anomaly-inflow data. 

Let us first discuss $d^1_{2,2}$, which describes the anomalies on 1-cells coming from SPTs on 2-cells. Just as in the previous example from Section~\ref{pmSpaceGrpEx}, we again have the problem that the little groups at 1-cells and 2-cells are not isomorphic. However, just as in the previous example, we can use the corestriction maps
\begin{equation}
    \cor_G^{G_{ij}}\colon \cH^3(G,\Uone) \to  \cH^3(G_{ij},\Uone)
\end{equation}
to implement the folding trick. For example, the action of $\cor_G^{G_{12}}$ on the 3-cochain $\om^{(3)}$ is the 3-cochain $\cor_G^{G_{12}}\om^{(3)}$ with
\begin{equation}\label{2dreflectionCorcochain}
    \cor_{G}^{G_{12}}\om^{(3)}(\,(g_1,M_y^{m_1}),\,(g_2,M_y^{m_2}),\,(g_3,M_y^{m_3})\,) =\prod_{k=0}^1 \rho_{M^k_y}^*\om^{(3)}(
    g_1,
    \rho_{M^{m_1}_y}(g_2),
    \rho_{M^{m_1+m_2}_y}(g_3)).
\end{equation}
Therefore, taking into account the relative orientations of the 1 and 2-cells, we find
\begin{align}
    d^1_{2,2}\left([\om^{(3)}]\right) = \left(
    \cor_{G}^{G_{12}}[\om^{(3)}],
    \cor_{G}^{G_{23}}[\om^{(3)}],
    \cor_{G}^{G_{13}}[\om^{(3)}]^{-1}
    \right)
    \in E^1_{1,2} = \prod_{i<j}\cH^3(G_{ij},\Uone).
\end{align}

The first differential $d^1_{1,1}$ describes the anomalies on 0-cells coming from SPTs on 1-cells. It must also take into account that the little groups at 1 and 0-cells are not isomorphic. This can also be done using the folding trick, where 1-cells sharing a common 0-cell on their boundary are folded onto each other. We claim that just as before, the folding trick is mathematically implemented by the corestriction maps
\begin{align}
    \cor_{G_{ij}}^{G_{k}}\colon \cH^2(G_{ij},\Uone) \to  \cH^2(G_{k},\Uone),
\end{align}
where ${k=i}$ or $j$. For example, $\cor_{G_{12}}^{G_{1}}$ maps the 2-cocycle $\om_{12}^{(2)}$ to a 2-cocycle satisfying
\begin{align}
    \cor_{G_{12}}^{G_1}\om_{12}^{(2)}(
    (g_1,R^{n_1},M_y^{m_1}),
    (g_2,R^{n_2},M_y^{m_2})
    ) =
    \prod_{k=0}^{3} 
        \om^{(2)}_{12}((\rho_{R^k} (g_1),M^{m_1}_y),(\rho_{R^{\ell}}(g_2),M_y^{m_2})),
\end{align}
where ${\ell\equiv (-1)^{m_1} (k+n_1)\bmod 4}$. $\cor_{G_{12}}^{G_1}\om_{12}^{(2)}$ is a product of the four 2-cocycles, which are the respective SPT data of the 1-cells ${[12]}$, ${R[12]}$, ${R^2[12]}$, and ${R^3[12]}$ folded onto each other. 
The fact that $\ell$ in the group element $\rho_{R^{\ell}}(g_2)$ depends on $k$, which was not the case in~\eqref{1dreflectionCorcochain} and~\eqref{2dreflectionCorcochain}, arises because $G_{12}$ is a non-normal subgroup of $G_1$.
The cohomology class of this 2-cocycle describes the total inflow anomaly at the 0-cell $[1]$ from the four 1-cells $[12]$, $R[12]$, $R^2[12]$, and $R^3[12]$. There is also the inflow anomaly at $[1]$ arising from the four 1-cells $[13]$, $R[13]$, $R^2[13]$, and $R^3[13]$ given by $\cor_{G_{13}}^{G_1}[\om_{13}^{(2)}]$. Taking into account the relative orientation of $[12]$ and $[13]$ with respect to $[1]$, the total inflow anomaly at $[1]$ is ${\cor_{G_{12}}^{G_1}[\om_{12}^{(2)}]^{-1}\, \cor_{G_{13}}^{G_1}[\om_{13}^{(2)}]^{-1}}$. Repeating this for all three of the 0-cells in the fundamental domain, we find
\begin{equation}
\begin{aligned}
    &d^1_{1,1}\left([\om^{(2)}_{12}], [\om^{(2)}_{23}], [\om^{(2)}_{13}]\right) = \bigg(
    \cor_{G_{12}}^{G_1}[\om_{12}^{(2)}]^{-1}\, \cor_{G_{13}}^{G_1}[\om_{13}^{(2)}]^{-1},
    \,
    \cor_{G_{12}}^{G_2}[\om_{12}^{(2)}] \,\cor_{G_{23}}^{G_2}[\om_{23}^{(2)}]^{-1},\\
    &\hspace{300pt}
    \cor_{G_{23}}^{G_3}[\om_{23}^{(2)}]\, \cor_{G_{13}}^{G_3}[\om_{13}^{(2)}]
    ~~\bigg).
\end{aligned}
\end{equation}

\section{Group cohomology and (co)restriction maps }\label{cohomologyApp}

In this Appendix, we review group cohomology and restriction and corestriction maps for group cohomology. We will prioritize working explicitly with cochains over abstract arguments. These explicit formulas are helpful to make the physical interpretation of restriction and corestriction maps in Appendix~\ref{d1 formula app} more transparent. The reader is referred to Ref.~\cite[Chapter II]{Weiss1969} for a more thorough introduction.

\subsection{Group cohomology}

Let us begin by recalling basic definitions of group cohomology. Consider a group $G$ and a left $G$-module $A$. As a $G$‑module, $A$ is an abelian group equipped with a left action of $G$. We denote the action of ${g\in G}$ on ${a\in A}$ by ${g\triangleright a\in A}$. It satisfies\footnote{In this Appendix, we use additive notation for $A$. In the main text, however, the underlying abelian group of $A$ is always $\Uone$ and we use multiplicative notation.}
\begin{equation}
    g\triangleright (a_1 + a_2) = g\triangleright a_1 + g\triangleright a_2,\qquad
    1 \triangleright a = a,\qquad gh\triangleright a = g\triangleright (h\triangleright a).
\end{equation}
Given $G$ and $A$, we can define the group of $n$-cochains\footnote{Strictly speaking, what we call cochains in this Appendix are inhomogeneous cochains.}
\begin{equation}
    \cC^n(G,A) = \mathrm{Map}(G^n,A).
\end{equation}
An element ${\om\in \cC^n(G,A)}$ is called an $n$-cochain, and is a map ${\om\colon G^n \to A}$. Therefore, $\om$ maps ${(g_1,\cdots, g_n)\in G^n}$ to $\om(g_1, \cdots, g_n)\in A$, which can be added together and acted on by ${g\in G}$:
\begin{align}
    &\om(h_1,\cdots, h_n) + \om(g_1,\cdots, g_n)\in A,\\
    &g\triangleright \om(g_1,\cdots, g_n)\in A.
\end{align}
The group structure on $\cC^n(G, A)$ is inherited from $A$. In particular, the group multiplication of two $n$-cochains ${\om_1,\om_2\in \cC^n(G,A)}$ is denoted by ${\om_1+\om_2\in \cC^n(G,A)}$ and satisfies
\begin{equation}
    (\om_1+\om_2)(g_1,\cdots, g_n) = \om_1(g_1,\cdots, g_n)+ \om_2(g_1,\cdots, g_n).
\end{equation}
Therefore, $\cC^n(G, A)$ is an abelian group. 

The groups of $n$-cochains fit into the cochain complex
\begin{equation}\label{groupcochaincomplex}
    \cdots\xrightarrow{\del^{n-1}} \cC^n(G,A)
    \xrightarrow{\del^{n}} \cC^{n+1}(G,A)
    \xrightarrow{\del^{n+1}} \cdots.
\end{equation}
As part of the definition of a cochain complex, the coboundary maps ${\del^n\colon \cC^n(G,A)\to \cC^{n+1}(G,A)}$ are group homomorphisms satisfying ${\del^{n+1}\circ \del^n = 0}$ (i.e., ${\del^{n+1}\circ \del^n}$ maps every n-cochain to the identity element of ${\cC^{n+2}(G,A)}$). Given an $n$-cochain $\om$, the $(n+1)$-cochain $\del^n\om$ maps ${(g_1,\cdots g_{n+1})}$ to
\begin{equation}
    \begin{aligned}
        \del^n\om(g_1,\cdots, &~g_{n+1}) = g_1\triangleright \om(g_2,\cdots , g_{n+1})
        +\sum_{k=1}^n (-1)^k \om(g_1,\cdots, g_{k}g_{k+1}, \cdots, g_{n+1})\\
        &\hspace{160pt}+(-1)^{n+1} \om(g_1,\cdots, g_{n}).
    \end{aligned}
\end{equation}
It is straightforward to check using this explicit expression that ${\del^{n+1}\del^n\om(g_1,\cdots, g_{n+2}) = 0}$ as expected.

The kernel $\ker(\del^n)$ of $\del^n$ and the image $\mathrm{im}(\del^{n-1})$ of $\del^{n-1}$ are both subgroups of $\cC^n(G,A)$. In fact, ${\mathrm{im}(\del^{n-1})\subset \ker(\del^n)}$ because ${\del^{n}\circ\del^{n-1} = 0}$. An $n$-cochain ${\om\in\ker(\del^n)}$ satisfies ${\del^n\om = 0}$ and is called an $n$-cocycle. On the other hand, $n$-cochains in $\mathrm{im}(\del^{n-1})$ are called $n$-coboundaries and are each equal to ${\del^{n-1}\la}$ for some ${(n-1)}$-cochain $\la$. 

The cohomology of the cochain complex~\eqref{groupcochaincomplex} is called the group cohomology of $G$ with coefficients in $A$. In particular, the degree $n$ group cohomology is defined as the quotient group
\begin{equation}
    \cH^n(G,A) = \ker(\del^n)/\mathrm{im}(\del^{n-1}).
\end{equation}
The elements of ${\cH^n(G,A)}$ are cohomology classes $[\om]$ for some $n$-cocycle $\om$ with ${[\om] = [\om + \del^{n-1}\la]}$ for any ${(n-1)}$-cochain $\la$. The group law of $\cH^n(G,A)$ is ${[\om_1]+[\om_2] = [\om_1+\om_2]}$.

An $n$-cochain ${\om\in\cC^n(G,A)}$ is called a normalized cochain if ${\om(g_1,\cdots, g_n) = 0}$ whenever some ${g_i = 1}$. For an $n$-cocycle $\om\in \ker(\del^n)$, there is always some ${n}$-coboundary ${\del^{n-1}f}$ such that ${\om - \del^{n-1}f}$ is a normalized cochain. The ${(n-1)}$-cochain ${f = \sum_{j=1}^n (-1)^{j-1} f_j}$ is defined recursively: for ${(n-1)}$-cochains $f_j$ and $n$-cocycles $\om_j$ with ${\om_0 \equiv \om}$, 
\begin{equation}
    \begin{aligned}
        f_j(g_1, \cdots, g_{n-1}) & = \om_{j-1}(g_1, \cdots, g_{j-1}, 1, g_j, \cdots, g_{n-1}), \\
        \om_j & =\om_{j-1}+(-1)^{j} \del^{n-1} f_j .
    \end{aligned}    
\end{equation}
The normalized cochain ${\om - \del^{n-1}f}$ is precisely $\om_n$. Hence, every class ${[\om]\in \cH^n(G,A)}$ admits a normalized $n$-cocycle representative.

\subsection{(Co)restriction maps}\label{corestrictionmapsAppSubSec}

Consider the cohomology groups $\cH^n(G,A)$ for the group $G$ and $G$-module $A$. 
Suppose there is another group $G'$ and $G'$‑module $A'$; consider their cohomology groups $\cH^n(G',A')$. Given homomorphisms relating ${G,A}$ and ${G',A'}$, it is possible that they induce maps between $\cH^n(G,A)$ and $\cH^n(G',A')$. In this section, we review two such maps, known as restriction and corestriction maps. For both of these maps, $G' \equiv H$ will be a subgroup of $G$, and $A'\equiv A$ will be the $H$-module obtained by restricting the $G$‑action on $A$ to $H$.

\subsubsection*{Restriction maps}

The restriction map is denoted by $\res^G_H$ and is a map
\begin{equation}\label{restrictionmap}
    \res^G_H \colon \cC^n(G,A)\to \cC^n(H,A).
\end{equation}
It is constructed using the inclusion homomorphism
\begin{equation}
    \iota\colon H \hookrightarrow G.
\end{equation}
In particular, $\res^G_H$ maps an $n$-cochain ${\om\in \cC^n(G,A)}$ to the $n$-cochain ${\res^G_H\om = \iota^*\om = \om\circ\iota}$ using the pullback $\iota^*$ of $\iota$. The $n$-cochain $\res^G_H\om\in \cC^n(H,A)$ satisfies
\begin{equation}\label{ResGHcochainApp}
    \res^G_H\om(h_1,\cdots, h_n) = \om(\iota(h_1),\cdots,\iota(h_n)).
\end{equation}
From this expression, it is clear that $\res^G_H$ is a group homomorphism ${\cC^n(G,A)\to\cC^n(H,A)}$. Furthermore, for a subgroup ${K\subset H}$, it satisfies
\begin{equation}
    \res^H_K\circ \res^G_H = \res^G_K.
\end{equation}

The explicit cochain expression~\eqref{ResGHcochainApp} also makes it clear that the coboundary homomorphisms commute with $\res^G_H$. That is, letting ${\del^n_G\colon \cC^n(G,A)\to \cC^{n+1}(G,A)}$ and ${\del^n_H\colon \cC^n(H,A)\to \cC^{n+1}(H,A)}$, they satisfy
\begin{equation}\label{delResCommutr}
    \del_H^n \circ \res^G_H = \res^G_H \circ \del^n_G.
\end{equation}
This follows from the fact that $\del^n_G$ is computed using the $G$ group law and its action on $A$, which agree inside $H$. Because of~\eqref{delResCommutr}, the restriction map descends to cohomology:
\begin{equation}
    \res^G_H \colon \cH^n(G,A)\to \cH^n(H,A).
\end{equation}
The action of $\res^G_H$ on a cohomology class ${[\om]\in \cH^n(G,A)}$ is ${\res^G_H[\om] = [\res^G_H\om]\in \cH^n(H,A)}$.

\subsubsection*{Corestriction maps}

The corestriction map goes in the opposite direction of the restriction map~\eqref{restrictionmap}, and is denoted by
\begin{equation}\label{corestrictionmap}
    \cor^G_H \colon \cC^n(H,A)\to \cC^n(G,A).
\end{equation}
It is defined for any subgroup ${H\subset G}$ with finite index $[G:H]$ (i.e., the number of cosets of $H$ in $G$ is finite). 

To construct the corestriction map for an explicit cochain, we need a way to naturally turn an element of $G$ into an element of ${H\subset G}$. This is done using the right cosets
\begin{equation}
    H \backslash G = \{H\,g\mid g \in G\}.
\end{equation}
In particular, a right transversal of ${H\backslash G}$, which is a set of representatives ${\overline{c}\in G}$ for each right coset ${c\in H\backslash G}$, i.e., ${H\bar{c} = c}$. Given this transversal, for each coset $c$ we define a map ${\ga_c\colon G\to H}$ given by
\begin{equation}
    \ga_c(g) = \overline{c} g \,(\overline{cg})^{-1}.
\end{equation}
The reason why ${\ga_c(g)\in H}$ for all $c$ and $g$ is that $\bar{c} g$ and $\overline{cg}$ are both in $cg$, so ${H\, \overline{c}g = H\, \overline{cg}}$ and ${\overline{c}g = h\, \overline{cg}}$ for some ${h\in H}$. Using $\ga_c$ and choosing a transversal, the corestriction map sends ${\om\in \cC^n(H,A)}$ to
\begin{equation}\label{coresMapImageApp}
    \cor^G_H\om(g_1,\cdots, g_n) = \sum_{c\in H\backslash G}\overline{c}^{-1}\triangleright \om(\ga_{s_1}(g_1),\cdots,\ga_{s_{n}}(g_n)),
\end{equation}
where the shorthand ${s_i = c g_1 g_2\cdots g_{i-1}}$ with ${s_1 = c}$ for each $c$. Clearly,~\eqref{corestrictionmap} is a group homomorphism.

The expression~\eqref{coresMapImageApp} is cumbersome but explicit. Using it, one can show that
\begin{equation}\label{delcorCommutr}
    \del_G^n \circ \cor^G_H = \cor^G_H \circ \del^n_H.
\end{equation}
Therefore, the corestriction map descends to cohomology:
\begin{align}
        \cor^G_H &\colon \cH^n(H,A)\to \cH^n(G,A).
\end{align}
In particular, $\cor^G_H$ transforms a cohomology class ${[\om]\in\cH^n(H,A)}$ to ${ \cor^G_H[\om] = [ \cor^G_H\om]\in \cH^n(G,A)}$. As a map of group cohomology, it does not depend on the choice of transversal for ${H\backslash G}$, and, for a subgroup ${K\subset H}$ with ${[G:K]}$ finite, it satisfies
\begin{equation}
    \cor^G_H \circ \cor^H_K = \cor^G_K.
\end{equation}
That is, ${[\cor^G_H \cor^H_K\om] = [\cor^G_K\om]}$.

Throughout this paper, we are often interested in semi-direct product groups $G$ and trivial modules ${A = \Uone}$. Therefore, let us mention how the corestriction map specializes for this case. Consider the corestriction map $\cor_{G\rtimes K}^{G\rtimes H}$ where ${K\subset H}$ and the action of $H$ on $G$ is given by ${\rho\colon H\to\Aut(G)}$. The action of $K$ on $G$ is described by restricting $\rho$ to $K$. $\cor_{G\rtimes K}^{G\rtimes H}$ is a map from $\cH^{n}(G\rtimes K,\Uone)$ to $\cH^{n}(G\rtimes H,\Uone)$. At the cochain level, the image of ${\cor_{G\rtimes K}^{G\rtimes H}}$ follows from the general expression~\eqref{coresMapImageApp}. Using multiplicative notation for $\Uone$, we find that ${\cor_{G\rtimes K}^{G\rtimes H}}$ maps ${\om\in \cC^{n}(G\rtimes K,\Uone)}$ to
\begin{align}\label{general cor map semi direc prodcs}
    \cor_{G\rtimes K}^{G\rtimes H} \,\om( 
    \cdots,
    (g_{i},h_{i}),\,
    \cdots)
    =
    \prod_{c\in K \backslash H}
    \om (\, 
    \cdots,
    (\rho_{\,\overline{s}_i}(g_i),\ga_{s_i}(h_i))
    ,\cdots
    \,),
\end{align}
where the shorthand ${s_i = c h_1 h_2\cdots h_{i-1}}$ with ${s_1 = c}$ for each $c$ and ${\ga_c(h) = \overline{c} h \,(\overline{ch})^{-1}\in K}$.

\subsubsection*{Mixed properties}

The restriction and corestriction maps on group cohomology satisfy various important additional properties. Here, we give an overview of two that involve a mixture of both restriction and corestriction maps. 

The first arises when applying the restriction map followed by the corestriction map. That is,
\begin{equation}
    \cor^G_H\circ \res^G_H\colon \cH^n(G,A) \to \cH^n(G,A).
\end{equation}
Using the explicit cochain expressions above, one can show that
\begin{equation}\label{correseqAPP}
    \cor^G_H\circ \res^G_H = [G:H]\,\mathrm{id}_{\cH^n(G,A)} .
\end{equation}
Therefore, each cohomology class ${[\om]\in\cH^n(G,A)}$ satisfies ${\cor^G_H\res^G_H [\om] = [G:H]\,[\om]}$. Importantly, the equation~\eqref{correseqAPP} 
holds only in cohomology; at the cochain level, it may fail by a coboundary.

The second is applying the corestriction map and then the restriction map. In particular, given subgroups ${K,H \subset G}$, there is a map
\begin{equation}
    \res^G_K\circ\cor^G_H\colon \cH^n(H,A) \to \cH^n(K,A).
\end{equation}
Choosing representatives $\overline{d}$ for each double coset $d\in K\backslash G/H$, this map can be written as
\begin{equation}\label{doubleCosetFormula}
    \res^G_K \circ \cor^G_H 
    = 
    \sum_{d\in K\backslash G/H} 
    \cor^K_{K\cap\, \overline{d} H\overline{d}^{-1}} 
    \circ 
    \mathrm{c}_{\,\overline{d}} 
    \circ 
    \res^H_{H\cap \overline{d}^{-1}K\overline{d}},
\end{equation}
where $\mathrm{c}_{\overline{d}}$ is the conjugation isomorphism. For a subgroup ${E\subset G}$, the conjugation isomorphism with ${g\in G}$ maps ${\mathrm{c}_{g}\colon \cH^n(E,A) \to \cH^n(g E g^{-1},A)}$ and acts on a $n$-cochain as
\begin{equation}
    \mathrm{c}_{g}\om(e_1,\cdots, e_n) = g\triangleright \om(g^{-1} e_1 g, \cdots, g^{-1} e_n g).
\end{equation}
In~\eqref{doubleCosetFormula}, $\mathrm{c}_{\overline{d}}$ is the conjugation isomorphism for $E = H\cap \overline{d}^{-1}K\overline{d}$. The right‑hand side of~\eqref{doubleCosetFormula} is independent of the choice of representatives $\overline{d}$ at the level of cohomology. When ${K}$ is a normal subgroup of $H$ and $G$, the double-coset formula~\eqref{doubleCosetFormula} simplifies to
\begin{equation}
    \res^G_K \circ \cor^G_H 
    = 
    \sum_{gH\in G/H} 
    \mathrm{c}_{g} 
    \circ 
    \res^H_{K}
    .
\end{equation}
Furthermore, if ${K=H}$ and $H$ is a normal subgroup of $G$, then it further simplifies to
\begin{equation}\label{res cor normalizer}
    \res^G_H \circ \cor^G_H 
    = N_{G/H},
\end{equation}
where the norm map ${N_{G/H} \equiv \sum_{gH\in G/H}\mathrm{c}_g}$.

\section{Real-space constructions for modulated SPTs}\label{def net mod spt app}

In this Appendix, we review the real-space construction from \Rf{B250806604} of general modulated SPTs. 
The real-space construction starts with a cellulation ${\La}$ of a topological space $X_d$, which dictates which spatial symmetries $G_\mathrm{s}$ can protect an SPT by requiring that $G_\mathrm{s}$ is a subgroup of the automorphism group of $\La$. In addition to this spatial symmetry group, we also specify an internal symmetry group $G$.\footnote{The group $G$ can be the trivial group.} If the total symmetry $G_\mathrm{tot}$ is anomaly-free, then the real-space construction can produce a ${(d+1)}$D $G_\mathrm{tot}$-SPT. Since we assume $G$ is a modulated symmetry here, the total symmetry group is ${G_{\mathrm{tot}} \cong G \rtimes G_\mathrm{s}}$, where the action of $G_\mathrm{s}$ on $G$ is specified by a group homomorphism ${\rho\colon G_\mathrm{s} \to \Aut(G)}$.

A central ingredient in the real-space construction is the collection of ``little groups'' of the total symmetry group $G_{\mathrm{tot}}$. Each ${n}$-cell $c_{n}$ of $\La$ has a $c_{n}$-stabilizer subgroup of $G_\mathrm{tot}$, which we denote by $G_{c_{n}}$ and call the little group at $c_n$. $G_{c_{n}}$ is defined as the subgroup of $G_\mathrm{tot}$ that transforms every point on $c_{n}$ to itself. Explicitly, it is given by
\begin{equation}
    G_{c_n} = \{ g\in G_\mathrm{tot} \mid g\triangleright x = x \quad \forall x \in c_n\},
\end{equation}
where ${g\triangleright x}$ denotes the action of $g$ on a point ${x\in X_d}$.

For general $G_\mathrm{tot}$, the little groups $G_{c_n}$ satisfy several important properties:
\begin{enumerate}
    \item The little group of every $d$-cell is just the internal symmetry group: ${G_{c_{d}}\cong G}$.
    \item For every spatial transformation ${s\in G_{\mathrm{s}}}$, the little groups at $c_n$ and ${s\triangleright c_n}$ are isomorphic. In particular, ${G_{s\triangleright c_n} = s\,G_{c_n}\,s^{-1} \cong G_{c_n}}$.
    \item The little group at $c_n$ is a subgroup of each little group on the boundary of $c_n$. That is, ${G_{c_{n}}\subset G_{c_{n-1}}}$ if ${c_{n-1}\in \pp_n c_n}$. 
    \item The little groups at two $n$-cells that share a boundary are not necessarily isomorphic. That is, for an ${(n-1)}$-cell $c_{n-1}$ in $\pp_n c_n$ and $\pp_n \t{c}_n$, the little groups  ${G_{c_{n}}, G_{\t{c}_{n}}\subset G_{c_{n-1}}}$ but ${G_{c_{n}}\not\cong G_{\t{c}_{n}}}$ in general.
\end{enumerate}
Since we assume that $G$ is a modulated symmetry, each little group 
\begin{equation}\label{littleGroupSplit}
   G_{c_n} \cong G \rtimes P_{c_n}, 
\end{equation}
where ${P_{c_n}\subset G_\mathrm{s}}$. We will denote the group elements of $G_{c_n}$ by ${(g,p)\in G\rtimes P_{c_n} \cong G_{c_n}}$.

When $G_\mathrm{s}$ is a space group of a crystalline lattice, there is a canonical cellulation $\La$ formed by the Wyckoff positions of $G_\mathrm{s}$.\footnote{The Wyckoff positions for each space group are fully tabulated in physical dimensions. See, for instance, \Rf{APMC2006}.} In particular, the $n$-cells of this cellulation are the $n$-dimensional Wyckoff patches of $\La$, which are $n$-dimensional subspace of $X_d$ whose points have the same site-symmetry group.\footnote{The site-symmetry group of a point $x$ in space is the subgroup of $G_\mathrm{s}$ that leaves $x$ invariant.} Each $d$-cell of this cellulation $\La$ is a fundamental domain of $G_\mathrm{s}$ in space. Therefore, the entire data specifying a real-space construction is specified by the data for a single $d$-cell and the ${n<d}$ cells on its boundary.

Given the cellulation $\La$ and total symmetry group $G_\mathrm{tot}$, the real-space construction for a $G_\mathrm{tot}$-SPT begins by assigning an ${(n+1)}$D SPT to each $n$-cell of $\La$. From the viewpoint of an $n$-cell $c_n$, the total symmetry group is an internal $G_{c_n}$ symmetry, so the decoration of $c_n$ is specified by a cohomology class\footnote{Here we ignore beyond-group cohomology SPTs for simplicity.}
\begin{equation}\label{cSPT}
    [\om_{c_n}^{(n+1)}] \in \cH^{n+1}(G_{c_n}, \Uone).
\end{equation}
Importantly, the group cohomology $\cH^{n+1}(G_{c_n}, \Uone)$ can be nontrivial even when the internal symmetry group $G$ is trivial.

The real-space construction is powerful because the data specifying a $G_\mathrm{tot}$-SPT reduces to the data for SPTs of internal symmetries. To be compatible with the crystalline symmetries, this data must satisfy the covariance condition
\begin{equation}\label{cSPTcond}
    \hat{\rho}_{s}^{*}[\om^{(n+1)}_{c_n}]
    =
    [\om^{(n+1)}_{s\triangleright c_n}]^{o_{c_n}(s)},
\end{equation}
where the group isomorphism 
\begin{equation}
    \begin{aligned}
        \hat{\rho}_s\colon G_{c_n}&\to G_{s\triangleright c_n}\\
        (g,p)&\mapsto (\rho_s(g),sps^{-1}),
    \end{aligned}
\end{equation}
and ${o_{c_n}(s) = +1}$ (${-1}$) if $s$ is orientation preserving (reversing) with respect to $c_n$.

A real-space construction involving $(n+1)$D SPTs on $n$-cells that are specified by~\eqref{cSPT} and satisfy~\eqref{cSPTcond} produces a phase with $G_\mathrm{tot}$ symmetry. However, this phase is not necessarily an SPT. Indeed, the $G_\mathrm{tot}$ symmetry may be anomalously realized at a junction of constituent SPTs due to anomaly-inflow. In order for the phase to be an SPT, the inflow anomaly of ${(n+2)}$D SPTs at every ${n=0,1,\cdots, d-1}$ cell must cancel. Physically, this ensures that any zero modes, gaplessness, or topological orders within the real-space construction can be trivialized using symmetric perturbations. 

Let us consider the case where the crystalline symmetry is lattice translation symmetry. As in the main text, we denote the translation group by ${G_\mathrm{s} = \mathbb{T}_d}$ and its elements as ${\bm{v}\in\mathbb{T}_d}$.

For lattice translations, the little groups~\eqref{littleGroupSplit} simplify to ${G_{c_n} = G}$ for all cells $c_n$ and the covariance condition~\eqref{cSPTcond} becomes
\begin{equation}\label{translation SPTcond}
    \rho_{\bm{v}}^{*}[\om^{(n+1)}_{c_n}]
    =
    [\om^{(n+1)}_{\bm{v}\triangleright c_n}].
\end{equation}
The inflow anomaly at an ${(n-1)}$-cell arises from the ${(n+1)}$D constituent SPTs on neighboring ${n}$-cells. Including the orientation structure of $\La$, the total inflow anomaly at $c_{n-1}$ is ${\prod_{c_{n}}[\om^{(n+1)}_{c_{n}}]^{\si_{c_{n-1}}(\pp c_n)}}$, where ${\si_{c_{n-1}}(\pp c_n)}$ is the coefficient of $c_{n-1}$ in $\pp_n c_{n}$. Therefore, for the real-space construction to yield an SPT, the cellular SPT data must satisfy
\begin{equation}\label{app anom canc def net spt}
    \prod_{c_{n}}[\om^{(n+1)}_{c_{n}}]^{\si_{c_{n-1}}(\pp c_n)} = [1],
\end{equation}
for every ${(n-1})$-cell with ${1\leq n \leq d}$. If the cellular data satisfies~\eqref{translation SPTcond} and~\eqref{app anom canc def net spt}, then it corresponds to a translation symmetric modulated SPT. If it satisfies~\eqref{app anom canc def net spt} but not~\eqref{translation SPTcond}, then it still corresponds to an SPT but one that explicitly breaks lattice translation symmetry.

Using the cellular chain description from Sections~\ref{CellComplexStratAnomSec} and~\ref{modulated strat sym subsection}, the SPT data can be packaged as the ${\cH^{n+1}(G,\Uone)}$-valued cellular $n$-chain
${\om^{(n+1)}_{n\text{-chain}} := \sum_{c_n} [\om^{(n+1)}_{c_{n}}]\, c_n}$. In terms of these $n$-chains, the conditions~\eqref{translation SPTcond} and~\eqref{app anom canc def net spt} are, respectively,
\begin{align}
    \bm{v}\triangleright \om^{(n+1)}_{n\text{-chain}} &= \om^{(n+1)}_{n\text{-chain}},\\
    \pp_{n}\om^{(n+1)}_{n\text{-chain}} &= 0.
\end{align}
Therefore, $\{\om^{(n+1)}_{n\text{-chain}}\}_{n=0}^{d}$ corresponds to a modulated SPT if each $\om^{(n+1)}_{n\text{-chain}}$ is a $\mathbb{T}_d$-invariant $n$-cycle.

\section{Exponential SPT construction}\label{ExpSPTModelDerivation}

In this Appendix, we present a construction of the stabilizer code model~\eqref{exp sym spt stab code} from the main text. We do so by gauging. In particular, the SPT of the original model's symmetry will correspond to a spontaneous symmetry breaking phase (SSB) of the gauged model. The latter is much easier to construct a model for. Then, by gauging a discrete symmetry of the gauged model, we return to a model in an SPT phase of the original exponential symmetry. This relation between SPT and SSB phases goes back to the Kennedy-Tasaki transformation~\cite{Kennedy:1992ifl, Kennedy:1992tke}, and has been used to construct SPT models protected by various symmetries.

Recall from the main text that the exponential symmetry operators are
\begin{equation}\label{app: double exp sym ops2}
    U_{(1,0)} = \prod_{j} (X_j)^{a^{j}},\qquad U_{(0,1)} = \prod_{j} (Z^\dag_j \t{Z}_j)^{b^j},
\end{equation}
where $X_j$, $Z_j$, $\t{X}_j$, and $\t{Z}_j$ are $\Z_p$ qudit operators. A general investigation into gauging modulated symmetries was carried out in \Rf{PDL240612962}. Gauging the $\Z_p$ exponential symmetry generated by $U_{(1,0)}$ can be implemented through the gauging map~\cite{HW230201207, PDL240612962}
\begin{equation}\label{expKT}
\begin{aligned}
    X_j &\to Z_{j-1}^{a^{-1}}Z_j^\dag,
    \qquad
    Z_j Z_{j+1}^{-a^{-1}} \to X_j,
    \\
    \t{X}_j &\to \t{X}_j,
    \qquad\hspace{48.5pt}
    \t{Z}_j \to \t{Z}_j.
\end{aligned}
\end{equation}
This maps each quantum lattice system with the ${\Z_p\times \Z_p}$ exponential symmetry~\eqref{app: double exp sym ops2} to a different lattice system with dual ${\Z_p\times \Z_p}$ symmetry generated by
\begin{equation}\label{dual double exp sym ops}
    U^\vee_{(1,0)} = \prod_{j} (X_j)^{a^{-j}},\qquad 
    U^\vee_{(0,1)} =  \prod_{j} (X_j^{-ab(ab-1)^{-1}} \t{Z}_j)^{b^j}.
\end{equation}
Here, $U^\vee_{(1,0)}$ generates the dual symmetry that arises after gauging~\cite{PDL240612962} and $U^\vee_{(0,1)}$ is image of $U_{(0,1)}$ under the gauging map. The expression for $U^\vee_{(0,1)}$ is most easily found using that the symmetry operator $U_{(0,1)}$ can be written as ${U_{(0,1)} = \prod_{j} [(Z_jZ_{j+1}^{-a^{-1}})^{-ab(ab-1)^{-1}} \t{Z}_j]^{b^j}}$.

The SPT phase for the exponential symmetry~\eqref{app: double exp sym ops2} corresponds to the spontaneous symmetry breaking phase of~\eqref{dual double exp sym ops} with $U^\vee_{(0,1)}$ unbroken and $U^\vee_{(1,0)}$ completely broken. A model realizing this SSB pattern is 
\begin{equation}\label{HdualexpApp}
    H^\vee = -\sum_{j} \left(\t{A}_j\, \t{A}^{-a}_{j + 1}+ \t{B}_j + \mathrm{h.c.}\right),
\end{equation}
where
\begin{equation}
    \t{A}_j = Z_j^{\,(ab-1)}\t{X}^{-ab}_j , \qquad \t{B}_j = X^{\dag}_j \t{Z}^{(
    1-a^{-1}b^{-1})}_j.
\end{equation}
The operators $\t{A}_{j_1}$ and $\t{B}_{j_2}$ commute for all ${j_1}$ and ${j_2}$. Therefore, the ground state subspace satisfies ${\t{A}_j\t{A}_{j+1}^{-a} = \t{B}_j = 1}$ for all $j$. This constraint causes ${U^\vee_{(0,1)} = 1}$ in the ground state subspace because ${U^\vee_{(0,1)} = \prod_{j} \t{B}_j^{ab\,(ab-1)^{-1}\,b^{j}}}$. Therefore, the $\Z_p$ symmetry generated by $U^\vee_{(0,1)}$ is not spontaneously broken. On the other hand, because ${U_{(1,0)}^\vee \t{A}^{-(ab-1)^{-1}a^j}_j U_{(1,0)}^{\vee\,\dag} = \ee^{\frac{2\pi\ii}p} \t{A}^{-(ab-1)^{-1}a^{j}}_j}$, the constraint ${\t{A}_j\t{A}_{j+1}^{-a} = 1}$ causes the $\Z_p$ symmetry generated by $U^\vee_{(1,0)}$ to completely spontaneously break. The Hamiltonian $H^\vee$, therefore, realizes the desired SSB pattern.

To find the corresponding exactly solvable model for the SPT protected by~\eqref{app: double exp sym ops2}, we now gauge the $\Z_p$ exponential symmetry generated by $U^\vee_{(1,0)}$ to return to a model with the original exponential symmetry~\eqref{app: double exp sym ops2}. This is done using the gauging map
\begin{equation}
\begin{aligned}
    X_j &\to Z_{j-1}^{a}Z_j^\dag,
    \qquad
    Z_j Z_{j+1}^{-a} \to X_j,
    \\
    \t{X}_j &\to \t{X}^{a^{-1}b^{-1}}_j,
    \qquad\hspace{17pt}
    \t{Z}_j \to \t{Z}^{ab}_j,
\end{aligned}
\end{equation}
which is formed using the gauging map~\eqref{expKT} with $a$ replaced by $a^{-1}$ followed by the unitary transformation $(\t{X}_j,\t{Z}_j) \to (\t{X}^{a^{-1}b^{-1}}_j,\t{Z}^{ab}_j)$. The dual symmetry of~\eqref{dual double exp sym ops} after gauging is precisely the original symmetry~\eqref{app: double exp sym ops2}. Furthermore, this maps the Hamiltonian~\eqref{HdualexpApp} to
\begin{equation}
    H = -\sum_j \left(\t{X}^\dag_j\, X_j^{(ab-1)}\, \t{X}^{a}_{j+1} + Z_{j-1}^{-a} \, \t{Z}_j^{\,(ab-1)}\, Z_j + \mathrm{h.c.}\right) ,
\end{equation}
which is the same as the Hamiltonian~\eqref{exp sym spt stab code} considered in the main text.

\addcontentsline{toc}{section}{References}

\hypersetup{linkcolor=brn}

\bibliographystyle{ytphys}
\bibliography{local.bib} 

\end{document}

%% file: Preamble.tex
\usepackage{graphicx}
\usepackage{dcolumn}
\usepackage{bm}
\usepackage{multirow}

\usepackage{centernot}

\usepackage[normalem]{ulem}
\usepackage{cancel}
\usepackage[nosort]{cite}

\usepackage{amsmath}
\numberwithin{equation}{section}
\allowdisplaybreaks

\usepackage{amsthm}
\usepackage{amstext}
\usepackage{amssymb}
\usepackage{mathrsfs}
\usepackage{amsfonts}
\usepackage{amsbsy} 
\usepackage{tensor}
\usepackage{physics}

\usepackage{wasysym}

\usepackage{latexsym}
\usepackage[american]{babel}
\usepackage{bbm}
\usepackage[backref=page, colorlinks,citecolor=pastelblue,linkcolor=pastelblue,urlcolor=pastelblue,hyperindex]{hyperref}

\newcommand*{\figref}[2][]{%
  \hyperref[{#2}]{%
    \ref*{#2}%
    \ifx\\#1\\%
    \else
      \,#1%
    \fi
  }%
}

\usepackage[all,matrix,cmtip]{xy}

\usepackage{csquotes}
\MakeOuterQuote{"}

\usepackage[dvipsnames]{xcolor}

\definecolor{red}{rgb}{1,0,0}
\definecolor{blue}{rgb}{0,0,1}
\definecolor{dblue}{rgb}{0,0,0.4}
\definecolor{green}{rgb}{0,1,0}
\definecolor{black}{rgb}{0,0,0}
\definecolor{white}{rgb}{1,1,1}
\definecolor{niceBlue}{RGB}{20,10,237}
\definecolor{pastelblue}{RGB}{20,93,160}

\definecolor{brn}{rgb}{.8,.4,.0}
\definecolor{redo}{rgb}{1,.5,.0}
\definecolor{ddgrn}{rgb}{0,0.4,0}
\definecolor{dgrn}{rgb}{0,0.55,0}
\definecolor{dbl}{rgb}{0,0,0.5}

\usepackage[bbgreekl]{mathbbol}

\newcommand{\Z}{\mathbb{Z}}
\newcommand{\C}{\mathbb{C}}
\newcommand{\R}{\mathbb{R}}

\newcommand{\p}[1]{\prime\,}

\renewcommand{\t}[1]{\widetilde{#1}}

\newcommand{\ii}{\hspace{1pt}\mathrm{i}\hspace{1pt}}
\newcommand{\ee}{\hspace{1pt}\mathrm{e}}

\newcommand{\<}{\langle}
\renewcommand{\>}{\rangle}

\newcommand{\Rf}[1]{Ref.~\citenum{#1}}

\newcommand{\pp}{\partial}

\newcommand{\cor}{\mathrm{cor}}
\newcommand{\res}{\mathrm{res}}

\newcommand{\bpm}{\begin{pmatrix}}
\newcommand{\epm}{\end{pmatrix}}
\newcommand{\bmm}{\begin{matrix}}
\newcommand{\emm}{\end{matrix}}

\newcommand{\cC}{ {\cal C} }

\newcommand{\cH}{ {\cal H} }

\newcommand{\cO}{ {\cal O} }

\newcommand{\cZ}{ {\cal Z} } 

\usepackage{euscript}

\newcommand\scrH         {\mathscr{H}}

\newcommand\scrV         {\mathscr{V}}

\newcommand{\al}{\alpha} 
 
\newcommand{\del}{\delta}

\newcommand{\ga}{\gamma}

\newcommand{\la}{\lambda} 
\newcommand{\La}{\Lambda} 
\newcommand{\om}{\omega}

\newcommand{\si}{\sigma} 
\newcommand{\Si}{\Sigma}

\newcommand{\txti}[1]{\textit{#1}}

\newcommand{\Aut}{\mathrm{Aut}}

\newcommand{\Uone}{\mathrm{U}(1)}

\usepackage{tikz}
\usepackage{tikz-cd}
\usetikzlibrary{arrows}
\usetikzlibrary{intersections}
\usetikzlibrary{shapes.geometric}
\usetikzlibrary{decorations.pathmorphing, patterns,shapes}
\usetikzlibrary{decorations.markings}
\usetikzlibrary{calc}

\usepackage{enumerate}

\usepackage{hhline}

\usepackage{pifont}

\definecolor{anom}{HTML}{B1002C}
\definecolor{spt}{HTML}{0057B8}
\definecolor{topCell}{HTML}{f2f2f2}
\definecolor{site1}{HTML}{b3cde3}
\definecolor{site2}{HTML}{ccebc5}
\definecolor{site3}{HTML}{decbe4}
\definecolor{link1}{HTML}{ffffcc}
\definecolor{link2}{HTML}{fbb4ae}
\definecolor{link3}{HTML}{fed9a6}

\usepackage{xpatch}

\makeatletter
\patchcmd{\@ssect@ltx}
    {\addcontentsline{toc}{#1}{\protect\numberline{}#8}}
    {}
    {}
    {}
\makeatother